\PassOptionsToPackage{top=3cm,left=3cm,right=3cm,bottom=3cm}{geometry}
\documentclass[fleqn,11pt]{wlscirep}
\usepackage{main}

\makeatletter
\newcommand*{\rom}[1]{\expandafter\@slowromancap\romannumeral #1@}
\makeatother

\begin{document}

\doublespacing
\nolinenumbers

\title{\centering\LARGE\singlespacing{Generative Bayesian modeling to nowcast the effective reproduction number from line list data with missing symptom onset dates}}
% author list

\author[1,2*]{Adrian Lison}
\author[3]{Sam Abbott}
\author[4]{Jana Huisman}
\author[1,2]{Tanja Stadler}

\affil[1]{ETH Zurich, Department of Biosystems Science and Engineering, Zurich, Switzerland}
\affil[2]{SIB Swiss Institute of Bioinformatics, Lausanne, Switzerland}
\affil[3]{Centre for Mathematical Modelling of Infectious Diseases, London School of Hygiene and Tropical Medicine, London, UK}
\affil[4]{Massachusetts Institute of Technology, Department of Physics, Physics of Living Systems, Cambridge, MA, United States of America}
\affil[*]{Corresponding author: adrian.lison@bsse.ethz.ch}

\begin{abstract}\normalfont
The time-varying effective reproduction number $R_t$ is a widely used indicator of transmission dynamics during infectious disease outbreaks. Timely estimates of $R_t$ can be obtained from observations close to the original date of infection, such as the date of symptom onset. However, these data often have missing information and are subject to right truncation. Previous methods have addressed these problems independently by first imputing missing onset dates, then adjusting truncated case counts, and finally estimating the effective reproduction number. This stepwise approach makes it difficult to propagate uncertainty and can introduce subtle biases during real-time estimation due to the continued impact of assumptions made in previous steps. In this work, we integrate imputation, truncation adjustment, and $R_t$ estimation into a single generative Bayesian model, allowing direct joint inference of case counts and $R_t$ from line list data with missing symptom onset dates. We then use this framework to compare the performance of nowcasting approaches with different stepwise and generative components on synthetic line list data for multiple outbreak scenarios and across different epidemic phases. We find that under long reporting delays, intermediate smoothing, as is common practice in stepwise approaches, can bias nowcasts of case counts and $R_t$, which is avoided in a joint generative approach due to shared regularization of all model components. On incomplete line list data, a fully generative approach enables the quantification of uncertainty due to missing onset dates without the need for an initial multiple imputation step. In a real-world comparison using hospitalization line list data from the COVID-19 pandemic in Switzerland, we observe the same qualitative differences between approaches. The generative modeling components developed in this work have been integrated and further extended in the R package epinowcast, providing a flexible and interpretable tool for real-time surveillance.

\par
\end{abstract}

\flushbottom
\maketitle
\thispagestyle{empty}

\section*{Summary}
During an infectious disease outbreak, public health authorities require timely indicators of transmission dynamics, such as the effective reproduction number $R_t$. Since reporting data are delayed and often incomplete, statistical methods must be employed to obtain real-time estimates of case numbers and $R_t$. Existing methods involve separate steps for imputing missing data, adjusting for reporting delays, and estimating $R_t$. This stepwise approach impedes uncertainty quantification and can lead to inconsistent smoothing assumptions across steps. In this paper, we propose an alternative approach based on generative Bayesian modeling which integrates all steps into a single nowcasting model that can be directly fit to observed data. Using synthetic and real-world line list data, we demonstrate that the generative approach better captures uncertainty and avoids bias from inconsistent assumptions. The model components of our approach have been integrated into the R package epinowcast for easy use in practice.

\newpage

\sloppy
\raggedbottom

\section{Introduction}

The accurate and timely tracking of transmission dynamics is an important objective of infectious disease surveillance during epidemics. For instance, the effective reproduction number $R_t$ was used as a central indicator of transmission dynamics to guide and evaluate the public health response in many countries during the COVID-19 pandemic\cite{huisman_estimation_2022,abbott_estimating_2020,vegvari_commentary_2022,li_temporal_2020,banholzer_methodologies_2022}. $R_t$ can be estimated from clinical case data such as confirmed cases, hospitalizations or deaths under the assumption that the proportion of infections resulting in a respective case remains constant over time\cite{sherrattExploringSurveillanceData2021}. Tracking $R_t$ in near real-time is challenging, however. Because of substantial delays between infection and reporting of cases, case counts by date of report only provide a delayed and potentially distorted signal of transmission dynamics\cite{gostic_practical_2020}. Therefore, approaches to estimate $R_t$ are often based on case counts by date of symptom onset, which is ideally closer to the date of infection and independent of reporting delays\cite{cori_new_2013,huisman_estimation_2022}. Such data are typically obtained from a line list, which stores individual information about each case, \eg date of symptom onset, date of positive test, and date of report.

In this work, we address three important challenges that arise when tracking transmission dynamics from symptom onset data. First, information about the date of symptom onset is typically not available for all cases in the line list. Symptom onset information may be missing due to different reasons, including gaps in ascertainment or reporting, and asymptomatic infections\cite{costa-santos_covid-19_2021,pullano_underdetection_2021}. Since these factors can change over time, cases with missing symptom onset date cannot simply be excluded when inferring $R_t$ over time\cite{whiteReportingErrorsInfectious2010}. Second, in real-time surveillance, the delay between symptom onset and reporting means that cases with symptom onset close to the present may not have been reported in the line list yet, which is also known as right truncation\cite{van_de_kassteele_nowcasting_2019}. As a result, case counts by date of symptom onset will be subject to a downward bias towards the present, and a truncation adjustment is needed to correct for this bias\cite{kalbfleisch_statistical_2002,hohle_bayesian_2014,bastos_modelling_2019}. Third, when estimating $R_t$ from case counts by date of symptom onset, one needs to account for stochastic incubation periods and transmission noise to accurately relate to the time of infection\cite{cori_new_2013,parag_quantifying_2022}.

Previous methods have typically addressed the above challenges in three separate steps\cite{gunther_nowcasting_2021,salazar_near_2022,li_bayesian_2021}. Specifically, cases with missing symptom onset date are first completed via an imputation model fitted on the cases with known symptom onset date. Then, case counts by date of symptom onset are corrected for occurred-but-not-yet-reported cases via a truncation adjustment model fitted on previously observed cases and reporting delays. Finally, $R_t$ is inferred from the case counts by date of symptom onset via a non-parametric method for $R_t$ estimation that accounts for incubation periods and noise. In such a stepwise approach, estimates obtained in each step are treated as data during the subsequent step, meaning that information can only flow in one direction, such that knowledge and assumptions used in later steps cannot inform earlier steps. Moreover, the uncertainty involved in earlier steps is either neglected\cite{gunther_nowcasting_2021} or must be incorporated via a resampling scheme, where later steps are repeatedly applied to each sample from the current step\cite{salazar_near_2022}. The overall efficiency of the latter solution strongly depends on the computational effort required for downstream steps. For example, Li and White combined imputation, truncation adjustment, and $R_t$ estimation in a fully Bayesian approach, which is however still stepwise internally and required the use of simple estimators for truncation adjustment and $R_t$ estimation which can be directly computed as posterior predictive quantities\cite{li_bayesian_2021}.

In this paper, we explore an alternative to the existing stepwise approaches by integrating missing date imputation, truncation adjustment, and effective reproduction number estimation in a single generative model\cite{gelmanBayesianWorkflow2020}. In essence, we use Bayesian hierarchical modeling\cite{gelman_bayesian_2013} to describe how cases arise from an infection process and are reported with stochastic, time-varying delays and potentially missing symptom onset dates. This allows us to estimate the time series of symptom onsets and the effective reproduction number over time as parameters of a single, consistent nowcasting model, where the priors and model components for imputation, truncation adjustment, and $R_t$ estimation jointly inform each other. Moreover, when estimated in a fully Bayesian framework, the uncertainty from all components is naturally represented by the posterior distribution of model parameters\cite{zelner_accounting_2021}.

To compare the generative and stepwise approaches to nowcasting, we study their performance and qualitative behavior on synthetic line list data for multiple outbreak scenarios and across epidemic phases. We ensure comparability by using maximally similar modeling components in all approaches that reflect the assumptions of the synthetic data equally well. Thereby, we study sufficiently realistic nowcasting settings with time-varying reporting delays, weekday effects in reporting, and randomly missing symptom onset dates. We furthermore compare the generative and stepwise approaches on hospitalization line list data from the first and second wave of the COVID-19 epidemic in Switzerland. Based on our findings, we discuss the relative strengths and weaknesses of the different modeling approaches in real-time surveillance and implications for further model development in the future.

\section{Methods}
\begin{figure}[!tbh]
    \vspace{-1.5cm}
    \centering
    \includegraphics[width=\textwidth]{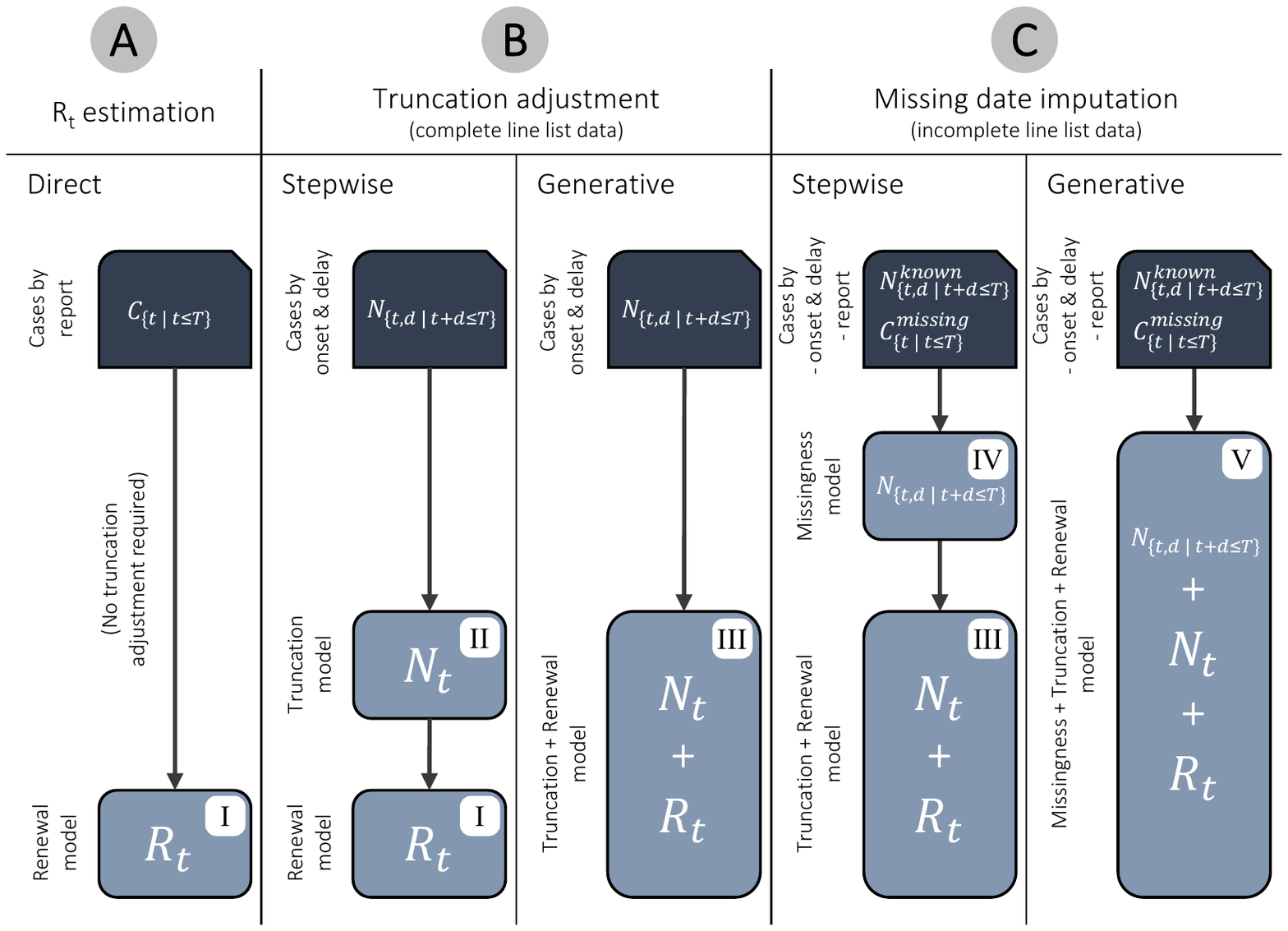}
    \caption{\textbf{Overview of different approaches for nowcasting the effective reproduction number $R_t$.} \textbf{(A)} Case counts by date of report are not biased by right truncation and can be used to estimate $R_t$ via a direct (no truncation adjustment) approach, assuming knowledge of the delay between infection and report. \textbf{(B)} Case counts by date of symptom onset are only delayed by the incubation period but subject to right truncation. The truncation can be adjusted for by using a stepwise (additional adjustment step) or generative (integrated truncation model) approach. \textbf{(C)} When line list data are incomplete, cases with missing onset date must be accounted for. This can be achieved using a stepwise (additional imputation step) or generative (integrated missingness model) approach. \textbf{(\rom{1}--\rom{5}}) Models used in the steps of the different approaches. To ensure comparability, the model components for $R_t$ estimation, truncation adjustment, and missing date imputation were designed to be maximally similar across all approaches (see \supp~\zref{appendix:model_specs} for full model definitions).}
    \label{fig:schematic}
\end{figure}

\Cref{fig:schematic} provides an overview of different stepwise and generative nowcasting approaches compared in this work. Our main nowcasting target is the so-called instantaneous effective reproduction number $R_t$, which describes the expected number of secondary infections caused by a primary infection if conditions remained as on date (index) $t$. We also nowcast the number of cases by date of symptom onset $N_t$, which can help to identify biases in $R_t$ estimation and is often of interest by itself. The data are individual cases in a line list with a date of report and a (potentially missing) date of symptom onset. These cases can be aggregated into different counts, such as $C_t$ (number of cases reported on date $t$) or $N_{t,d}$ (number of cases with symptom onset on date $t$ and a reporting delay of $d$ days). Importantly, the counts $N_{t,d}$ are subject to right truncation, as we can only observe cases with $t+d \leq T$ until the present date $T$. Moreover, cases with missing symptom onset date can only be counted by report date, as $C_t^{\text{missing}}$. In the following, we develop our generative model for nowcasting $R_t$ by starting with a model for direct $R_t$ estimation using only counts by date of report $C_t$. We then gradually increase the complexity to account for right-truncated case counts by date of symptom onset and for incomplete line list data. At each stage, we first present a stepwise approach similar to existing nowcasting methods, and then show how the additional step can be directly integrated into the generative model. 

\subsection{(A) Effective reproduction number estimation}\label{sec:R_estimation}
There exist various mechanistic and non-mechanistic approaches to estimate $R_t$, most of which rely on the renewal equation\cite{fraserEstimatingIndividualHousehold2007,champredon_equivalence_2018} as a mathematical foundation. For example, the non-parametric method by Cori et al.\cite{cori_new_2013} implemented in the widely used package EpiEstim allows to compute Bayesian $R_t$ estimates efficiently from a time series of case counts. To account for the delay between infection and reporting of cases however, these estimates must be shifted back in time by the mean delay, or one must first infer the time series of infections in an intermediate step using deconvolution\cite{gostic_practical_2020, huisman_estimation_2022}. This and other limitations have motivated the recent development of Bayesian methods which model infections and $R_t$ as latent variables. Here we use a semi-mechanistic renewal model similar to existing tools\cite{EpiNow2,epidemia} to estimate $R_t$ directly from observed case counts (\Cref{fig:schematic}A). This allows us to specify a single generative model for our data, which can then be adapted to right-truncated and incomplete line lists.

Specifically, we model $C_t$, the number of cases by date of report, as Poisson distributed with rate $\lambda_t$, the expected number of cases reported on date $t$. We link $\lambda_t$ to the number of infections on date $t$, denoted $I_t$, via convolution over previous dates, \ie
\begin{linenomath*}\begin{equation}\label{eq:cases_report}
C_t | \lambda_t \sim \text{Poisson}(\lambda_t), \quad \lambda_t = \sum_{d=0}^D I_{t-d} \, \rho_{t-d} \, p_d
\end{equation}\end{linenomath*} where $\rho_{t-d}$ is an ascertainment proportion, \ie the proportion of infections at time $t-d$ that eventually become reported cases, and $p_d$ the probability of report $d$ days after infection, with an assumed maximum delay $D$. We here let $\rho_t=\rho$ for all $t$, \ie we assume that the ascertainment proportion is constant over time. Importantly, both $\rho$ and $p$ are assumed to be known, and $p$ is typically composed of i) a delay from infection to symptom onset (\ie the incubation period), and ii) a delay from symptom onset to reporting. The latter is specific to each reporting system and should be estimated from line list data. We here follow the common practice of using the empirical distribution of delays observed in the line list as a proxy for the reporting delay. This naive approach ignores the right truncation of observations and assumes a fixed delay over time, and we assess the bias resulting from this simplification on synthetic data. We remark that a rough estimate of $\rho$ is often sufficient for estimating $R_t$, as constant-in-time multipliers cannot bias $R_t$ estimates and will distort uncertainty quantification only in extreme cases of misspecification\cite{whiteReportingErrorsInfectious2010}. If $\rho_t$ changes over time however, $R_t$ estimates will be biased, with slow, constant changes introducing small but continuous bias, and incidental abrupt changes introducing bigger but only temporary bias. In real-world settings, both of these types of changes can occur.

The number of infections $I_t$ is modeled via a renewal process as in previous work\cite{abbott_estimating_2020,bhatt_semi-mechanistic_2020,flaxmanEstimatingEffectsNonpharmaceutical2020,tehEfficientBayesianInference2022}, \ie
\begin{linenomath*}\begin{equation}\label{eq:renewal}
I_t|\iota_t \sim \text{Poisson}(\iota_t), \quad \iota_t = E[I_t] = R_t \sum_{s=1}^G \psi_s \, I_{t-s},
\end{equation}\end{linenomath*} where $\psi$ represents an assumed intrinsic generation time distribution\cite{champredon_intrinsic_2015} and $G$ is the maximum generation time (see \supp~\zref{appendix:seeding} for the modeling of initial infections). Note that we here use a stochastic formulation, \ie we account for variance in the offspring distribution by sampling realized infections $I_t$ as independently Poisson distributed given $\iota_t$\cite{banholzerEstimatingEffectsNonpharmaceutical2021,bhatt_semi-mechanistic_2020}.
As a smoothing prior for $R_t$ over time, we use a first-order random walk. To ensure $R_t>0$, we apply a softplus link to the random walk, \ie
\begin{linenomath*}\begin{equation}\label{eq:R_random_walk}
\text{softplus}^{-1}(R_t)|R_{t-1} \sim N(\text{softplus}^{-1}(R_{t-1}),{\sigma_R}^2),
\end{equation}\end{linenomath*} where $\text{softplus}(x) = \frac{\log(1+e^{kx})}{k}$ is the softplus function with sharpness parameter $k$. We choose $k=4$ to ensure that the random walk is effectively on the unit scale for all realistic values of $R_t$, thereby modeling changes in $R_t$ over time as symmetric (see \supp~\zref{appendix:link_functions}). This reflects our expectation that interventions and behavioral changes have roughly symmetric effects on $R_t$\cite{sharma_how_2020}, but other link functions such as log\cite{EpiNow2} or generalized logit\cite{epidemia} are also possible. We use weakly informative priors for $R_1$ and $\sigma_{R}$ (see \supp~\zref{appendix:random_walk_model}), to allow the intercept and variance of the random walk to be estimated from the data.

Using Eq. (\ref{eq:cases_report})--(\ref{eq:R_random_walk}) (full model in \supp~\zref{appendix:modelRt}), we can compute the likelihood $P\left(C_{\{t | t \leq T\}} \;\middle|\; R_{\{t | t \leq T\}}, \theta\right)$ with nuisance parameters $\theta$ and, together with corresponding priors (\supp~\zref{appendix:priors}), sample from the posterior $P\left(R_{\{t | t \leq T\}}, \theta \;\middle|\; C_{\{t | t \leq T\}}\right)$ using Markov chain Monte Carlo (MCMC, see \Cref{sec:implementation}) to obtain Bayesian estimates of $R_t$ over time.

\subsection{(B) Truncation adjustment}\label{sec:truncation_adjustment}
The direct estimation of $R_t$ from case counts by date of report $C_t$ requires a deconvolution of $C_t$ with potentially long and difficult-to-estimate delays between infection and reporting. An alternative is to estimate $R_t$ from case counts by date of symptom onset $N_t$ (\Cref{fig:schematic}B), such that the delay distribution in \Cref{eq:cases_report} is simply the incubation period, which is shorter and generally independent of the reporting system. On the present date $T$ however, only $N_{t,d}$ with $t+d \leq T$ are known, such that the real-time count $\sum_{d=0}^{T-t} N_{t,d}$ is a downward-biased estimate of $N_t$. This bias must be corrected by estimating the number of symptom onsets which have occurred but are not yet reported, also known in the statistical literature as ``nowcasting'' in the narrow sense\cite{kalbfleisch_statistical_2002}. Conceptually this task is related to time-to-event analysis\cite{seamanEstimatingTimetoeventDistribution2022}, as it requires estimating the distribution of reporting delays from right truncated observations $N_{t,d}$ to predict the not-yet-reported cases.

\subsubsection*{Stepwise approach}\label{sec:trunc_stepwise}
To nowcast $R_t$, existing methods follow a stepwise approach\cite{gunther_nowcasting_2021,salazar_near_2022,li_bayesian_2021}, where first $N_t$ is estimated using the observed $N_{t,d}$ in a truncation adjustment step, and then $R_t$ is estimated from the time series of $N_t$ (instead of $C_t$, as in \Cref{sec:R_estimation}). For the truncation adjustment step, we specify a model similar to Günther et al.\cite{gunther_nowcasting_2021}, which jointly combines Bayesian smoothing of the epidemic curve\cite{mcgough_nowcasting_2020,bastos_modelling_2019} with a discrete time-to-event model for the reporting delay distribution\cite{hohle_bayesian_2014}. The latter allows to estimate reporting delays that change over time (\eg due to changes in the reporting process or the burden on health authorities) and are subject to seasonality (\eg due to weekday effects), which is both observed in practice\cite{hohle_bayesian_2014}. For this, we define $p_{t,d}$ as the probability that a case with symptom onset on date $t$ is reported with a delay of $d \in [0, 1, \ldots D]$ days after symptom onset, \ie we assume that reporting cannot occur before symptom onset (no negative delays) and is at most $D$ days after symptom onset. In practice, it is important to assume a long enough maximum delay $D$ that sufficiently covers the true delay distribution. Cases with delays longer than $D$ should then be completely dropped from the analysis, as setting their delays to $D$ instead can lead to right truncation bias. Given the above definition, we assume that cases with symptom onset on date $t$ are Poisson distributed with rate $\lambda_t$, and their reporting delays are categorically distributed with event probabilities $p_{t,d}$. We can then model the observed numbers of cases with symptom onset on date $t$ and reporting delay $d$ as \begin{linenomath*}\begin{equation}\label{eq:Ntd_obs}
    N_{t,d} | \lambda_t, p_{t,d} \sim \text{Poisson}(\lambda_t \, p_{t,d}),
\end{equation}\end{linenomath*}
To model the delay probabilities $p_{t,d}$, we use a discrete time-to-event model in which we parameterize the hazard of reporting $h_{t,d}$ via a logistic regression\cite{hohle_bayesian_2014,stonerMultivariateHierarchicalFrameworks2020}, \ie
\begin{linenomath*}\begin{equation}\label{eq:hazard}
p_{t,d} = h_{t,d}\,\prod_{i=1}^{d} (1-h_{t,d-i}), \quad \text{logit}(h_{t,d}) = \gamma_d + z_t^\top \beta + w_{t+d}^\top \, \eta,
\end{equation}\end{linenomath*} where $\gamma_d$ is the logit baseline hazard for delay $d$, and $z_t$ and $w_{t+d}$ are vectors of covariates with coefficient vectors $\beta$ and $\eta$, respectively. This implements a non-parametric, proportional hazards model\cite{coxRegressionModelsLifeTables1972} for the reporting delay as in Günther et al., however, we here model both effects by date of symptom onset $z_t^\top \beta$ and effects by date of report $w_{t+d}^\top \, \eta$. We design $z_t$ to account for changes in the reporting delay of cases over time via a piecewise linear weekly change point model, and $w_{t+d}$ to account for differences in reporting between weekdays (see \supp~\zref{appendix:time-to-event_model} for details).
As a smoothing prior for $\lambda_t$ over time, we use a first-order random walk on the log scale, \ie
\begin{linenomath*}\begin{equation}\label{eq:lambda_prior}
    \log(\lambda_t)|\lambda_{t-1} \sim N(\log(\lambda_{t-1}),{\sigma_{\log(\lambda)}}^2).
\end{equation}\end{linenomath*}
This is a minimally informed prior assuming a stationary time series of symptom onsets which has often been used in earlier work\cite{mcgough_nowcasting_2020,bastos_modelling_2019,gunther_nowcasting_2021,bergstromBayesianNowcastingLeading2022}, and we test non-stationary exponential smoothing as an alternative prior in a supplementary analysis (\supp~\zref{appendix:ets_model}). Analogous to \Cref{sec:R_estimation}, we use weakly informative priors for $\log(\lambda_1)$ and $\sigma_{\log(\lambda)}$ (see \supp~\zref{appendix:random_walk_model}), to estimate the random walk intercept and variance from the data.

Given Eq. (\ref{eq:Ntd_obs})--(\ref{eq:lambda_prior}) (full model in \supp~\zref{appendix:modeltrunc}), we can compute the likelihood $P\left(N_{\{t,d | t+d \leq T\}} \;\middle|\; \lambda_{\{t | t \leq T\}}, p_{\{t,d | t \leq T, d \leq D\}}, \theta\right)$, and obtain posterior samples $\hat{\lambda}_{\{t | t \leq T\}}^{(i)},\, \hat{p}_{\{t,d | t \leq T, d \leq D\}}^{(i)} \sim P\left(\lambda_{\{t | t \leq T\}}, p_{\{t,d | t \leq T, d \leq D\}}, \theta \;\middle|\; N_{\{t,d | t+d \leq T\}}\right)$ via MCMC. These can then be used to produce samples of the not-yet-observed case counts $\hat{N}_{\{t,d | t+d>T\}}^{(i)}$ by drawing from $\text{Poisson}(\hat{\lambda}_t^{(i)}\hat{p}_{t,d}^{(i)})$, respectively. Samples from the nowcast for $N_t$ are then simply obtained by summing up the already observed and the sampled, not-yet-observed case counts, \ie $\hat{N}_t^{(i)} = \sum_{d=0}^{\min(T-t,D)} N_{t,d} + \sum_{d=T-t+1}^{D} \hat{N}_{t,d}^{(i)}$.

To obtain a nowcast for $R_t$, existing nowcasting approaches typically estimate $R_t$ from the nowcast for $N_t$ in a separate step. Since the nowcast for $N_t$ is subject to considerable uncertainty on dates close to the present, stepwise approaches must repeat the $R_t$ estimation step on many samples $\hat{N}_{\{t|t\leq T\}}^{(i)}$ from the truncation adjustment, and finally combine the resulting $R_t$ estimates. This ``resampling'' procedure is necessary to account for the uncertainty from the truncation adjustment but introduces computational overhead especially if the individual $R_t$ estimation steps are costly. In practice, stepwise approaches therefore either ignore uncertainty from the truncation adjustment\cite{hawrylukGaussianProcessNowcasting2021} or use simple, non-parametric methods such as EpiEstim to estimate $R_t$\cite{gunther_nowcasting_2021,salazar_near_2022}. To ensure a consistent comparison of approaches in this study, we fitted the generative model introduced in \Cref{sec:R_estimation} on 50 different samples $\hat{N}_{\{t|t\leq T\}}^{(i)}$ using MCMC, which incurred comparatively high computational cost. We additionally produced nowcasts using EpiEstim for $R_t$ estimation in a supplementary analysis (\supp~\zref{appendix:epiestim} and \zref{appendix:implementation}).

\subsubsection*{Generative approach}\label{sec:trunc_generative}
As an alternative to the stepwise approach, we here propose to integrate $R_t$ estimation and truncation adjustment in a single generative model. This is achieved by replacing \Cref{eq:lambda_prior} in the truncation adjustment model with
\begin{linenomath*}\begin{equation}\label{eq:lambda_convolution}
\lambda_t = \sum_{s=0}^L I_{t-s}\, \rho_{t-s} \, \tau_s,
\end{equation}\end{linenomath*}
where $\rho_{t-s}$ is again an assumed ascertainment proportion, $\tau$ is the incubation period distribution, $L$ is the maximum incubation period, and $I_t$ is the number of infections on date $t$. Analogous to \Cref{sec:R_estimation}, infections $I_t$ are then modeled via a renewal model using \Cref{eq:renewal} and \Cref{eq:R_random_walk}. By doing so, we replace the non-parametric smoothing prior for $\lambda_t$ with a model of symptom onsets that arise from latent infections generated through a renewal process. This yields a single hierarchical model based on Eq. (\ref{eq:renewal}) -- (\ref{eq:hazard}), and (\ref{eq:lambda_convolution}) (full model in \supp~\zref{appendix:modeltruncRt}), under which we can directly compute the joint likelihood
\begin{linenomath*}\begin{equation*}
P\left(N_{\{t,d | t+d \leq T\}} \;\middle|\; \lambda_{\{t | t \leq T\}}, p_{\{t,d | t \leq T, d \leq D\}}, R_{\{t | t \leq T\}}, \theta\right),
\end{equation*}\end{linenomath*} and obtain posterior samples for $R_t$ and $N_t$ as described before. The resulting generative approach allows to nowcast $R_t$ and $N_t$ in a single step, while quantifying uncertainty both from truncation adjustment and $R_t$ estimation. Moreover, by explicitly linking the time series of symptom onsets to the underlying infection dynamics, we use the renewal model as a structural prior for the temporal correlation of $N_t$. This way, epidemic modeling is not only used to estimate $R_t$ but also to regularize the nowcast for $N_t$, thereby avoiding the additional non-parametric smoothing of $\lambda_t$ in the stepwise approach.

\subsection{(C) Missing date imputation}
In practice, line list data are often incomplete, \ie the date of symptom onset can be missing for a substantial proportion of cases. Ignoring incomplete cases and computing nowcasts as described in \Cref{sec:truncation_adjustment} only using the complete cases $N_{t,d}^{\text{known}}$ will underestimate $N_t$ and bias nowcasts of $R_t$ if the proportion of cases with missing onset date changes over time. It is therefore important to also include the incomplete cases, which are only available as case counts $C_t^{\text{missing}}$ by date of report. Here we discuss two stepwise approaches of imputing missing onset dates and then show how a missingness model can be directly integrated into the generative nowcasting model (\Cref{fig:schematic}C).

\subsubsection*{Stepwise approach}
A simple but potentially biased approach is to treat the reporting delays of cases in the line list as independent and identically distributed. Under this assumption, missing onset dates can be imputed by simply subtracting a random delay from the date of report of each incomplete case, \ie each case counted in $C_{t}^{\text{missing}}$. The random delays could for example be drawn from the empirical delay distribution of delays from complete cases in the line list. However, even if there was no right truncation or changes in the reporting delay over time, such a procedure will yield biased results during an epidemic wave. This is because subtracting delays from the date of report implicitly conditions the delay distribution on the date of report\cite{gostic_practical_2020,petermann_pitfall_2020}. The resulting distribution, also known as backward delay distribution, depends on the shape of the past epidemic curve and therefore varies during an epidemic. A more accurate approach is therefore to only assume that cases with the same date of report have identically distributed delays, and to estimate the so-called backward delay probability $p^{\leftarrow}_{t,d}$ that a case reported on date $t$ had its symptom onset $d$ days ago\cite{gunther_nowcasting_2021,salazar_near_2022,li_bayesian_2021}. We here implement this approach by modeling the number of cases with known symptom onset date as \begin{linenomath*}\begin{equation}
(N_{t,0}^{\text{known}},\dots,N_{t-D,D}^{\text{known}}) \sim \text{Multinom}(p^{\leftarrow}_{t,0},\dots,p^{\leftarrow}_{t,D}),
\end{equation}\end{linenomath*} and parameterizing $p^{\leftarrow}_{t,d}$ using an identical discrete time-to-event model as specified in \Cref{eq:hazard}, but indexed by date of report instead of date of symptom onset (\supp~\zref{appendix:modelmiss}). We remark that by conditioning on the date of report, the above model is not biased by right truncation, but potentially by left truncation (see \supp~\zref{appendix:backward_delays}). We compute the likelihood $P\left(N_{\{t,d | t+d \leq T\}}^{\text{known}} \;\middle|\; p^{\leftarrow}_{t,d}, \theta\right)$ to obtain posterior samples from $P\left(p^{\leftarrow}_{t,d}, \theta \;\middle|\; N_{\{t,d | t+d \leq T\}}^{\text{known}}\right)$ via MCMC, and use the latter to impute missing onset dates by drawing random delays from the corresponding backward delay distribution. Importantly, using observed delays to impute missing onset dates implies a missing-at-random assumption, \ie reporting delays are assumed to be independent of whether cases in the line list are complete or incomplete. Moreover, both the independent imputation and the backward-delay imputation approach are stepwise in that they add an additional imputation step prior to nowcasting. As a result, the imputed symptom onset dates are treated as observed data during the following step(s), which creates the same obstacle to uncertainty quantification as described in \Cref{sec:trunc_stepwise}. Thus, to account for uncertainty of the imputation, a multiple imputation scheme must be used, where multiple imputed data sets are created and a separate nowcast is estimated for each data set\cite{salazar_near_2022}. With more complex nowcasting models however, multiple imputation is often impractical and therefore omitted\cite{gunther_nowcasting_2021}, as fitting to a large number of imputed data sets would be prohibitively computationally expensive. Similarly, we only assessed single imputations due to resource constraints in this study.

\subsubsection*{Generative approach}
As an alternative, we propose to directly account for missing symptom onset dates in the generative nowcasting model described in \Cref{sec:truncation_adjustment}. For this, we assume that symptom onsets on date $t$ become known with probability $\alpha_t$ and missing with probability $1-\alpha_t$. The case counts with known and missing symptom onset date $t$ and delay $d$ are then modeled as
\begin{linenomath*}\begin{equation}\label{eq:Ntd_known}
N_{t,d}^{\text{known}} | \lambda_t, p_{t,d}, \alpha_t \sim \text{Poisson}(\lambda_t \, p_{t,d} \, \alpha_t), \quad N_{t,d}^{\text{missing}} | \lambda_t, p_{t,d}, \alpha_t \sim \text{Poisson}(\lambda_t \, p_{t,d} \, (1-\alpha_t)),
\end{equation}\end{linenomath*}
which again assumes that symptom onset dates are missing at random, \ie independent of their reporting delay. While the $N_{t,d}^{\text{missing}}$ are not directly observed in the line list, we know that $C_{t}^{\text{missing}}=\sum_{d=0}^{\min(D,t-1)} N_{t-d,d}^{\text{missing}}.$ Hence, $C_{t}^{\text{missing}}$ is the sum of Poisson distributed case numbers from previous days with corresponding delays, and it is also Poisson distributed with
\begin{linenomath*}\begin{equation}\label{eq:Ct}
C_{t}^{\text{missing}} | \lambda, p, \alpha \sim \text{Poisson}\left(\sum_{d=0}^{\min(D,t-1)} \lambda_{t-d} \, p_{t-d,d} \, (1-\alpha_{t-d})\right).
\end{equation}\end{linenomath*} Note that under this model, only observations $C_{t}^{\text{missing}}$ for $t>D$ should be used to avoid bias from left-truncation, or the latent parameters $\lambda$, $p$, and $\alpha$ must be further extended into the past (\supp~\zref{appendix:missing_onsets_model}). As a minimally informed, stationary smoothing prior for $\alpha_t$ over time, we use a first-order random walk on the logit scale, \ie
\begin{linenomath*}\begin{equation}\label{eq:alpha_smoothing}
\text{logit}(\alpha_t)|\alpha_{t-1} \sim N(\text{logit}(\alpha_{t-1}),{\sigma_{\text{logit}(\alpha)}}^2),
\end{equation}\end{linenomath*}
and estimate $\alpha_1$ and $\sigma_{\text{logit}(\alpha)}$ with weakly informed priors as before. This assumes that the probability of missing symptom onset dates varies only by date of symptom onset, but more sophisticated models could also account for weekday effects or effects by date of report. Assuming $N_{t,d}^{\text{known}}$ and $C_{t}^{\text{missing}}$ as independent given $\lambda$, $p$, and $\alpha$, we can replace \Cref{eq:Ntd_obs} in the existing hierarchical nowcasting model described in \Cref{sec:truncation_adjustment} with Eq. (\ref{eq:Ntd_known}), (\ref{eq:Ct}), and (\ref{eq:alpha_smoothing}) (full model in \supp~\zref{appendix:modelmisstruncR}). Under the resulting model, we can compute the joint likelihood
\begin{linenomath*}\begin{equation*}
P\left(N_{\{t,d | t+d \leq T\}}^{\text{known}}, C_{\{t | t \leq T\}}^{\text{missing}} \;\middle|\; \lambda_{\{t | t \leq T\}}, p_{\{t,d | t \leq T, d \leq D\}}, \alpha_{\{t | t \leq T\}}, R_{\{t | t \leq T\}}, \theta\right)
\end{equation*}\end{linenomath*}
and obtain posterior samples for $R_t$ and $N_t$ using MCMC as before. Therefore, the model can be jointly fit to incomplete line list data without the need for an additional imputation step. This also accounts for the uncertainty resulting from missing onset dates without the use of a multiple imputation scheme. In comparison to stepwise approaches, the generative approach requires estimating the probability of missing onset dates over time, but in turn avoids estimating backward-delay distributions. Moreover, using estimates for $\alpha_t$, separate nowcasts for $N_t^{\text{known}}$ and $N_t^{\text{missing}}$ can be easily obtained.

\subsection{Modeling of overdispersion}
As found in earlier work\cite{gunther_nowcasting_2021}, the modeling of overdispersed case counts can improve nowcasting performance in real-world settings. When nowcasting COVID-19 hospitalizations in Switzerland, we therefore model observed case counts as Negative Binomial instead of Poisson distributed in all our models, \eg we use
\begin{linenomath*}\begin{equation}
N_{t,d} | \lambda_t, p_{t,d}, \phi \sim \text{NegBin}(\lambda_t \, p_{t,d}, \phi),
\end{equation}\end{linenomath*}
instead of Eq.~\ref{eq:Ntd_obs}, where the inverse of $\phi$ defines the level of overdispersion, with $Var[N_{t,d} | \lambda_t, p_{t,d}, \phi] = (\lambda_t \, p_{t,d}) (1+\frac{\lambda_t \, p_{t,d}}{\phi})$. We place a prior on $\frac{1}{\sqrt{\phi}}$ to regularize the overdispersion (\supp~\zref{appendix:priors}) and use a pooled overdispersion parameter for all observations.

\subsection{Implementation and estimation}\label{sec:implementation}
We implemented all models used in the different nowcasting approaches (\Cref{fig:schematic}) as probabilistic programs in Stan\cite{stan_development_team_stan_2022} (\supp~\zref{appendix:implementation}). This allowed us to estimate all parameters in a fully Bayesian framework with Markov chain Monte Carlo (MCMC) using cmdstan version 2.30.0 via cmdstanr version 0.5.0\cite{cmdstanr}. To provide regularization, we used weakly informative priors on the model parameters (see \supp~\zref{appendix:priors} for details). For each estimation, the default configuration of the No-U-Turn sampler was run, \ie four chains with 1,000 warm-up and 1,000 sampling iterations each. After model fitting, we computed common Bayesian model diagnostics to check for convergence (Gelman-Rubin convergence diagnostic $\hat{R}$)\cite{Gelman.1992} and sufficient sample sizes (ESS ratio)\cite{geyer_mcmc_2011} (\supp~\zref{appendix:diagnostics}). We used the R package EpiEstim\cite{cori_new_2013} for $R_t$ estimation with the method by Cori et al., and the R package scoringutils\cite{scoringutils} for computation of weighted interval scores. All processing steps and analyses were performed in R 4.1.0, and the models and code for this study are available from zenodo at \url{https://doi.org/10.5281/zenodo.8279676}.

\subsection{Comparison of approaches using synthetic data}\label{sec:methods_sim}
We used synthetic data to compare the different stepwise and generative nowcasting approaches. To simulate synthetic data, we generated infections through a stochastic renewal process over a period of 200 days in which $R_t$ follows a piecewise linear time series with prespecified change points. The simulation matches the generative models described in \Cref{sec:R_estimation,sec:trunc_generative}, but simulates the reporting of individual cases and therefore does not assume independent observation noise. We simulated two different scenarios, representing a first and second epidemic wave, respectively (\supp~\zref{appendix:sim_infections}). The first wave scenario starts with a small number of seeding infections and strong transmission ($R_t=2$), followed by a shift to suppression ($R_t=0.8$). The second wave scenario starts with moderate case numbers and controlled transmission ($R_t=1$), followed by a resurgence ($R_t=1.4$) and finally decline ($R_t=0.8$). For each scenario, we produced 50 simulation runs with different realizations of the infection process. We parameterized the generation interval and the incubation period distributions based on estimates for COVID-19 from the literature\cite{linton_incubation_2020,hartInferenceSARSCoV2Generation2022}, and supplied the same distributions to the nowcasting models. To mimic the ascertainment of hospitalization data, we let infections become reported hospitalizations with probability $\rho = 2\%$. For each simulation run, we obtained a synthetic line list by simulating the symptom onset and day of report of each case. The baseline reporting delay was specified as a discretized gamma distribution with a mean of 9 days and standard deviation of 8 days, which is in line with the delay between symptom onset and reporting of hospitalization during the first year of the COVID-19 pandemic in Switzerland (\supp~Figure~\zref{fig:switzerland_emp_delay}). We used a piecewise linear model to simulate changes in the reporting hazard over time, with random changes in the trend every four weeks, which differed across simulation runs. We also included weekday effects with the odds of reporting being at the baseline Wednesday -- Friday, 70\% lower on Saturdays, 80\% lower on Sundays, 10\% higher on Mondays, and 5\% higher on Tuesdays. Details of the simulation of infections and reporting are provided in \supp~\zref{appendix:synthetic_data}.

For each simulation run and scenario, nowcasts were obtained for selected weeks in different phases of the epidemic wave (\Cref{fig:wave1_feather_renewal_Nt}--\ref{fig:wave1_feather_miss_Rt}, \supp~\zref{appendix:priors}) by applying the nowcasting approaches at different lags of the respective week (at the end of the week, one week after, two weeks after). This allowed us to assess changes in performance as more data becomes available. We fitted the models using always the last three months of line list data until the date of the nowcast. 
Complete line list data, \ie without missing symptom onset dates, were used to compare the direct $R_t$ estimation with the stepwise and generative truncation adjustment approaches. Incomplete line list data were used to compare the stepwise and the generative missing date imputation approaches. Here we randomly selected cases to have a missing symptom onset date with a probability varying over time between 20\% and 60\% according to a random piecewise linear function, which differed across simulation runs.

To evaluate nowcasting performance in each of the 50 simulation runs, we scored the nowcasts for each week in each phase with respect to $N_t$ and $R_t$ against the simulated ground truth. As a performance measure, we used the weighted interval score (WIS), which is a quantile-based proper scoring rule that approximates the continuous ranked probability score (CRPS). The CRPS generalizes the absolute error to probabilistic nowcasts. For an observation $y$, \ie the true $N_t$ or $R_t$, and a predictive distribution $F$, \ie the probabilistic nowcast for $N_t$ or $R_t$, the weighted interval score is computed as
\begin{linenomath*}\begin{equation}
\text{WIS}(F,y) = \frac{1}{2K+1} \left(|y - F_{0.5}| + \sum_{k=1}^{K} (\alpha_k \, \text{IS}_{\alpha_k}(F, y)) \right),
\end{equation}\end{linenomath*}
where $|y - F_{0.5}|$ is the absolute error of the median, $\alpha_k \in \{\alpha_1, \dots, \alpha_K\}$ represent different central prediction intervals at coverage level $(1-\alpha_k)$, and $\text{IS}_{\alpha}(F,y)$ is the interval score for a specific interval, defined as
\begin{linenomath*}\begin{equation}
\text{IS}_{\alpha}(F,y) = \underbrace{\vphantom{\frac{2}{\alpha}}(F_{1-\frac{\alpha}{2}}-F_{\frac{\alpha}{2}})}_{\text{dispersion}} + \underbrace{\frac{2}{\alpha} (y-F_{1-\frac{\alpha}{2}}) \mathds{1}_{y>F_{1-\frac{\alpha}{2}}}}_{\text{underprediction}} + \underbrace{\frac{2}{\alpha} (F_{\frac{\alpha}{2}}-y) \mathds{1}_{y<F_{\frac{\alpha}{2}}}}_{\text{overprediction}}.
\end{equation}\end{linenomath*}
That is, the interval score can be decomposed into three penalty components, where the dispersion component encourages sharpness, and the under- and overprediction components encourage calibration of the nowcast. We used $K=11$ different $\alpha_k$, corresponding to the 10\%, 20\%, $\dots$, 90\%, 95\%, and 98\% credible intervals (CrI) for $N_t$ and $R_t$, respectively. To obtain an overall score $\overline{\text{WIS}}$ for each evaluated week, we took the arithmetic mean of scores from each day of the week, which again yields a proper score. We furthermore computed the relative contribution of each penalty component (dispersion, over-, and underprediction) to the respective $\overline{\text{WIS}}$. In the results, we report the average $\overline{\text{WIS}}$ over all 50 simulation runs, as well as the percentage of runs where each approach performed best.

\subsection{Application to COVID-19 in Switzerland}\label{sec:methods_real}
We also compared the direct, stepwise, and generative nowcasting approaches during the COVID-19 pandemic in Switzerland. For this, we used anonymized line list data of cases reported to the Federal Office of Public Health (FOPH) from January~01, 2020 -- March~31, 2021. Because in Switzerland the date of symptom onset for COVID-19 was only consistently recorded for hospitalized patients, we only used cases of hospitalization from the line list. As date of report, the date when a hospitalization was recorded by FOPH was used. The symptom onset date was either reported together with the clinical record of hospitalization or slightly before (due to a previous test result which was later merged), but not updated afterward. Negative delays were therefore assumed to be data entry errors, and the respective symptom onset dates were set to missing.

We produced retrospective nowcasts for the number of hospitalizations by date of symptom onset $N_t$ (including cases with and without known symptom onset) and the effective reproduction number $R_t$ during different phases of the first and second wave (before, at, and after peak, respectively). We assumed a maximum delay of 8 weeks, which covered 97.03\% of reporting delays, and removed the small percentage of cases with larger delays from the analysis. When modeling weekday effects on the reporting delay, we coded national public holidays in Switzerland as Sundays. To account for the gradual replacement of the SARS-CoV-2 wild-type by the alpha variant from December 2020~--~April 2021, we used different incubation period distributions ($\tau^{\text{wild-type}}=\Gamma(\alpha = 2.74, \beta = 0.52)$\cite{linton_incubation_2020}, $\tau^{\text{alpha}}=\Gamma(\alpha = 3.08, \beta = 0.63)$\cite{manicaEstimationIncubationPeriod2023}) and generation interval distributions ($\psi^{\text{wild-type}}=\Gamma(\alpha = 1.43, \beta = 0.29)$\cite{hartInferenceSARSCoV2Generation2022}, $\psi^{\text{alpha}}=\Gamma(\alpha = 1.75, \beta = 0.38)$\cite{hartGenerationTimeAlpha2022}) over time according to a logistic growth transition (\supp~\zref{appendix:epi_params}).

On real-world data, the nowcasting performance with regard to $N_t$ can be evaluated in retrospect by observing the number of cases with a certain onset date after a sufficiently long delay. This is however only possible for cases with known onset date, \ie $N_t^{\text{known}}$. To obtain a proxy ground truth for $N_t^{\text{missing}}$, we used nowcasts at large lags beyond the respective date (7--14 days more than the maximum delay). These nowcasts are informed by fully reported case counts and thus no longer subject to right truncation. The resulting ``consolidated'' estimates of the true $N_t$ were used to evaluate the performance of the different real-time nowcasts with regard to $N_t$, using the same weighted interval scoring rule as described for the synthetic data. Since the effective reproduction number $R_t$ cannot be observed in a real-world setting, and corresponding estimates are subject to considerable uncertainty, we refrained from assessing the performance of $R_t$ nowcasts and focused on a qualitative comparison instead.

\section{Results}

\subsection{Synthetic data: $R_t$ estimation and truncation adjustment}
We used complete synthetic line list data, \ie without missing symptom onset dates, to compare i) the direct $R_t$ estimation approach, ii) the stepwise nowcasting approach with an additional truncation adjustment step, and iii) the generative nowcasting approach with a joint truncation and renewal model. 

\subsubsection{Nowcasts of the number of cases by symptom onset date $N_t$}
\begin{figure}[!tbh]
    \vspace{-1.6cm}
    \centering
    \includegraphics[width=\textwidth]{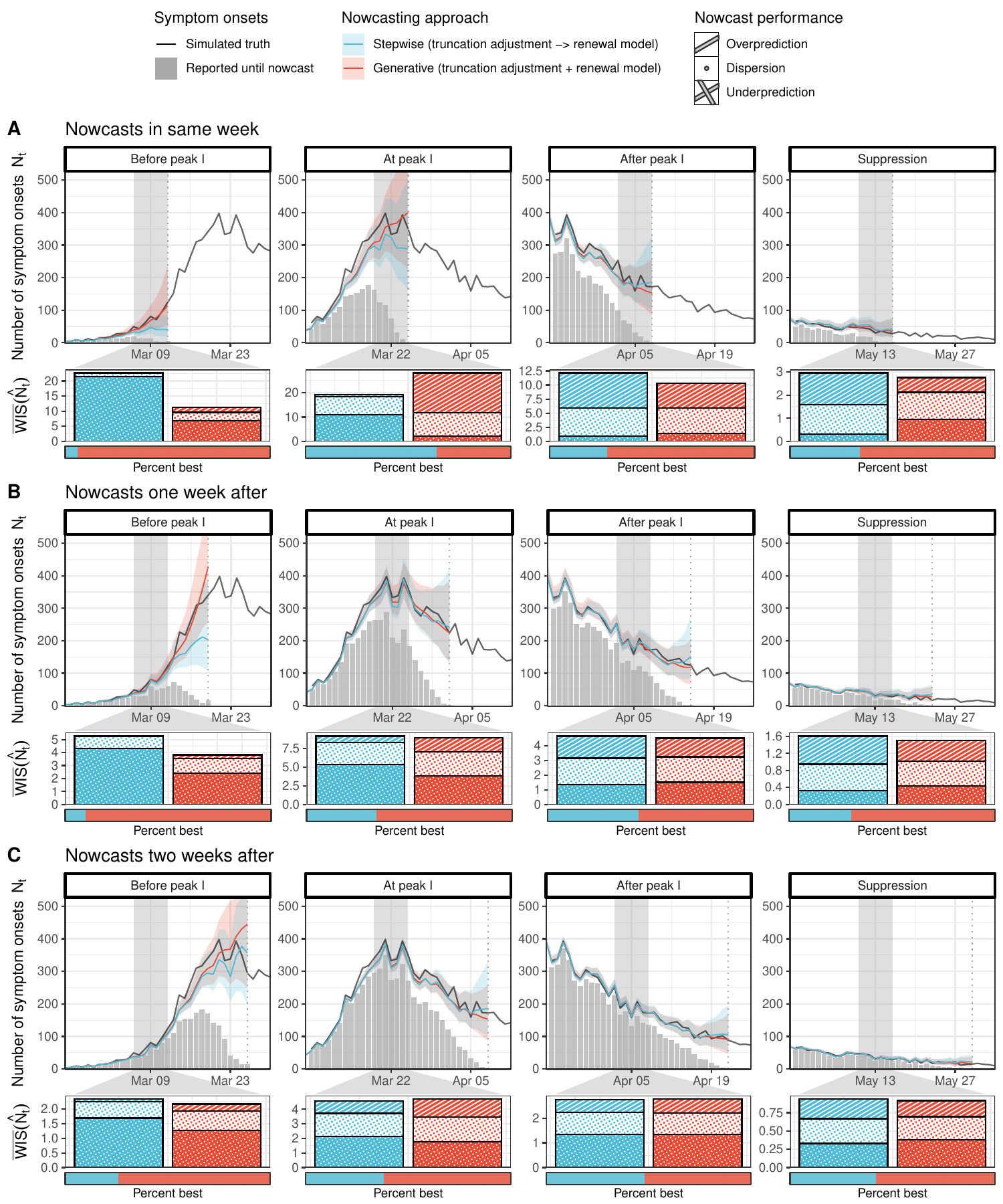}
    \caption{\textbf{Nowcasts of $N_t$ on line list data of a simulated first wave scenario using different approaches of adjusting for right truncation.} Shown are the true number of cases by symptom onset date $N_t$ (black), the number of cases reported until the nowcast date (grey bars), and point nowcasts with 95\% credible intervals (CrI) in four different phases of the epidemic wave, obtained through i) a stepwise approach using cases by date of symptom onset with a truncation adjustment step (blue), and ii) a generative approach using cases by date of symptom onset with an integrated truncation and renewal model (red). The direct approach using cases by date of report does not produce nowcasts of $N_t$ (not shown). Shown below each phase is the weighted interval score (WIS, lower is better) for $N_t$ nowcasts of each approach during a selected week (grey shade) over 50 scenario runs. Colored vertical bars show average scores, decomposed into penalties for underprediction (crosshatch), dispersion (circles), and overprediction (stripes). The horizontal proportion bar below shows the percentages of runs where each approach performed best. Results are shown for nowcasts made at different lags from the selected week (vertical dotted lines), \ie at the end of the selected week (top row), one week later (middle row), and two weeks later (bottom row).}
    \label{fig:wave1_feather_renewal_Nt}
\end{figure}

\Cref{fig:wave1_feather_renewal_Nt} shows the simulated true $N_t$ and corresponding nowcasts for an exemplary simulation run in different phases of the first wave scenario (corresponding results for the second wave scenario are shown in \supp~Fig.~\zref{fig:wave2_feather_renewal_Nt}). The baseline simulated reporting delay had a mean of 9 days and standard deviation of 8 days, and the maximum assumed delay was $D=56$ days. The nowcasts were conducted using the different approaches and at different lags (nowcast at end of same week, one week after, two weeks after). The direct $R_t$ estimation approach did not produce nowcasts for $N_t$, as it directly uses counts by date of report $C_t$ instead. 

Compared to $C_t$, which includes reporting delays, the time series of $N_t$ tracked transmission dynamics in a more timely and regular manner (\supp~Fig.~\zref{fig:sim_wave1}-\zref{fig:sim_wave2}). In a real-time setting however, the reported number of symptom onsets can have a substantial downward bias towards the present (\Cref{fig:wave1_feather_renewal_Nt}, grey bars), because a proportion of symptom onsets that occur close to the present are not yet reported at the time of nowcasting. This truncation bias was adjusted for both by the stepwise and generative approach, however, nowcasts differed between the two approaches especially on dates close to the present. 

While the generative approach predicted exponentially growing or declining case counts before and after the peak of the wave, the stepwise approach predicted case counts to remain at the recent level in all phases. This qualitative difference was strongest for nowcasts conducted at short lags, where nowcasting uncertainty was especially high, and declined with longer lags, as more data became available for the respective week. 

Below each lag and phase, \Cref{fig:wave1_feather_renewal_Nt} and \supp~Fig.~\zref{fig:wave2_feather_renewal_Nt} show the weighted interval score (WIS) for each approach across 50 simulation runs. The generative approach achieved systematically better scores in phases with exponential growth or decline, while the stepwise approach had a bias towards underprediction of $N_t$ during exponential growth and towards overprediction during decline. This further highlights the difference in smoothing assumptions of the two approaches. While the renewal model component in the generative approach predicted a continuation of recent infection dynamics, the minimally informed random walk prior in the stepwise approach predicted stationary case counts. Notably, at the peak of the wave, this led the generative approach to overpredict case counts, however only at short lags (same week nowcast), when little signal on the change in transmission dynamics was available. Overall, the WIS for both approaches decreased substantially with larger lags, and nowcasts conducted two weeks after the evaluation period show little difference in performance between the stepwise and the generative approach. 

\subsubsection{Nowcasts of the effective reproduction number $R_t$}
\begin{figure}[!tbh]
    \vspace{-1.6cm}
    \centering
    \includegraphics[width=\textwidth]{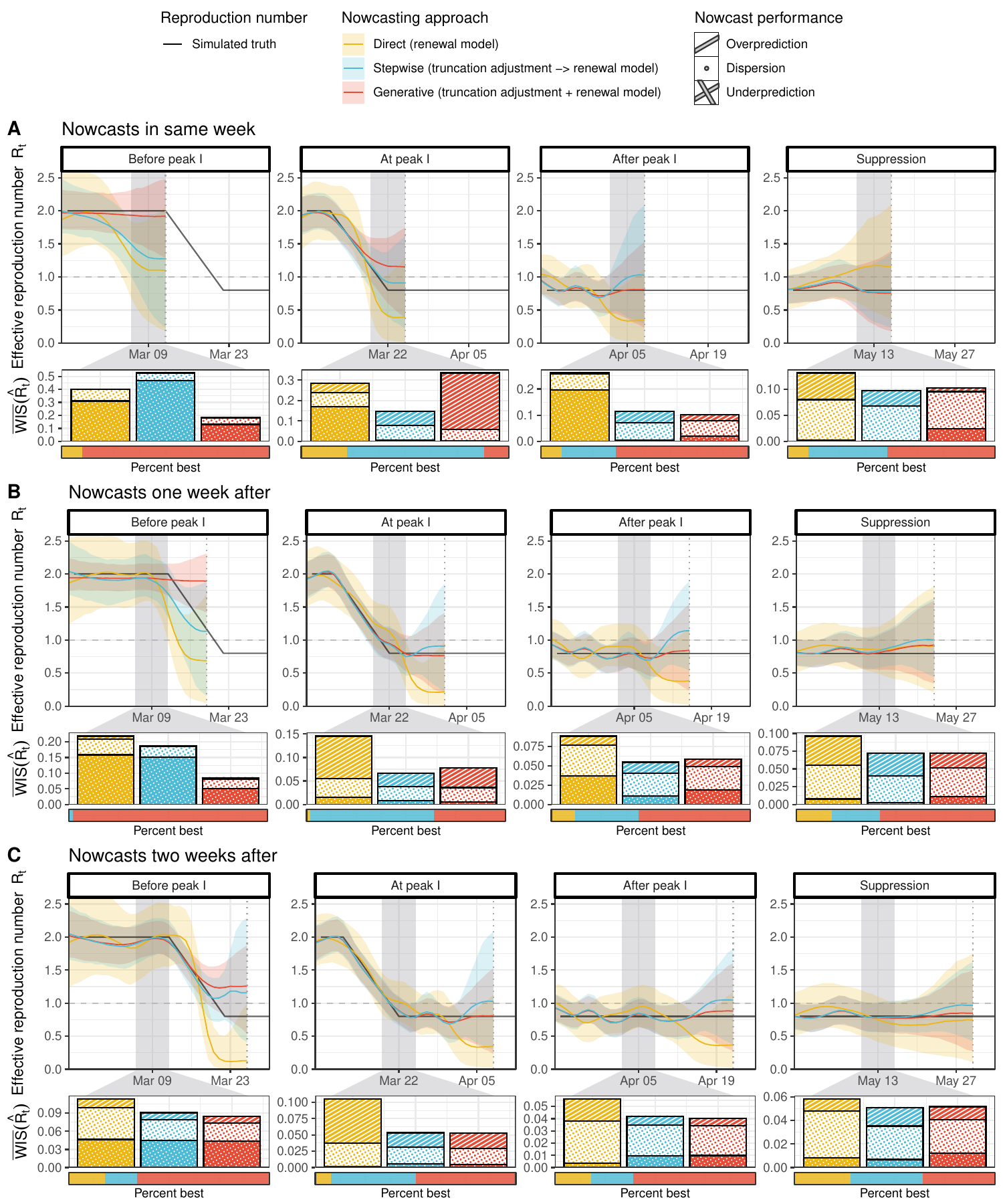}
    \caption{\textbf{Nowcasts of $R_t$ on line list data of a simulated first wave scenario using different approaches of adjusting for right truncation.} Shown are the true $R_t$ (black) and point nowcasts with 95\% credible intervals (CrI) in four different phases of the epidemic wave, obtained through i) a direct approach using cases by date of report with no truncation adjustment (yellow), ii) a stepwise approach using cases by date of symptom onset with a truncation adjustment step (blue), and iii) a generative approach using cases by date of symptom onset with an integrated truncation and renewal model (red). Shown below each phase is the weighted interval score (WIS, lower is better) for $R_t$ nowcasts of each approach during a selected week (grey shade) over 50 scenario runs. Colored bars show average scores, decomposed into penalties for underprediction (crosshatch), dispersion (circles), and overprediction (stripes). The horizontal proportion bar below shows the percentages of runs where each approach performed best. Results are shown for nowcasts made at different lags from the selected week (vertical dotted lines), \ie at the end of the selected week (top row), one week later (middle row), and two weeks later (bottom row).}
    \label{fig:wave1_feather_renewal_Rt}
\end{figure}

\Cref{fig:wave1_feather_renewal_Rt} shows results for nowcasting $R_t$ in different phases of the first wave scenario (corresponding results for the second wave scenario are shown in \supp~Fig.~\zref{fig:wave2_feather_renewal_Rt}). Here we found distinct biases in the direct and the stepwise approach. The direct $R_t$ estimation approach, which used the empirical distribution of delays between symptom onset and reporting in the line list to model observed cases $C_t$, showed a strong downward bias of $R_t$ nowcasts during the first wave, where few cases had been observed initially. This reflects the right truncation of line list data, which generally leads to an underestimation of reporting delays and therefore of infections in the recent past. In the second wave scenario, the effect of right truncation was smaller. However, due to the long delay between infection and reporting in case counts $C_t$, changes in $R_t$ were detected comparatively late by the direct $R_t$ estimation approach. The stepwise truncation adjustment approach showed a different form of bias, \ie nowcasts tended to $R_t=1$ towards the present, which translated into under- or overprediction of $R_t$ depending on the epidemic phase. Importantly, this was observed although the $R_t$ estimation step of the stepwise approach used a smoothing prior which expects $R_t$ to remain at its current level. This indicates that the $R_t$ estimation step was also impacted by the stationary smoothing assumption in the previous truncation adjustment step, \ie the constant $N_t$ nowcasts biased the downstream $R_t$ estimates towards equilibrium-level transmission. The generative approach did not show such a bias and generally achieved the smallest WIS among the three approaches. Where no sufficient signal for a change in transmission dynamics was available, $R_t$ nowcasts from the generative approach remained at the current level. At the peak of the epidemic wave, when transmission declined from $R_t=2$ to $R_t=0.8$, this led to an overprediction of $R_t$ for nowcasts in the same week. For all approaches, performance scores improved with larger lags, however, the inferior performance of the direct approach in the first wave scenario was pronounced even for nowcasts at a two-week lag.

\subsubsection{Additional comparisons}
% Subsection smoothing prior ETS
In the above comparisons, a random walk prior on the expected number of symptom onsets was used in the truncation adjustment step of the stepwise approach. While this smoothing prior has been a default choice in existing approaches\cite{mcgough_nowcasting_2020,bastos_modelling_2019,gunther_nowcasting_2021,bergstromBayesianNowcastingLeading2022}, we also implemented a non-stationary, exponential smoothing prior (\supp~\zref{appendix:ets_model}) to test whether the observed bias in nowcasts remained. We found that non-stationary smoothing in the stepwise approach reduced the bias of both $N_t$ and $R_t$ nowcasts and improved the WIS, but a slight bias remained during phases of exponential growth or decline (\supp~Fig.~\zref{fig:wave1_feather_renewal_ets_Nt}--\zref{fig:wave2_feather_renewal_ets_Rt}). We also tested the use of a non-stationary smoothing prior on $R_t$ for the generative approach, which led to  slightly better performance of $R_t$ nowcasts when transmission was changing.

% Subsection EpiEstim
As it is widely used in practice, we also implemented a stepwise approach where the $R_t$ estimation step is performed using the non-parametric method by Cori et al. via the package EpiEstim (\supp~\zref{appendix:epiestim}), instead of our semi-mechanistic renewal model described in \Cref{sec:R_estimation}. Since the $R_t$ estimation step does not influence the previous truncation adjustment step, nowcasts for $N_t$ with EpiEstim were identical compared to when using a semi-mechanistic renewal model. The nowcasts for $R_t$ showed the same bias towards $R_t=1$, however, they were often more volatile and at the same time considerably less uncertain (\supp~Fig.~\zref{fig:wave1_feather_epiestim_Rt}--\zref{fig:wave2_feather_epiestim_Rt}) when using EpiEstim in the $R_t$ estimation step. This overconfidence generally led to higher WIS values of the stepwise approach with EpiEstim. We note that when using EpiEstim, nowcasts for $R_t$ could only be obtained at a lag of 8 days or more, as $R_t$ estimates must be shifted backward in time to account for the incubation period and to center the 7-day smoothing window used for EpiEstim\cite{gostic_practical_2020}.

\FloatBarrier

\subsection{Synthetic data: Missing symptom onset date imputation}
To compare the different stepwise and generative approaches to impute missing onset dates, we used the same synthetic line list data as before, but simulated symptom onset dates to be missing with a time-varying probability $1-\alpha_t$ (\Cref{sec:methods_sim}). We tested i) a stepwise approach using independent imputation for the imputation step, ii) a stepwise approach using backward(-delay) imputation for the imputation step, and iii) a generative approach using an integrated missingness model. All three approaches used a generative model for truncation adjustment and $R_t$ estimation as defined in \Cref{sec:trunc_generative}, allowing us to specifically assess differences resulting from the imputation approach.

\begin{figure}[!tbh]
    \vspace{-1.6cm}
    \centering
    \includegraphics[width=\textwidth]{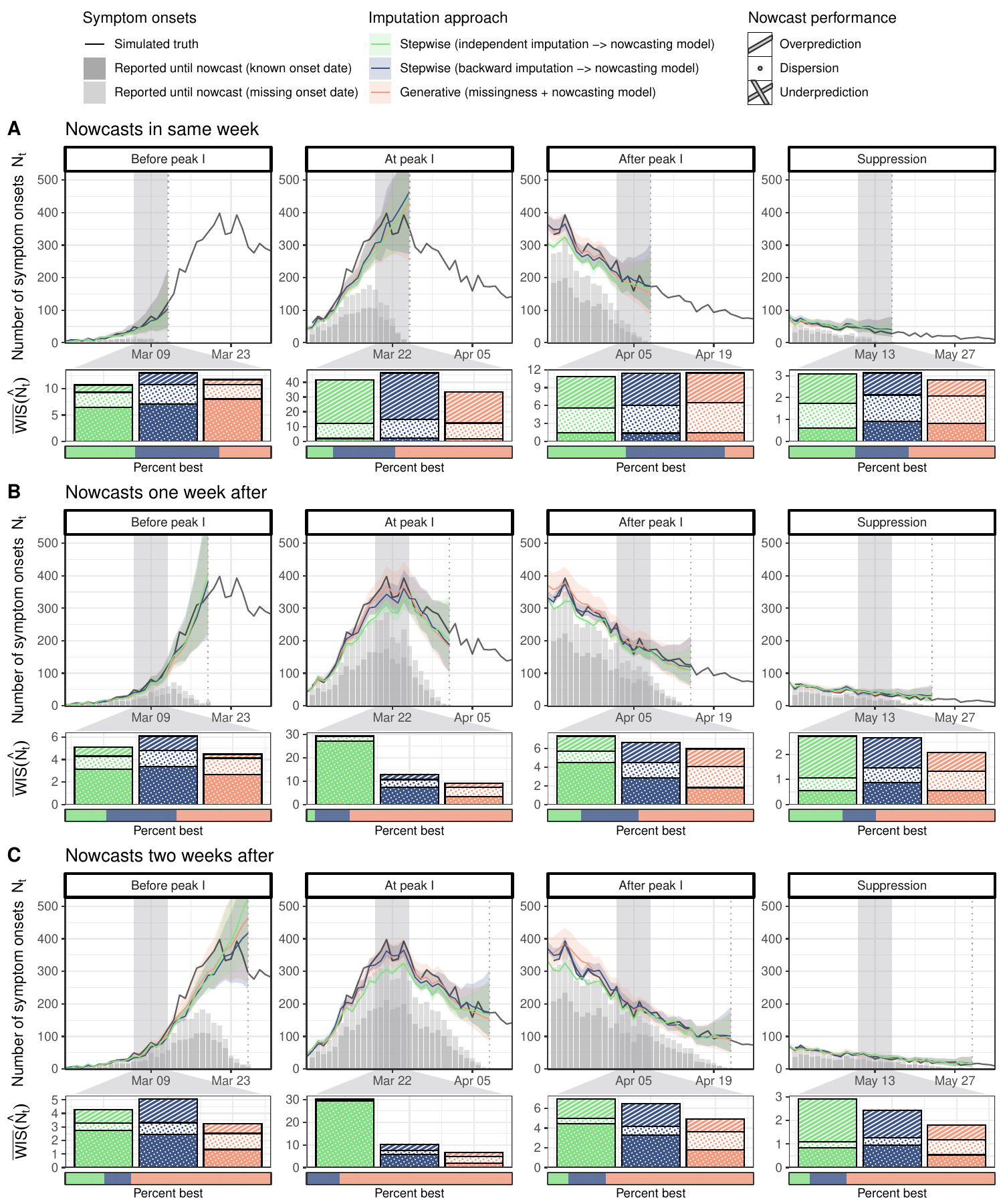}
    \caption{\textbf{Nowcasts of $N_t$ on incomplete line list data of a simulated first wave scenario using different approaches to account for missing onset dates.} Shown are the true number of cases by symptom onset date $N_t$ (black), the number of cases reported until the nowcast date (dark grey bars for known, light grey bars for missing onset dates), and point nowcasts with 95\% credible intervals (CrI) in four different phases of the epidemic wave, obtained through i) a stepwise approach using a forward imputation step (green), ii) a stepwise approach using a backward imputation step (blue), and iii) a generative approach using an integrated missingness model (red). All approaches used a generative model for nowcasting. Shown below each phase is the weighted interval score (WIS, lower is better) for $N_t$ nowcasts of each approach during a selected week (grey shade) over 50 scenario runs. Colored bars show average scores, decomposed into penalties for underprediction (crosshatch), dispersion (circles), and overprediction (stripes). The horizontal proportion bar below shows the percentages of runs where each approach performed best. Results are shown for nowcasts made at different lags from the selected week (vertical dotted lines), \ie at the end of the selected week (top row), one week later (middle row), and two weeks later (bottom row).}
    \label{fig:wave1_feather_miss_Nt}
\end{figure}

\Cref{fig:wave1_feather_miss_Nt} shows the simulated true $N_t$ and nowcasts for an exemplary simulation run and performance scores over 50 runs in different phases of the first wave scenario (second wave scenario shown in \supp~Fig.~\zref{fig:wave2_feather_miss_Nt}). A strong downward bias in the number of symptom onsets reported until the date of the nowcast was observed both for cases with known and with missing onset date (\Cref{fig:wave1_feather_miss_Nt}, grey and light grey bars). $N_t$ nowcasts conducted at the end of the same week were highly similar for all three approaches, however, differences were observed for longer lags, \ie one and two weeks after. Here, nowcasts from the generative approach consistently achieved a better WIS than the stepwise approaches. Nowcasts from the stepwise approach with independent imputation, which ignores the conditioning on the date of report, tended to underpredict the peak of the wave, and this bias increased with longer lags, as more data became available. In contrast, the stepwise approach with backward-delay imputation was generally unbiased, with approximately equal penalties for over- and under-prediction. However, as uncertainty from the imputation step was ignored, this approach yielded overconfident nowcasts at larger lags. Nowcasts from the generative approach generally had the widest uncertainty intervals.

\begin{figure}[!tbh]
    \vspace{-1.6cm}
    \centering
    \includegraphics[width=\textwidth]{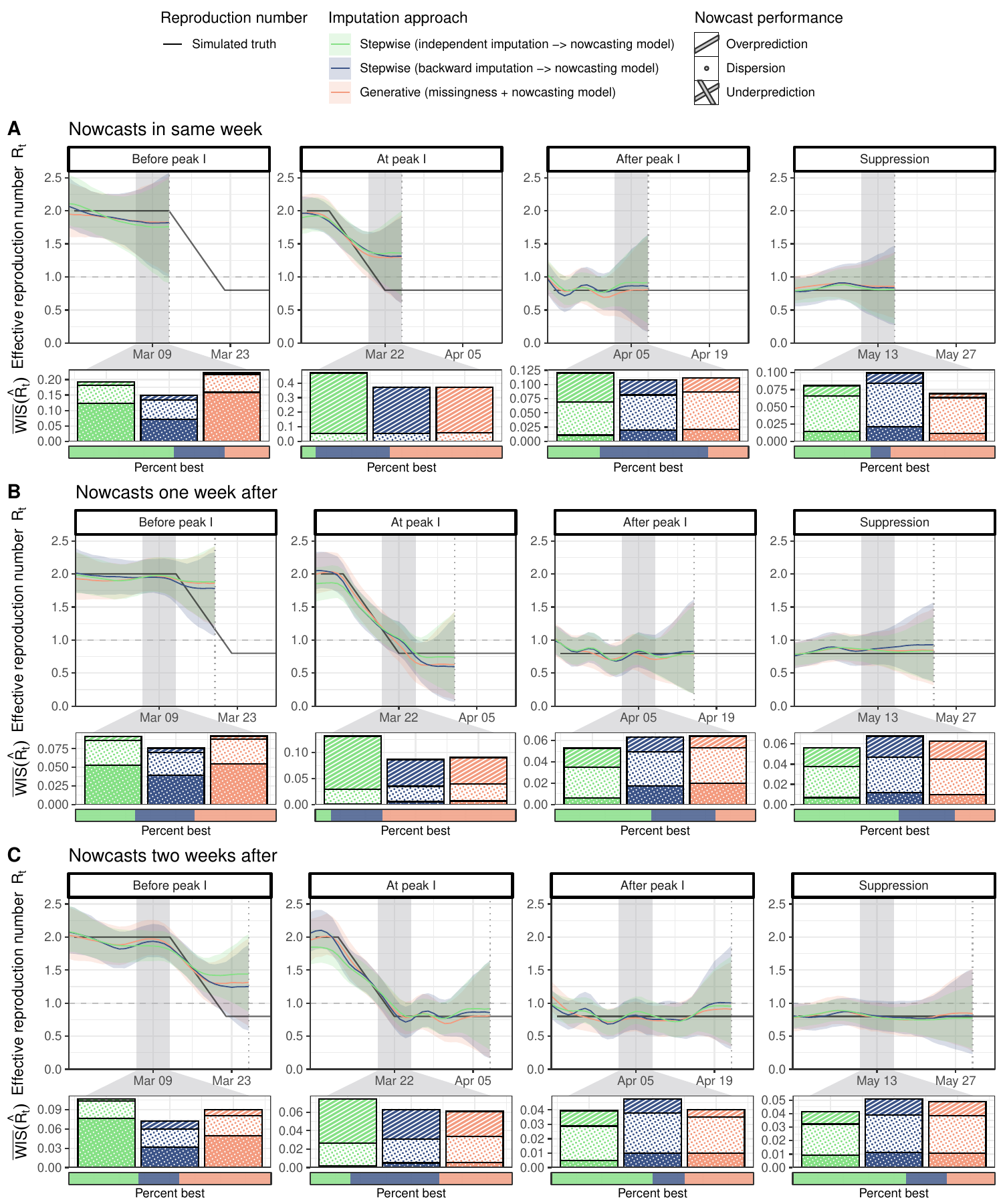}
    \caption{\textbf{Nowcasts of $R_t$ on incomplete line list data of a simulated first wave scenario using different approaches of accounting for missing onset dates.} Shown are the true $R_t$ (black) and point nowcasts with 95\% credible intervals (CrI) in four different phases of the epidemic wave, obtained through i) a stepwise approach using a forward imputation step (green), ii) a stepwise approach using a backward imputation step (blue), and iii) a generative approach using an integrated missingness model (red). All approaches used a generative model for nowcasting. Shown below each phase is the weighted interval score (WIS, lower is better) for $R_t$ nowcasts of each approach during a selected week (grey shade) over 50 scenario runs. Colored bars show average scores, decomposed into penalties for underprediction (crosshatch), dispersion (circles), and overprediction (stripes). The horizontal proportion bar below shows the percentages of runs where each approach performed best. Results are shown for nowcasts made at different lags from the selected week (vertical dotted lines), \ie at the end of the selected week (top row), one week later (middle row), and two weeks later (bottom row).}
    \label{fig:wave1_feather_miss_Rt}
\end{figure}

\Cref{fig:wave1_feather_miss_Rt} and \supp~Fig.~\zref{fig:wave2_feather_miss_Rt} show the corresponding results for nowcasts of $R_t$ in the first and second wave scenario, respectively. The three approaches to missing date imputation produced mostly similar $R_t$ nowcasts both at shorter and longer lags, but we observed a small bias of the stepwise approach with independent imputation around the change point of $R_t$ due to the underprediction of the peak in $N_t$. However, the differences in performance between the approaches were not consistent across phases and comparatively small with regard to the overall WIS, which was dominated by the uncertainty and error from the truncation adjustment and $R_t$ estimation.

\FloatBarrier

\subsection{COVID-19 hospitalizations in Switzerland}
We used line list data of hospitalized patients who tested positive for SARS-CoV-2 during the COVID-19 pandemic in Switzerland, reported between January~01, 2020~--~March 31, 2021, to retrospectively nowcast the number of hospitalizations by date of symptom onset and the effective reproduction number. Overall, the line list contained 25,831 cases of hospitalization, with symptom onset missing for 33\% of the cases (\supp~Fig.~\zref{fig:switzerland_date_of_report}). 137 cases had a symptom onset date later than the date of report, in which case the symptom onset date was likely a data entry error and set to missing (see \Cref{sec:methods_real}). Our assumed maximum delay of $D=56$ days covered 97\% of observed reporting delays (\supp~Fig.~\zref{fig:switzerland_emp_delay}), and cases with larger delays were removed from the analysis. We assumed different incubation period and generation time distributions for the wildtype and alpha variants of SARS-CoV-2 and accounted for the introduction and spread of the alpha variant during the second wave in Switzerland (see \Cref{sec:methods_real}). Nowcasts were conducted using i) the direct $R_t$ estimation approach, ii) a fully stepwise approach (using a backward-delay imputation step, a truncation adjustment step, and an $R_t$ estimation step), and iii) a fully generative approach (using a joint missingness, truncation, and renewal model).

\begin{figure}[!tbh]
    \vspace{-2cm}
    \centering
    \includegraphics[width=\textwidth]{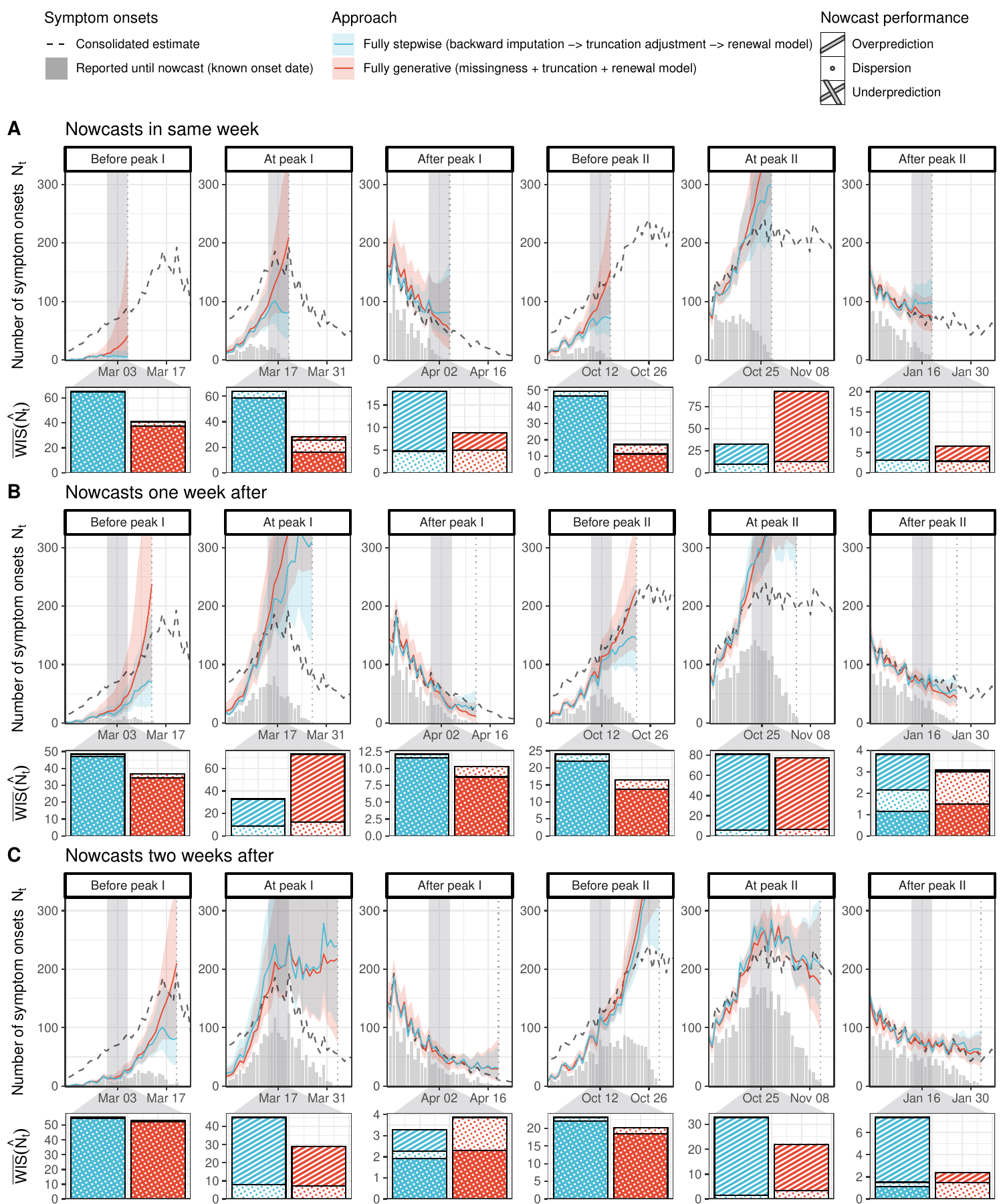}
    \caption{\textbf{Nowcasts of $N_t$ on incomplete hospitalization line list data during the COVID-19 pandemic in Switzerland.} Shown are nowcasts with 95\% credible interval (CrI) for the number of hospitalizations with COVID-19 by date of symptom onset $N_t$ in different phases of the first and second wave, obtained through i) a fully stepwise approach using cases by date of symptom onset with a backward imputation step and a truncation adjustment step (blue), and ii) a fully generative approach using an integrated missingness, truncation and renewal model (red). The direct approach using cases by date of report does not produce nowcasts of $N_t$ (not shown). Also shown are the number of cases by symptom onset date reported until the respective nowcast date (grey bars), and consolidated point estimates (7--14 days after maximum delay, averaged over all models) of the true $N_t$ (black dashed lines). Shown below each phase is the weighted interval score (WIS, lower is better) for $N_t$ nowcasts of each approach during a selected week (grey shade) evaluated on the consolidated point estimates. The scores (colored bars) are decomposed into penalties for underprediction (crosshatch), dispersion (circles), and overprediction (stripes). Results are shown for nowcasts made at different lags from the selected week (vertical dotted lines), \ie at the end of the selected week (top row), one week later (middle row), and two weeks later (bottom row).}
    \label{fig:switzerland_feather_Nt}
\end{figure}

\begin{figure}[!tbh]
    \vspace{-1.7cm}
    \centering
    \includegraphics[width=\textwidth]{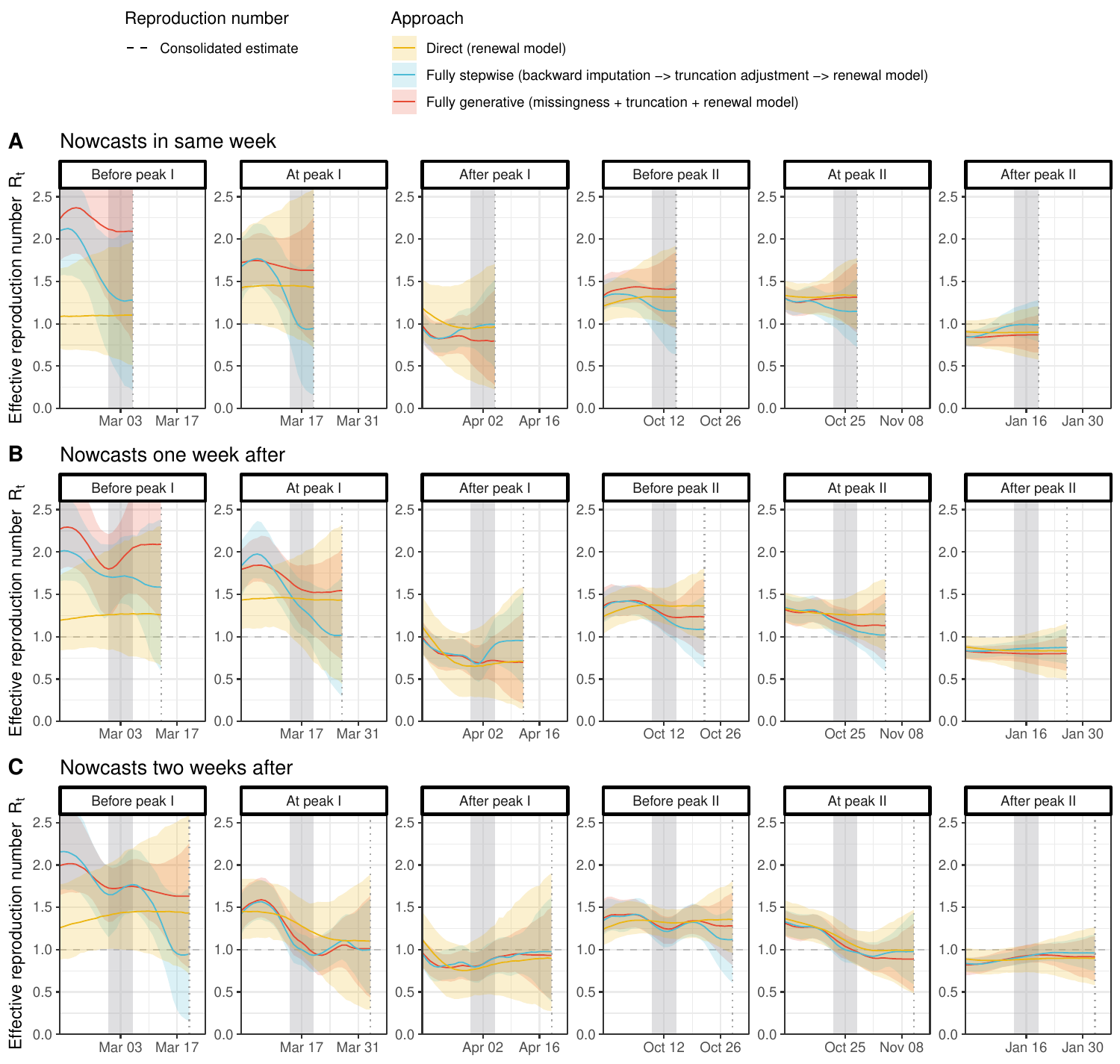}
    \caption{\textbf{Nowcasts of $R_t$ on incomplete hospitalization line list data during the COVID-19 pandemic in Switzerland.} Shown are nowcasts with 95\% credible interval (CrI) for the effective reproduction number $R_t$ in different phases of the first and second wave, obtained from hospitalization line list data through i) a direct approach using cases by date of report with no truncation adjustment (yellow), ii) a fully stepwise approach using cases by date of symptom onset with a backward imputation step and a truncation adjustment step (blue), and iii) a fully generative approach using cases by date of symptom onset with an integrated missingness, truncation, and renewal model (red). Results are shown for nowcasts made at different lags from the selected week (vertical dotted lines), \ie at the end of the selected week (top row), one week later (middle row), and two weeks later (bottom row).}
    \label{fig:switzerland_feather_Rt}
\end{figure}

\Cref{fig:switzerland_feather_Nt} shows nowcasts of $N_t$, \ie the number of hospitalizations by date of symptom onset, in different phases of the first and second COVID-19 wave in Switzerland. In contrast to nowcasts by date of hospitalization, these estimates include cases that have developed symptoms but are not yet hospitalized, and thus present a more timely indicator of the underlying infection dynamics. Here we found the same qualitative differences between the generative and the stepwise approach as on synthetic data. In particular, nowcasts from the stepwise approach remained mostly stationary on days close to the nowcast date, while nowcasts from the generative approach extrapolated the recent trend of exponential growth or decline. To assess real-time nowcasting performance with regard to $N_t$, we used consolidated estimates based on nowcasts lagged by 7--14 days beyond the maximum delay as a proxy ground truth (see \Cref{sec:methods_real}). Based on the consolidated estimates, the generative approach achieved a better WIS in all phases except at the peak of the wave. Nowcasts from the stepwise approach were slightly sharper than from the generative approach at larger lags, likely because of the neglected uncertainty from the missing onset date imputation. Of note, both nowcasting approaches underpredicted $N_t$ in the early phase of the epidemic in Switzerland, presumably due to an underestimation of the reporting delay based on very few initial hospitalization cases. 
\Cref{fig:switzerland_feather_Rt} shows corresponding nowcasts for the hospitalization-based $R_t$. Generally, we found $R_t$ nowcasts at a lag below one week to be largely dominated by the smoothing prior of the nowcasting model. This is partly to be expected, as transmission dynamics are additionally delayed by the incubation period, which was on average $\approx 5$ days in our case, thus real-time estimates for $R_t$ are generally less informed than for $N_t$. Of note, nowcasts using the direct $R_t$ estimation approach, which is based on the more delayed case counts $C_t$, were dominated by the smoothing prior even longer, until a lag of two weeks or more. $R_t$ nowcasts from the fully stepwise and fully generative approach mostly agreed after a lag of two weeks but showed systematic differences at shorter lags. Specifically, as in our results on synthetic data, nowcasts from the stepwise approach tended toward $R_t=1$, while nowcasts from the generative approach predicted $R_t$ to remain at its current level. Nowcasting uncertainty was mostly similar between the stepwise and the generative approach.

\FloatBarrier

\section{Discussion}
In this paper, we developed a fully generative model to nowcast the effective reproduction number $R_t$ from right-truncated line list data with missing symptom onset dates. Previous methods have addressed this task by separating missing onset date imputation, truncation adjustment, and $R_t$ estimation into consecutive, independent steps\cite{gunther_nowcasting_2021,salazar_near_2022,li_bayesian_2021}. Here we proposed to unify these tasks in a single generative model that can be fit directly to observed data. With such a model, all quantities of interest and their associated uncertainty can be estimated jointly in a Bayesian framework. This is in contrast to stepwise approaches, which face difficulties in propagating uncertainty between consecutive steps and often require resampling schemes that can restrict the individual components. Moreover, information can only flow in one direction, such that priors and model structure from later steps cannot inform earlier steps. Instead, our generative model enables shared regularization between imputation, truncation adjustment, and $R_t$ estimation, thereby eliminating the need for non-parametric smoothing at intermediate steps.

To compare the generative and the stepwise approaches with regard to performance and qualitative behavior, we applied them to synthetic line list data for different outbreak scenarios and to real-world data of hospitalizations during the COVID-19 pandemic in Switzerland. We found that under realistic delays between symptom onset and reporting of hospitalized cases, nowcasts for days close to the present are based on few reported cases and thus can be substantially influenced by prior assumptions about the smoothness and shape of the epidemic curve. In nowcasting practice, it is therefore important to be aware that under long reporting delays, real-time nowcasts can effectively become weakly informed forecasts due to a lack of data signal.

% Main findings regarding Rt estimation and truncation adjustment
When nowcasting $R_t$, estimates can be obtained either from case counts by date of report $C_t$, or from case counts by date of symptom onset $N_t$. The approach using $C_t$ can be more direct as it requires no truncation adjustment of case counts, but it depends on external estimates of the reporting delay distribution to infer the time series of infections. Our results show that the resulting $R_t$ estimates can still be substantially biased when using naive empirical estimates of the reporting delay. On the other hand, approaches using cases by date of symptom onset are more complex as they must adjust for the right truncation of the real-time case count. Existing methods conduct this truncation adjustment in a separate step, and we found that this can systematically over- and underpredict case counts depending on the current epidemic phase. Specifically, we demonstrated that when modeling the expected number of cases via a stationary random walk - as customary in earlier studies\cite{gunther_nowcasting_2021,mcgough_nowcasting_2020,salazar_near_2022,bastos_modelling_2019,bergstromBayesianNowcastingLeading2022} - nowcasts will be biased during periods of exponential growth or decline. Notably, our results show that the stationary smoothing can even overwrite prior assumptions in the subsequent $R_t$ estimation step and therefore bias the $R_t$ estimates towards $R_t=1$.
In our proposed generative approach, the renewal model assumed for $R_t$ estimation also serves as an epidemiologically motivated smoothing prior for the case counts. As a result, the generative model achieved better nowcasts of $N_t$ and $R_t$ in phases of epidemic growth or decline by extrapolating the current growth dynamics. At the immediate peak of the epidemic curve, this led to a higher chance of overpredicting $N_t$, which is a generally known trade-off in renewal modeling but at the same time the preferred minimal assumption absent a sufficient signal of changes in transmission\cite{bosse_comparing_2022}. Overall, our findings highlight the importance of adequate smoothing priors for nowcasting under long delays, and demonstrate how model misspecification can impact downstream estimates in stepwise modeling pipelines.

% Main findings regarding missingness model
When nowcasting from incomplete line list data with missing symptom onset dates, we found that even with up to 60\% of missing data, informative nowcasts of the total number of symptom onset dates could be made. To account for the missing onset dates, we implemented two stepwise approaches, i.e. with an additional i) independent imputation step and ii) backward-delay imputation step, as well as a generative approach that uses an integrated missingness model. As multiple imputation was not computationally feasible, the stepwise approaches only used a single imputation and thereby ignored uncertainty arising from the missing dates. In contrast, our generative model jointly fits to the complete and incomplete cases in the line list and naturally accounts for uncertainty arising from missing onset dates. Notably, we find that for short nowcasting lags, when few cases have been reported and overall nowcasting uncertainty is high, performance differences between the approaches were mostly negligible. At longer lags, however, the stepwise approach with independent imputation showed bias from misspecified delays especially around the peak of the epidemic curve. In contrast, the stepwise approach with backward-delay imputation correctly modeled delays during imputation and was thus unbiased but overconfident as it did not account for imputation uncertainty. $N_t$ nowcasts using the generative approach with an integrated missingness model had the highest uncertainty and achieved superior performance at longer lags. The resulting performance differences were however mostly with regard to $N_t$, and no clear differences with regard to $R_t$ could be identified, where the overall uncertainty is generally high. In the setting studied here, it thus seems that the uncertainty from missing onset dates may be small compared to the uncertainty from reporting delays and infection dynamics.
In our model, we specified $\alpha_t$, the probability for symptom onset dates to be known, with respect to the onset date $t$. It should be noted that in settings where symptom onset information for already reported cases is entered retrospectively in the line list, also known as backfilling, this probability may also depend on the date of report. That is, the probability for symptom onset dates to be known could be systematically lower for cases reported closer to the present. In such a setting, our generative model may still correctly account for cases with missing onset date, however, this could require a different prior for $\alpha_t$ which adequately represents a downward trend towards the present. When backfilling is the main source of missingness, it may also be preferable to index $\alpha_t$ by the date of report instead.

% Extensions and modifications
In our models, we used a common but simplistic non-parametric smoothing prior, \ie a first-order random walk, to smooth the time-varying parameters. In practice, a large variety of smoothing priors can be used for this purpose, with the potential of improving the performance of each approach. In a supplementary analysis, we found that when the stepwise approach is used with a non-stationary, exponential smoothing prior\cite{hyndman_forecasting_2018}, the tendency to over- or underpredict case counts during phases of epidemic growth or decline is strongly reduced. Nevertheless, some bias remained and it should be noted that the use of more complex smoothing priors may also introduce other, more subtle biases to $R_t$ estimation. At the same time, non-stationary smoothing priors such as innovations state space models\cite{hyndman_forecasting_2018} or Gaussian processes\cite{abbott_estimating_2020,rasmussen_gaussian_2006} can also be employed to smooth $R_t$ in a generative model, potentially enabling tighter credible intervals and better capturing changes in transmission dynamics. Moreover, additional knowledge or data could be integrated in such priors, \eg to provide more information on time-varying reporting delays or to account for population mobility and non-pharmaceutical interventions\cite{leung_real-time_2021,lison_estimating_2022}. In the latter case, a generative approach offers the advantage of modeling population-level effects on disease transmission directly via $R_t$ instead of via the expected number of symptom onsets $\lambda_t$, thereby ensuring temporal accuracy of the effect model. Moreover, due to the joint inference approach, better estimation of reporting delays may also improve the handling of missing onset dates, and more accurate models of transmission dynamics may also improve nowcasts of case counts. However, as the suitability of each option will vary with the specific setting and application, it is important that practical tools for real-time surveillance allow for flexible model specification. In line with the generative approach described in this paper, we have contributed a missingness and renewal model component to the open-source R package \emph{epinowcast}\cite{epinowcast}.

% Overall limitations
We note several general limitations of the methods used in this work. First, to model changes in the reporting delay over time, we have used a time-to-event model formulation in which effects by date of onset and date of report are applied proportionally on the logit scale to the hazard for each delay\cite{gunther_nowcasting_2021}. This assumption reduces model complexity, but limits the ability to represent certain types of distributional shifts over time and excludes negative delays. Second, we treated reported symptom onset dates as exact, while in reality there may be uncertainty and inconsistency in how onset dates are ascertained. We also note that line lists can include patients with asymptomatic infection, in which case the imputed symptom onset dates are only hypothetical events that serve as a proxy for the date of infection. This may be the case, for example, for patients who are hospitalized for other reasons and become infected in hospital, but will only bias $R_t$ estimates if the proportion of infections in hospital to the total number of infections changes over time. Third, like previous methods, our models encode a missing-at-random assumption for cases with missing symptom onset date. If missingness correlates with reporting delay, this assumption will be violated and could lead to bias in imputation and truncation adjustment. It is therefore recommendable to conduct sensitivity analyses if additional data about the presence of symptoms is available. In theory, such information could also be integrated into our framework to explicitly model correlation between reporting delays and missingness. Fourth, we assumed that the distributions for the generation interval and the incubation period are exactly known, which is seldom the case in practice. However, both in the generative and the stepwise approach, it is possible to account for uncertainty in these parameters\cite{abbott_estimating_2020}. Fifth, we assumed a fixed ascertainment proportion $\rho_t=\rho$ over time to ensure identifiability. Under this assumption, exact knowledge of the ascertainment proportion is not essential because a misspecification of $\rho$ does not bias $R_t$ and distorts uncertainty quantification only in extreme cases. It thus allows us to obtain $R_t$ estimates from hospitalization data that are indicative of population-wide transmission, but will lead to bias in time periods where $\rho_t$ changes. This is particularly important when there are differences in reporting delays or ascertainment proportions for cases from distinct subpopulations, \eg age groups. In such a case, it can be necessary to add group structure to the renewal and reporting delay components of the model\cite{vandekassteeleNowcastingNumberNew2019,epidemia,epinowcast}, to avoid bias from differing transmission dynamics between the subpopulations. Moreover, time-varying ascertainment proportions may also estimated by fitting to several data sources, \eg hospitalizations and deaths, simultaneously. Sixth, we have only tested a simplistic version of the direct $R_t$ estimation approach, with naive external estimates of the reporting delay. In practice, reporting delays may be externally estimated using more sophisticated methods that account for right truncation and censoring, and the $R_t$ estimation model can account for weekday effects and other modifiers of case counts\cite{EpiNow2}. Last, in our assessments using synthetic data, we stratified the performance of different approaches by epidemic phase. While this is helpful for understanding the behavior of a model, we do not recommend this approach for model selection in real-time surveillance, as it is generally difficult to identify the current epidemic phase.

% Conclusion / broader picture
We conclude by noting that the generative model proposed in this work naturally encompasses other forms of $R_t$ estimation. For example, in settings where no line list data is available, $R_t$ can only be estimated from the time series of reported cases as described in \Cref{sec:R_estimation}. However, this scenario corresponds to a special case in the generative model where all symptom onset dates are missing and a strong prior on the reporting delay is used. Similarly, $R_t$ is sometimes only estimated retrospectively, when case data are already fully reported, with some onset dates missing. This corresponds to a case in our model where all days of interest are sufficiently far in the past, beyond the maximum reporting delay. In addition, we have here used the example of nowcasting by date of symptom onset, but our framework is similarly applicable to other dates of reference, such as the date of positive test. 

By describing how cases arise from an infection process and are reported with time-varying stochastic delays and partially missing dates, generative modeling offers an interpretable framework for the development of user-friendly yet extensible nowcasting tools. Together with up-to-date, de-identified line list data with information on reporting delays, such tools are essential for real-time surveillance during epidemic outbreaks.

\newpage

%TC:ignore
\bibliography{references.bib}

\newpage

\section*{Declarations}

\subsection*{Ethics approval}

Ethics approval was not required for this study. 

\subsection*{Competing interests}

Tanja Stadler is president of the Swiss ``Scientific Advisory Panel COVID-19'' (\url{https://science-panel-covid19.ch/}). All other authors declare no competing interests.

\subsection*{Author Contributions}
\textbf{Conceptualization:} Adrian Lison, Tanja Stadler

\noindent\textbf{Data curation:} Adrian Lison

\noindent\textbf{Formal Analysis:} Adrian Lison

\noindent\textbf{Funding acquisition:} Tanja Stadler

\noindent\textbf{Investigation:} Adrian Lison

\noindent\textbf{Methodology:} Adrian Lison

\noindent\textbf{Project administration:} Adrian Lison

\noindent\textbf{Resources:} Tanja Stadler

\noindent\textbf{Software:} Adrian Lison, Sam Abbott

\noindent\textbf{Supervision:} Tanja Stadler

\noindent\textbf{Validation:} Adrian Lison, Jana Huisman

\noindent\textbf{Visualization:} Adrian Lison

\noindent\textbf{Writing – original draft:} Adrian Lison

\noindent\textbf{Writing – review \& editing:} Adrian Lison, Sam Abbott, Jana Huisman, Tanja Stadler

\subsection*{Acknowledgments}
We thank Felix Günther for insightful discussions about the modeling of missing data. We thank the Federal Office of Public Health (FOPH) in Switzerland for their collaboration and the provision of COVID-19 line list data for this study.

\subsection*{Data availability} 
The simulated line list data and corresponding nowcasts, as well as all statistical models and reproducible analysis scripts are available from zenodo at \url{https://doi.org/10.5281/zenodo.8279676}. Real-world line list data of COVID-19 hospitalizations in Switzerland cannot be shared publicly due to data protection regulations. The line list data are available under terms of data protection upon request from the Swiss Federal Office of Public Health (FOPH).

%TC:endignore
\end{document}

\doublespacing
\nolinenumbers

\title{\centering\LARGE\singlespacing{Supporting information:\\Generative Bayesian modeling to nowcast the effective reproduction number from line list data with missing symptom onset dates}}

% author list

\author[1,2*]{Adrian Lison}
\author[3]{Sam Abbott}
\author[4]{Jana Huisman}
\author[1,2]{Tanja Stadler}

\affil[1]{ETH Zurich, Department of Biosystems Science and Engineering, Zurich, Switzerland}
\affil[2]{SIB Swiss Institute of Bioinformatics}
\affil[3]{Centre for Mathematical Modelling of Infectious Diseases, London School of Hygiene and Tropical Medicine, London, UK}
\affil[4]{Massachusetts Institute of Technology, Department of Physics, Physics of Living Systems, Cambridge, MA, United States of America}
\affil[*]{Corresponding author: adrian.lison@bsse.ethz.ch}

\flushbottom
\maketitle
\thispagestyle{empty}

%\newpage

\sloppy
\raggedbottom

\appendix
\tableofcontents

%\listoffigures
%\listoftables

\newpage
\section{Details of modeling components}
The nowcasting approaches compared in this work use a varying number of models with different complexity. For instance, a fully stepwise approach for incomplete line list data uses a model for imputation, a model for truncation adjustment, and a model for $R_t$ estimation, while a fully generative approach only fits a single, joint model. At the same time, some models share certain characteristics, such as estimating a reporting delay distribution. In the following, we present these shared characteristics and discuss their modeling in detail.

\subsection{Time series smoothing}\label{appendix:smoothing}
All of our models include one or several latent, time-varying parameters, for example the effective reproduction number $R_t$, the expected number of symptom onsets $\lambda_t$, or the probability of a symptom onset date to become known $\alpha_t$. When we do not further decompose these parameters via a mechanistic model, we instead specify a non-parametric time series smoothing prior for them. This allows us to estimate the parameters from observed data while providing some regularization through the smoothing structure of the prior.

\subsubsection{Random walk model}\label{appendix:random_walk_model}
In our main analysis, we use the simple and popular random walk model\cite{mcgough_nowcasting_2020,bastos_modelling_2019,gunther_nowcasting_2021},
$$f(\theta_t)|\theta_{t-1} \sim N(f(\theta_{t-1}),{\sigma_{f(\theta)}}^2),$$
where $\theta$ is a time-varying parameter to estimate and $f$ is a link function, \eg a log link. We further use diffuse hyperpriors for the intercept $\theta_1$ and the variance ${\sigma_{f(\theta)}}^2$ of the random walk (see \Cref{appendix:priors}), allowing these to be simultaneously estimated from the data. In the context of nowcasting, where comparatively few data are available towards the present, this means that uncertainty about the most recent values of $\theta$ also depends on the estimated variance of the random walk. Intuitively, this reflects that expectations are less certain when an observed process exhibited high variance in the past.

\subsubsection{Exponential smoothing}\label{appendix:ets_model}
We tested the use of a non-stationary smoothing prior as an alternative to the simple random walk prior for $\lambda_t$ and $R_t$ in the stepwise approach and for $R_t$ in the generative approach. Here we used an innovations state space model with additive errors that implements an exponential smoothing prior according to Holt's linear trend method with dampening\cite{hyndmanForecastingPrinciplesPractice2018}. The time-varying parameter of interest, $\theta$, was modeled recursively as
\begin{align*}
    f(\theta_t) &= l_{t-1} + \phi' \, b_{t-1} + \epsilon_t \\
    l_t &= l_{t-1} + \phi' \, b_{t-1} + \alpha' \, \epsilon_t \\
    b_t &= \phi' \, b_{t-1} + \beta' \, \epsilon_t \\
    \epsilon_t &\sim N(0,{\sigma_{f(\theta)}}^2),
\end{align*}
where $l_t$ is the level, $b_t$ the trend, and $\epsilon_t$ the additive error at time $t$, respectively. Moreover, $\alpha'$ is a level smoothing parameter, $\beta'$ a trend smoothing parameter, and $\phi'$ a dampening parameter. As before, we use hyperpriors for the intercepts $l_1$ and $b_1$ and for the variance ${\sigma_{f(\theta)}}^2$ of the innovations (see \Cref{appendix:priors}), allowing these to be estimated from the data. We also use weakly informative priors for the smoothing parameters $\alpha'$ and $\beta'$ (\Cref{appendix:priors}). Due to identifiability issues, the dampening parameter was fixed at $\phi'=0.9$, meaning that the strength of the trend halves approximately every week of predicting into the future.

\subsubsection{Link functions}\label{appendix:link_functions}
We use different link functions for the time-varying parameters. For $\lambda_t$ in the stepwise truncation adjustment approach, we use a log link, which means that increments have a multiplicative effect on the expected number of symptom onsets, which is in line with an exponential growth model on $\lambda_t$, where the increments can be interpreted as instantaneous growth rates. For $\alpha_t$, we use a logit link, which ensures that $0<\alpha_t<1$. For $R_t$, we use a softplus link\cite{dugasIncorporatingSecondOrderFunctional2000}, \ie $f(x) = \text{softplus}^{-1}(x)$ with $\text{softplus}(x)=\frac{\log(1+e^{kx})}{k}$, where $k$ is a sharpness parameter. We choose $k=4$ to obtain a function that behaves effectively like $f(x)=x$ for practically relevant values of $R_t$ while ensuring $R_t>0$. Therefore, increments have approximately additive effects on $R_t$, which we slightly prefer as a default smoothing assumption, as it avoids asymmetric uncertainty intervals for $R_t$ towards the present and allows to specify an easily interpretable prior for the random walk variance. Such a model may for example be derived by assuming additive effects of interventions and behavioral changes\cite{sharma_how_2020}. However, if a maximum effective reproduction number can be assumed, a viable alternative may be the use of a generalized logit link\cite{epidemia}.

\subsection{Renewal modeling}

\subsubsection{Effective reproduction number}
In all our models, $R_t$ specifically refers to the \textit{instantaneous} effective reproduction number, \ie the expected number of secondary infections an individual would cause if conditions remained as they were at time $t$\cite{gostic_practical_2020,fraserEstimatingIndividualHousehold2007}. It is equal to the expected number of infections at time $t$ divided by the total infectiousness at time $t$, which depends on previous infections weighted by the generation interval distribution\cite{cori_new_2013}. $R_t$ is thus a measure of the overall per-person transmission at time $t$, including effects of population immunity as well as interventions and behavioral changes. In renewal modeling, $R_t$ is inferred as a latent parameter of the discrete renewal equation\cite{bhatt_semi-mechanistic_2020}
$$\iota_t = E[I_t] = R_t \sum_{s=1}^G \psi_s \, I_{t-s},$$
where $I_{t-s}$ is the number of infections that occurred $s$ days before $t$, and $\psi$ the generation interval distribution with maximum generation time $G$.

\subsubsection{Seeding of infections}\label{appendix:seeding}
The renewal equation is only properly defined for times $t>G$, since new infections may be generated from previous infections up to $G$ days before. The number of initial infections for $t \leq G$ must therefore be modeled using a non-mechanistic prior (also known as seeding when at the start of a new epidemic). We here use a random walk for $\iota_t$ on the log scale, \ie
$$\log(\iota_t)|\iota_{t-1} \sim N(\log(\iota_{t-1}),{\sigma_{\log(\iota)}}^2),$$
with hyperpriors for the intercept and random walk variance as described in \Cref{appendix:priors}. 

\subsubsection{Realized infections}
To account for noise in the infection process, we assume that the number of realized infections is Poisson distributed with rate $\iota_t$ (given through the renewal equation), \ie
$I_t|\iota_t \sim \text{Poisson}(\iota_t)$.
However, discrete latent parameters cannot be sampled using the Hamiltonian MCMC No-U-Turn sampler in Stan. Similar to earlier work\cite{banholzerEstimatingEffectsNonpharmaceutical2021,bhatt_semi-mechanistic_2020}, we thus use the Normal distribution as a continuous approximation of the Poisson distribution, \ie
$I_t|\iota_t \sim \text{Normal}(\iota_t, \sqrt{\iota_t})$.

\subsubsection{EpiEstim}\label{appendix:epiestim}

The package EpiEstim implements a Bayesian framework by Cori et al.\cite{cori_new_2013} to directly compute $R_t$ estimates from the time series of cases by symptom onset date $N_t$. Under this framework, using a $\text{Gamma}(a,b)$ prior for $R_t$ and a smoothing window of $\tau$ days, the posterior distribution of $R_t$ is $$R_t \sim \text{Gamma}\left(a + \sum_{k=0}^{\tau-1} N_{t-k},\; \frac{1}{\frac{1}{b} + \sum_{k=0}^{\tau-1} \sum_{s=1}^{G} N_{t-k-s} \psi_{s}}\right).$$

\subsection{Modeling of reporting delays}
\subsubsection{Time-to-event model and reporting effects}\label{appendix:time-to-event_model}
We express the reporting delay probabilities $p_{t,d}$ via a discrete time-to-event model\cite{hohle_bayesian_2014,gunther_nowcasting_2021}. That is, we model the so-called hazard of reporting $h_{t,d}$, \ie the conditional probability of a case with symptom onset on date $t$ to be reported with a delay of $d$ days, given that it is not reported earlier. We use logistic regression to parameterize $h_{t,d}$, \ie
$$\text{logit}(h_{t,d}) = \gamma_d + z_t^\top \beta + w_{t+d}^\top \, \eta.$$
Here, $\gamma_d$ is the logit baseline hazard for delay $d$. Since $\gamma_d$ is estimated as an independent parameter, we obtain a non-parametric model for the reporting delay distribution with flexible shape. To account for differences in reporting over time, $z_t^\top \beta$ and $w_{t+d}^\top \, \eta$ are added as additional predictors. Here, $z_t$ is a vector of covariates dependent on symptom onset date $t$, and $w_{t+d}$ is a vector of covariates dependent on the report date, which is $d$ days after symptom onset date $t$. Moreover, $\beta$ and $\eta$ are vectors of coefficients to be estimated. This specification implements a proportional hazards model\cite{coxRegressionModelsLifeTables1972} for the reporting delay, where $z_t^\top \beta$ describes effects on the hazard by date of symptom onset, and $w_{t+d}^\top \, \eta$ describes effects by date of report.

We design $z_t$ to account for changes in the reporting delay of cases over time via a piecewise linear change point model\cite{gunther_nowcasting_2021}, \ie
$$z_t = (z_{t,1}, z_{t,2}, \dots, z_{t,n}), \; z_{t,i} = \max(0, \min(iK-(T-t), K)),$$
where the $i$ represent $n = ((T-1)~\text{div}~K)+1$ different change points, spaced at $K=7$ days distance and indexed backward in time from the present (the first changepoint is $K$ days before the present). Note that as the most distant changepoint is not necessarily preceded by $K$ days, it is instead specified as $z_{t,n} = \max(0, \min((n-1)K + ((T-1) \bmod K) -(T-t), ((T-1) \bmod K)))$.

We design $w_{t+d}$ to indicate different weekdays using dummy encoding, \ie
$$w_{t+d} = (w_{t+d,1}, w_{t+d,2}, \dots, w_{t+d,6}),\; w_{t+d,i} = \mathds{1}_{\text{weekday}(t+d)=i},$$
where $i=1$ represents Mondays and $i=6$ represents Saturdays (Sundays represent the intercept). When applying our models to real-world data, public holidays were coded as Sundays.

Given the hazard $h_{t,d}$, the probability for a case with symptom onset on date $t$ to be reported with delay $d$ is\cite{hohle_bayesian_2014} $$p_{t,d} = h_{t,d}\,\prod_{i=1}^{d} (1-h_{t,d-i}) = h_{t,d} (1-\sum_{i=1}^{d} p_{t,d-i}).$$

\subsubsection{Backward delays}\label{appendix:backward_delays}
When modeling backward delays, we use a model analogous to the one for forward delays. For a case with date of report $t'$ and reporting delay $d$, the backward reporting hazard is then
$$\text{logit}(h^{\leftarrow}_{t',d}) = \gamma_d + z_{t'-d}^\top \beta + w_{t'}^\top \, \eta.$$
That is, the effects by date of symptom onset and by date of report are indexed relative to the current date of report. Importantly, if data are left-censored, \ie only cases with date of symptom onset $t \geq 1$ are available, then the backward delay must only be estimated for $t' > D$, otherwise it will be biased by left-truncation.

\subsection{Modeling of missing symptom onset dates}\label{appendix:missing_onsets_model}

By introducing the time-varying parameter $\alpha_t$, \ie the probability of a case to be complete, that is to have a known onset date, we can explicitly distinguish between case counts with known and with missing symptom onset date in our model ($N_{t,d}^{\text{known}}$ and $N_{t,d}^{\text{missing}}$). We here index $\alpha_t$ by the date of symptom onset $t$, but it is in theory also possible to use the date of report $t'=t+d$ instead. In the latter case (date of report), we would have
\begin{equation*}
N_{t,d}^{\text{known}} | \lambda_t, p_{t,d}, \alpha_{t+d} \sim \text{Poisson}(\lambda_t \, p_{t,d} \, \alpha_{t+d}), \quad N_{t,d}^{\text{missing}} | \lambda_t, p_{t,d}, \alpha_{t+d} \sim \text{Poisson}(\lambda_t \, p_{t,d} \, (1-\alpha_{t+d})).
\end{equation*}
Which version is to be preferred will depend on assumptions about the process leading to missing onset dates. We only expect larger differences when $a_t$ is specified via a more complex time series model which \eg accounts for weekday effects or trends towards the present.

In the joint model for missing onset dates, truncation adjustment, and $R_t$ estimation as specified in \Cref{appendix:modelmisstruncR}, only observations $C_{t}^{\text{missing}}$ for $t>D$ can be used. This is because the counts $C_{t}^{\text{missing}}$ with $t\leq D$, potentially include cases with symptom onset before $t=1$. Ignoring these cases would thus lead to bias from left truncation. If $C_{t}^{\text{missing}}$ with $t\leq D$ should be included as observations as well, it is possible to extend the latent parameters $\lambda$, $p$, and $\alpha$ up to $D$ days further into the past. If $\lambda_t$ is modeled via a generative renewal model as in \Cref{appendix:modelmisstruncR}, this means that all corresponding parameters such as $I_t$, $\iota_t$, and $R_t$ must also be extended further into the past. An important disadvantage of this approach is that these extended parameters are potentially informed by very little data. In particular, the reporting delay probabilities $p_{t,d}$ for $t<1$ are not directly informed by any observations in the line list, meaning that weekday effects but no trends over time can be modeled for these delays.

\section{Full model definitions}\label{appendix:model_specs}

To realize the different approaches compared in this study, we implemented (\modelR) a model for reproduction number estimation, (\modeltrunc) a model for truncation adjustment, (\modeltruncR) a joint model for truncation adjustment and reproduction number estimation, (\modelimpute) a model for missing onset date imputation using backward delays, and (\modelmisstruncR) a joint model for missing onset date, truncation adjustment, and reproduction number estimation. In the following, we provide concise, formal definitions of each model. We refer the reader to the main text for detailed explanations of the notation and variables.

\subsection{Model \rom{1}: $R_t$ estimation}\label{appendix:modelRt}

\begin{align*}
    & \textbf{Model} \\
    & C_t | \lambda_t \sim \text{Poisson}(\lambda_t) \;/\; \text{NegBin}(\lambda_t) & \text{cases by report date} \\
    & \lambda_t = \sum_{d=0}^D I_{t-d} \, \rho_{t-d} \, p_d & \text{expected cases} \\
    & I_t | \iota_t \sim \text{Poisson}(\iota_t) & \text{realized infections} \\
    & \log(\iota_t)|\iota_{t-1} \sim N(\log(\iota_{t-1}),{\sigma_{\log(\iota)}}^2) \;|\; t \leq G & \text{expected infections (seeding)} \\
    & \iota_t = E[I_t] = R_t \sum_{s=1}^G \psi_s \, I_{t-s} \;|\; t > G & \text{expected infections (renewal)} \\
    & \text{softplus}(R_t)|R_{t-1} \sim N(\text{softplus}(R_{t-1}),{\sigma_R}^2) & \text{reproduction number} \\
    \\
    & \textbf{Fixed parameters} \\
    & \rho_t = \rho & \text{ascertainment proportion} \\
    & p = (p_0, p_1, \dots, p_D) & \text{reporting delay distribution} \\
    & \psi = (\psi_1, \psi_2, \dots, \tau_G) & \text{generation time distribution} \\
    \\
    & \textbf{Parameters with priors (see \ref{appendix:priors} for priors)} \\
    & \frac{1}{\sqrt{\phi}} & \text{overdispersion} \\
    & \log(\iota_1), \sigma_{\log(\iota)} & \text{random walk seeding phase} \\
    & R_1, \sigma_R & \text{random walk reproduction number}
\end{align*}

\subsection{Model \rom{2}: Truncation adjustment}\label{appendix:modeltrunc}
\begin{align*}
    & \textbf{Model} \\
    & N_{t,d} | \lambda_t, p_{t,d}, \phi \sim \text{Poisson}(\lambda_t \, p_{t,d}) \;/\; \text{NegBin}(\lambda_t \, p_{t,d}, \phi) & \text{cases by onset date \& delay} \\
    & p_{t,d} = h_{t,d}\,\prod_{i=1}^{d} (1-h_{t,d-i}) & \text{reporting delay probabilities} \\
    & \text{logit}(h_{t,d}) = \gamma_d + z_t^\top \beta + w_{t+d}^\top \, \eta & \text{reporting hazards} \\
    & \log(\lambda_t)|\lambda_{t-1} \sim N(\log(\lambda_{t-1}),{\sigma_{\log(\lambda)}}^2) & \text{expected onsets} \\
    \\
    & \textbf{Covariates} \\
    & z_t = (z_{t,1}, z_{t,2}, \dots, z_{t,n})  & \text{date of symptom onset covariates} \\
    & w_{t+d} = (w_{t+d,1}, w_{t+d,2}, \dots, w_{t+d,6}) & \text{date of report covariates}  \\
    \\
    & \textbf{Parameters with priors (see \ref{appendix:priors} for priors)} \\
    & \frac{1}{\sqrt{\phi}} & \text{overdispersion} \\
    & \gamma_d & \text{baseline reporting hazard} \\
    & \beta_i, \eta_i & \text{reporting delay effects} \\
    & \log(\lambda_1), \sigma_{\log(\lambda)} & \text{random walk expected onsets}
\end{align*}

\subsection{Model \rom{3}: Joint truncation adjustment and $R_t$ estimation}\label{appendix:modeltruncRt}
\begin{align*}
    & \textbf{Model} \\
    & N_{t,d} | \lambda_t, p_{t,d}, \phi \sim \text{Poisson}(\lambda_t \, p_{t,d}) \;/\; \text{NegBin}(\lambda_t \, p_{t,d}, \phi) & \text{cases by onset date \& delay} \\
    & p_{t,d} = h_{t,d}\,\prod_{i=1}^{d} (1-h_{t,d-i}) & \text{reporting delay probabilities} \\
    & \text{logit}(h_{t,d}) = \gamma_d + z_t^\top \beta + w_{t+d}^\top \, \eta & \text{reporting hazards} \\
    & \lambda_t = \sum_{s=0}^L I_{t-s} \, \rho_{t-s} \, \tau_s & \text{expected symptom onsets} \\
    & I_t|\iota_t \sim \text{Poisson}(\iota_t) & \text{realized infections} \\
    & \log(\iota_t)|\iota_{t-1} \sim N(\log(\iota_{t-1}),{\sigma_{\log(\iota)}}^2) \;|\; t \leq G & \text{expected infections (seeding)} \\
    &\iota_t = E[I_t] = R_t \sum_{s=1}^G \psi_s \, I_{t-s} \;|\; t > G & \text{expected infections (renewal)} \\
    &\text{softplus}(R_t)|R_{t-1} \sim N(\text{softplus}(R_{t-1}),{\sigma_R}^2) & \text{reproduction number} \\
    \\
    & \textbf{Covariates} \\
    & z_t = (z_{t,1}, z_{t,2}, \dots, z_{t,n})  & \text{date of symptom onset covariates} \\
    & w_{t+d} = (w_{t+d,1}, w_{t+d,2}, \dots, w_{t+d,6}) & \text{date of report covariates}  \\
    \\
    & \textbf{Fixed parameters} \\
    & \rho_t = \rho & \text{ascertainment proportion} \\
    & \tau = (\tau_0, \tau_1, \dots, \tau_L) & \text{incubation period distribution} \\
    & \psi = (\psi_1, \psi_2, \dots, \tau_G) & \text{generation time distribution} \\
    \\
    & \textbf{Parameters with priors (see \ref{appendix:priors} for priors)} \\
    & \frac{1}{\sqrt{\phi}} & \text{overdispersion} \\
    & \gamma_d & \text{baseline reporting hazard} \\
    & \beta_i, \eta_i & \text{reporting delay effects} \\
    & \log(\iota_1), \sigma_{\log(\iota)} & \text{random walk seeding phase} \\
    & R_1, \sigma_R & \text{random walk reproduction number}
\end{align*}

\subsection{Model \rom{4}: Missing onset date imputation (backward-delay)}\label{appendix:modelmiss}

\begin{align*}
    & \textbf{Model} \\
    & (N_{t,0}^{\text{known}},\dots,N_{t-D,D}^{\text{known}}) \sim \text{Multinom}(p^{\leftarrow}_{t,0},\dots,p^{\leftarrow}_{t,D}) & \text{complete cases by onset date \& delay} \\
    & p^{\leftarrow}_{t,d} = h^{\leftarrow}_{t,d}\,\prod_{i=1}^{d} (1-h^{\leftarrow}_{t,d-i}) & \text{backward reporting delay probabilities} \\
    & \text{logit}(h^{\leftarrow}_{t',d}) = \gamma_d + z_{t'-d}^\top \beta + w_{t'}^\top \, \eta & \text{backward reporting hazards} \\
    \\
    & \textbf{Covariates} \\
    & z_t = (z_{t,1}, z_{t,2}, \dots, z_{t,n})  & \text{date of symptom onset covariates} \\
    & w_{t+d} = (w_{t+d,1}, w_{t+d,2}, \dots, w_{t+d,6}) & \text{date of report covariates} \\
    \\
    & \textbf{Parameters with priors (see \ref{appendix:priors} for priors)} \\
    & \gamma_d & \text{backward baseline reporting hazard} \\
    & \beta_i, \eta_i & \text{backward reporting delay effects} \\
    \\
    & \textbf{Posterior predictive samples} \\
    & p^{\leftarrow^{(i)}}_{t,d} \sim P\left(p^{\leftarrow}_{t,d}, \theta \;\middle|\; N_{\{t,d | t+d \leq T\}}^{\text{known}}\right) \\
    & d_{t}^{\leftarrow^{(i)}} \sim \text{Categorical}(p^{\leftarrow}_{t,0}, p^{\leftarrow}_{t,1}, \dots, p^{\leftarrow}_{t,D})
\end{align*}

\subsection{Model \rom{5}: Joint missing onset date, truncation adjustment, and $R_t$ estimation}\label{appendix:modelmisstruncR}

\begin{align*}
    & \textbf{Model} \\
    & N_{t,d}^{\text{known}} | \lambda_t, p_{t,d}, \alpha_t \sim \genfrac{}{}{0pt}{}{\text{Poisson}(\lambda_t \, p_{t,d} \, \alpha_t)\;/\hfill}{\text{NegBin}(\lambda_t \, p_{t,d} \, \alpha_t, \phi)\hfill}  & \genfrac{}{}{0pt}{}{\hfill\text{complete cases}}{\hfill\text{by onset date \& delay}} \\
    & C_{t}^{\text{missing}} | \lambda, p, \alpha \sim \genfrac{}{}{0pt}{}{\text{Poisson} (\lambda_t^{\text{mising}})\;/\hfill}{\text{NegBin}(\lambda_t^{\text{mising}}, \phi)\hfill} & \text{incomplete cases by report date} \\
    & \lambda_t^{\text{mising}} = \sum_{d=0}^{\min(D,t-1)} \lambda_{t-d} \, p_{t-d,d} \, (1-\alpha_{t-d}) & \genfrac{}{}{0pt}{}{\hfill\text{expected incomplete cases}}{\hfill\text{by report date}} \\
    & p_{t,d} = h_{t,d}\,\prod_{i=1}^{d} (1-h_{t,d-i}) & \text{reporting delay probabilities} \\
    & \text{logit}(h_{t,d}) = \gamma_d + z_t^\top \beta + w_{t+d}^\top \, \eta & \text{reporting hazards} \\
    & \text{logit}(\alpha_t)|\alpha_{t-1} \sim N(\text{logit}(\alpha_{t-1}),{\sigma_{\text{logit}(\alpha)}}^2) & \text{probability of known onset date} \\
    & \lambda_t = \sum_{s=0}^L I_{t-s} \, \rho_{t-s} \, \tau_s & \text{expected symptom onsets} \\
    & I_t|\iota_t \sim \text{Poisson}(\iota_t) & \text{realized infections} \\
    & \log(\iota_t)|\iota_{t-1} \sim N(\log(\iota_{t-1}),{\sigma_{\log(\iota)}}^2) \;|\; t \leq G & \text{expected infections (seeding)} \\
    &\iota_t = E[I_t] = R_t \sum_{s=1}^G \psi_s \, I_{t-s} \;|\; t > G & \text{expected infections (renewal)} \\
    &\text{softplus}(R_t)|R_{t-1} \sim N(\text{softplus}(R_{t-1}),{\sigma_R}^2) & \text{reproduction number} \\
    \\
    & \textbf{Covariates} \\
    & z_t = (z_{t,1}, z_{t,2}, \dots, z_{t,n})  & \text{date of symptom onset covariates} \\
    & w_{t+d} = (w_{t+d,1}, w_{t+d,2}, \dots, w_{t+d,6}) & \text{date of report covariates}  \\
    \\
    & \textbf{Fixed parameters} \\
    & \rho_t = \rho & \text{ascertainment proportion} \\
    & \tau = (\tau_0, \tau_1, \dots, \tau_L) & \text{incubation period distribution} \\
    & \psi = (\psi_1, \psi_2, \dots, \tau_G) & \text{generation time distribution} \\
    \\
    & \textbf{Parameters with priors (see \ref{appendix:priors} for priors)} \\
    & \frac{1}{\sqrt{\phi}} & \text{overdispersion} \\
    & \gamma_d & \text{baseline reporting hazard} \\
    & \beta_i, \eta_i & \text{reporting delay effects} \\
    & \text{logit}(\alpha_1), \sigma_{\text{logit}(\alpha)} & \genfrac{}{}{0pt}{}{\hfill\text{random walk for}}{\hfill\text{probability of known onset}}\\
    & \log(\iota_1), \sigma_{\log(\iota)} & \text{random walk seeding phase} \\
    & R_1, \sigma_R & \text{random walk reproduction number}
\end{align*}

\subsection{Priors}\label{appendix:priors}

\Cref{tbl:prior_choices} provides an overview of the priors used for parameters in different model components. Priors were specified broadly to provide regularization to the estimation without distorting results, and the same priors were used across all models. Note that for the random walk models on time-varying parameters such as $\lambda_t$, $R_t$ or $\alpha_t$, we used a hyperprior that allowed the random walk variance to be simultaneously estimated from the data.

For the baseline hazard in the reporting delay model, we provided a constant, diffuse prior across all delays, corresponding to a piecewise exponential distribution in expectation, but with sufficient uncertainty to accommodate various delay distributions. When modeling symptom onsets in model \modeltrunc, the prior for the intercept of $\lambda$ was centered on a rough empirical estimate of the expected number of symptom onsets at $t=1$, \ie $\hat{\lambda}_1 = \sum_{d=0}^D (N_{1,d}^{\text{known}} + \frac{1}{D} C_{1+d}^{\text{missing}})$. Similarly, for the renewal model used in models \modelR, \modeltruncR, and \modelmisstruncR, the intercept of infections in the seeding phase was given a prior centered on $\hat{\iota}_1 = \frac{1}{\rho} \hat{\lambda}_1 $, with ascertainment proportion $\rho$. Such use of data in the prior only served to constrain the infection time series to the correct order of magnitude. When modeling the effective reproduction number, the use of a softplus link meant that $R_t$ is essentially described via a random walk on the unit scale. Here we chose a wide variance prior that would theoretically allow fast changes in $R_t$, such as dropping from 4 to 1 within one week. 

\renewcommand{\arraystretch}{1.2}
\begin{table}[H]
\caption{\textbf{Prior choices for model parameters.}}
\centering
\footnotesize
\begin{tabular}{p{1.3cm}p{4.5cm}p{3.7cm}p{4.5cm}p{1.7cm}}
\toprule
Parameter & Description & Prior & Interpretation & Models \\ 
\midrule
\noalign{\vskip 2mm}
\multicolumn{3}{l}{\textbf{Symptom onsets and transmission}} & \\
$\log(\lambda_1)$ & Intercept for $\log$ symptom onsets & $\text{Normal}(\mu = \hat{\lambda}_1, \sigma = 0.5)$ & Empirical estimate of number of cases with $\times 3$ uncertainty margin & \modeltrunc \\

$\sigma_{\log(\lambda)}$ & Standard deviation of random walk on $\log(\lambda_t)$ & $\text{Normal}^+(\mu = \frac{1}{20}, \sigma = \frac{1}{40})$ & Daily relative change in cases between $\pm20\%$ & \modeltrunc \\

$\log(\iota_1)$ & Intercept for $\log$ infections in seeding phase & $\text{Normal}(\mu = \hat{\iota}_1, \sigma = 0.5)$ & Empirical estimate of number of infections with $\times 3$ uncertainty margin & \modelR, \modeltruncR, \modelmisstruncR \\

$\sigma_{\log(\iota)}$ & Standard deviation of random walk on $\log(\iota_t)$ & $\text{Normal}^+(\mu = \frac{1}{20}, \sigma = \frac{1}{40})$ & Daily relative change in infections between $\pm20\%$ & \modelR, \modeltruncR, \modelmisstruncR \\

$R_1$ & Intercept for effective reproduction number & $\text{Normal}(\mu = 1, \sigma = 0.8)$ & Reproduction number initially between 0 and 3 & \modelR, \modeltruncR, \modelmisstruncR \\

$\sigma_R$ & Standard deviation of random walk on $R_t$ & $\text{Normal}^+(\mu = 0, \sigma = 0.1)$ & Weekly absolute increase or decrease of $R_t$ between $\pm3$ & \modelR, \modeltruncR, \modelmisstruncR \\

\noalign{\vskip 2mm}
\multicolumn{2}{l}{\textbf{Reporting delays}} & & \\
$\gamma_d$ & Baseline logit reporting hazard for a delay of $d$ days & \makecell{$\text{Normal}(\mu = \mu_\gamma, \sigma = \sigma_\gamma),$ \\ $\mu_\gamma = \text{logit}(1-0.01^{1/D}),$ \\$ \sigma_\gamma = \frac{1}{2} (\text{logit}(0.98) - \mu_\gamma)$}  & Piecewise exponential reporting delay with $P(d=0)$ between 2\% and 98\% and expectation that $P(d<D) = 99\%$ & \modeltrunc, \modeltruncR, \modelimpute, \modelmisstruncR \\

$\beta_i$ & Linear trend of the logit reporting hazard during week $i$ & $\text{Normal}(\mu = 0, \sigma = 0.1)$ & Odds for daily reporting differ between weeks maximally by a factor of 4 & \modeltrunc, \modeltruncR, \modelimpute, \modelmisstruncR \\

$\eta_i$ & Effect of weekday $i$ on logit reporting hazard & $\text{Normal}(\mu = 0, \sigma = 0.75)$ & Odds for daily reporting differ between weekdays maximally by a factor of 20 & \modeltrunc, \modeltruncR, \modelimpute, \modelmisstruncR \\

\noalign{\vskip 2mm}
\multicolumn{2}{l}{\textbf{Missing onset dates}} & & \\
$\text{logit}(\alpha_1)$ & Intercept for probability of known onset date & $\text{Normal}(\mu = 0, \sigma = 2)$ & Probability $\alpha_1$ between 2\% and 98\% & \modelmisstruncR \\

$\sigma_{\text{logit}(\alpha)}$ & Standard deviation of random walk on $\text{logit}(\alpha_t)$ & $\text{Normal}^+(\mu = 0, \sigma = 0.5)$ & Odds for missing onset dates differ between consecutive days maximally by a factor of 3 & \modelmisstruncR \\

\noalign{\vskip 2mm}
\multicolumn{2}{l}{\textbf{Observations}} & & \\
$\frac{1}{\sqrt{\phi}}$ & Overdispersion parameter for case counts & $\text{Normal}^+(\mu = 0, \sigma = 1)$ & Generic overdispersion prior for negative binomial\cite{stan_development_team_prior_2020} & \makecell{\modelR, \modeltrunc, \modeltruncR, \modelmisstruncR \\(optional)} \\

\noalign{\vskip 2mm}
\multicolumn{2}{l}{\textbf{Exponential smoothing}} & & \\
$\alpha'$ & Smoothing parameter for level & $\text{Beta}(\alpha = 5, \beta = 5)$ & the last 3--20 days of observations (including the current day) determine over 99\% of the current level & \makecell{\modelR, \modeltrunc, \modeltruncR, \modelmisstruncR \\(ETS version)} \\
$\beta'$ & Smoothing parameter for trend & $\text{Beta}(\alpha = 5, \beta = 5)$ & the last 3--20 days of observations (including the current day) determine over 99\% of the current trend & \makecell{\modelR, \modeltrunc, \modeltruncR, \modelmisstruncR \\(ETS version)} \\

\bottomrule
\multicolumn{4}{l}{$\text{Normal}^+$: Normal distribution truncated from below at zero}
\end{tabular}
\label{tbl:prior_choices}
\end{table}

\section{Implementation and estimation}

\subsection{Implementation details}\label{appendix:implementation}

We implemented models \modelR--\modelmisstruncR~as probabilistic programs in Stan\cite{stan_development_team_stan_2022}. The implementation of model \modeltrunc~was adapted from the largely similar model by Günther et al.\cite{gunther_nowcasting_2021}, available at \url{https://github.com/FelixGuenther/nc_covid19_bavaria} and improved for efficiency. In particular, we used log scale (or logit scale) versions of the parameters $\lambda_t$, $p_{t,d}$, and $\alpha_t$ and propagated them until the final likelihood to reduce computations and to improve numerical stability. We used non-centered parameterizations for the noise components of the random walk and innovations state space models.

For the stepwise approaches, posterior samples were extracted from the fitted model for the previous step and used as data input for the subsequent step. In the case of imputation, we used only a single random sample of the posterior distribution of imputed symptom onset dates from model \modelimpute~to compute the imputed case counts $\hat{N}_{t,d}$, which were then used as input for model \modeltrunc. In the case of reproduction number estimation, we extracted 50 time series of $\hat{N}_t$ drawn from the posterior predictive distribution of symptom onsets from model \modeltrunc~and fitted model \modelR~(or EpiEstim) separately on each time series. The resulting samples of the posterior for $R_t$ from the 50 different runs of model \modelR~were then combined into an overall posterior distribution of $R_t$.

In a supplementary analysis, we also estimate $R_t$ with EpiEstim instead of model \modelR. For this, we applied EpiEstim with a centered smoothing window of $\tau=7$ days separately for each time series draw of $\hat{N}_t$ from the fitted model \modeltrunc, obtaining estimates for the posterior mean and standard deviation of $R_t$ over time. For each draw, we then sampled a trajectory for $R_t$ by drawing from a Gamma distribution with the corresponding mean and standard deviation at each time step. The resulting samples of $R_t$ were then combined into an overall posterior distribution for $R_t$. Finally, because $R_t$ estimates obtained on the time series of symptom onsets instead of infections lag the actual transmission dynamics\cite{gostic_practical_2020}, we shifted the estimates of $R_t$ back in time by the mean incubation period as well as by half of the smoothing window.

\subsection{Sampling}\label{appendix:sampling}
For sampling of Stan models, we used cmdstan version 2.30.0 via cmdstanr version 0.5.0\cite{cmdstanr} with the default configuration of the No-U-Turn sampler, \ie four chains with 1,000 warm-up and 1,000 sampling iterations each\cite{stan_development_team_stan_2022}. To achieve a faster warm-up, important parameters of the model were initialized using crude empirical estimates from the data. For example, the baseline reporting hazard was initialized using an empirical estimate of the delay distribution ignoring changes over time, weekday effects, and truncation, and the time series of infections was initialized by shifting the time series of observed symptom onsets backward in time by the mean incubation period.
 
\subsection{Diagnostics}\label{appendix:diagnostics}
Each fitted model was automatically checked for ineffective sampling, indicated by low effective sample sizes (ESS < 400)\cite{geyer_mcmc_2011}, and for convergence problems, indicated by divergent transitions\cite{betancourt_conceptual_2018}, a low Bayesian fraction of missing information (E-BFMI < 0.2)\cite{betancourt_diagnosing_2016}, or high values of the Gelman-Rubin convergence diagnostic ($\hat{R} > 1.01$)\cite{Gelman.1992}. The checks indicated sufficient effective sampling sizes and good convergence and mixing of chains for all fitted models.

\section{Comparison on synthetic line list data}\label{appendix:synthetic_data}
\subsection{Simulation}
To obtain a synthetic line list from a single simulation run, we first simulated a stochastic trajectory of infections according to a specified transmission scenario (first or second wave), and then sampled cases with date of symptom onset and reporting delay that are recorded in the line list. Finally, to simulate incomplete line list data, we randomly marked symptom onset dates as missing for some cases. The reporting delays and probabilities of missingness were simulated to change randomly over time and differ across simulation runs. In the following, we provide details for each step of the simulation.

\subsubsection{Simulation of infections}\label{appendix:sim_infections}
Infections were simulated by applying a stochastic renewal model to a given transmission scenario. A transmission scenario in our simulation is characterized by a piecewise linear time series for $R_t$ and a seeding rate of initial infections.
The first wave scenario starts with a seeding rate $\mu_{\text{seed}}=0.5$ and $R_t=2$. Transmission stays at $R_t=2$ until day 70, then drops linearly to $R_t = 0.8$ within 10 days, then stays at $R_t = 0.8$ for the remaining time until day 200. The second wave scenario starts with a much higher seeding rate of $\mu_{\text{seed}}=1000$ and equilibrium-level transmission, \ie $R_t=1$. Here the seeding does not necessarily represent imported cases, it is only used to initialize the simulation of a second wave during ongoing transmission at the epidemic threshold. Transmission stays at $R_t=1$ until day 70, then linearly increases to $R_t=1.4$ within 10 days, remains at $R_t=1.4$ for 20 days, and finally decreases to $R_t=0.7$ within 30 days and remains at $R_t=0.7$ for the remaining time until day 200.

For all simulations, we used a generation interval distribution $w$ that follows a discretized Gamma distribution with shape 1.43 and rate 0.29 (\ie mean 4.9 and standard deviation 4.1 days)\cite{hartInferenceSARSCoV2Generation2022} and is truncated at a maximum generation time of 21 days. To obtain a simulated trajectory of infections, we sampled initial infections during the first 3 weeks as independently Poisson distributed with rate  $\mu_{\text{seed}}$ on each day. Then, we recursively applied the renewal equation $$ \iota_t = E[I_t] = R_t \sum_{s=1}^t \psi_s \, I_{t-s},$$ while sampling realized infections $I_t|\iota_t \sim \text{Poisson}(\iota_t)$ at each time step, which can be interpreted as realizing an age-dependent branching process\cite{bhatt_semi-mechanistic_2020}. The final trajectory of infections was then used to simulate a synthetic line list.

\subsubsection{Simulation of line list}\label{appendix:sim_linelist}
Not all infections become reported cases. To mimic the generation of hospitalization line list data, we sampled infections only with a probability of 2\% as cases in the line list. For all simulations, we assumed the incubation period to be independent of the generation interval and Gamma distributed with a mean of 5.3 days and a standard deviation of 3.2 days\cite{linton_incubation_2020}. We drew a random incubation period for each case to simulate its date of symptom onset.

To obtain the corresponding date of report, we drew a reporting delay from a discrete delay distribution. As the baseline delay, we used a discretized lognormal distribution with a mean of 9 days and a standard deviation of 8 days. To simulate differences in the delay distribution over time, we applied the same discrete time-to-event model as used in the nowcasting approaches. Specifically, we determined $\gamma_d$, the logit baseline hazard for reporting with $d$ days delay from the above assumed baseline delay distribution. Then, for symptom onset on date $t$ and reporting delay $d$, we added a proportional-hazards trend effect $\Tilde{z}_t$ and a reporting weekday effect $\Tilde{w}_{t+d}$ to the logit baseline hazard $\gamma_d$, such that the overall hazard is $\text{logit}(h_{t,d}) = \gamma_d + \Tilde{z}_t + \Tilde{w}_{t+d}$. The trend effect $\Tilde{z}_t$ was simulated as a random piecewise linear function with changepoints every 4 weeks, where the values at the changepoints are drawn from a geometric random walk constrained between $\pm0.3$, corresponding to a roughly 30\% increase or decrease in the hazard odds of reporting. The reporting weekday effect $\Tilde{w}_{t+d}$ was chosen such that the hazard odds of reporting are 70\% lower on Saturdays, 80\% lower on Sundays, 10\% higher on Mondays and 5\% higher on Tuesdays. 

To simulate incomplete line list data with missing symptom onset dates, we selected cases with symptom onset on date $t$ to have a missing onset date with probability $1-\Tilde{\alpha}_t$, and let $1-\Tilde{\alpha}_t$ vary according to a random piecewise linear function with changepoints every 4 weeks, where the values at the changepoints were drawn from a geometric random walk constrained between 20\% and 60\% missingness.

\begin{figure}[!tbh]
    \centering
    \includegraphics[width=\textwidth]{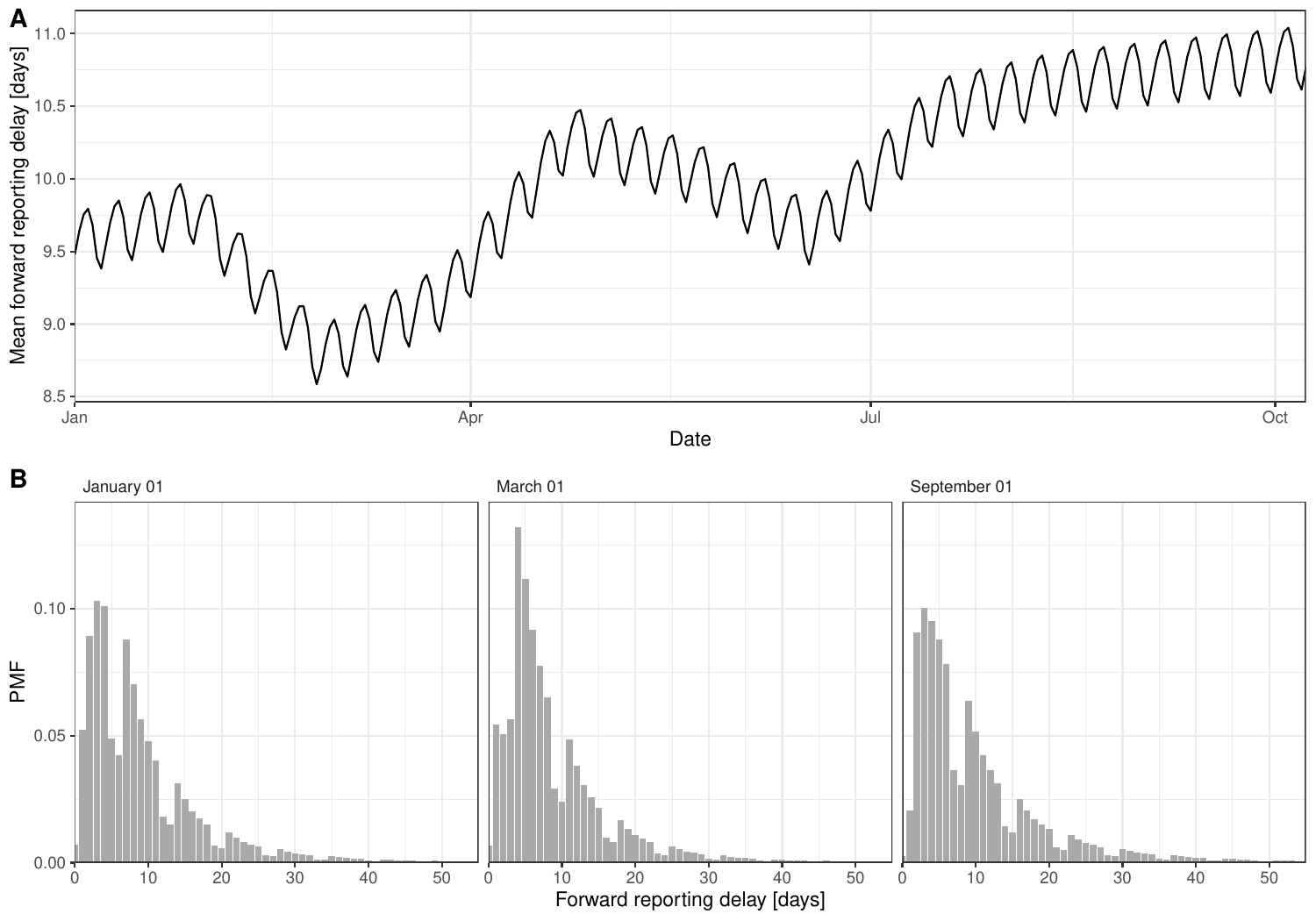}
    \caption{\textbf{Example of a simulated, time-varying delay distribution.} (A) Simulated mean forward reporting delay over time, including both weekly seasonality and a linear time trend with changepoints every 4 weeks. (B) Simulated reporting delay distributions shown at selected points in time. The smaller hazard for reporting on weekends shows as a pattern in the forward delay distribution.}
    \label{fig:sim_delay_dists}
\end{figure}

\begin{figure}[!tbh]
    \centering
    \includegraphics[width=\textwidth]{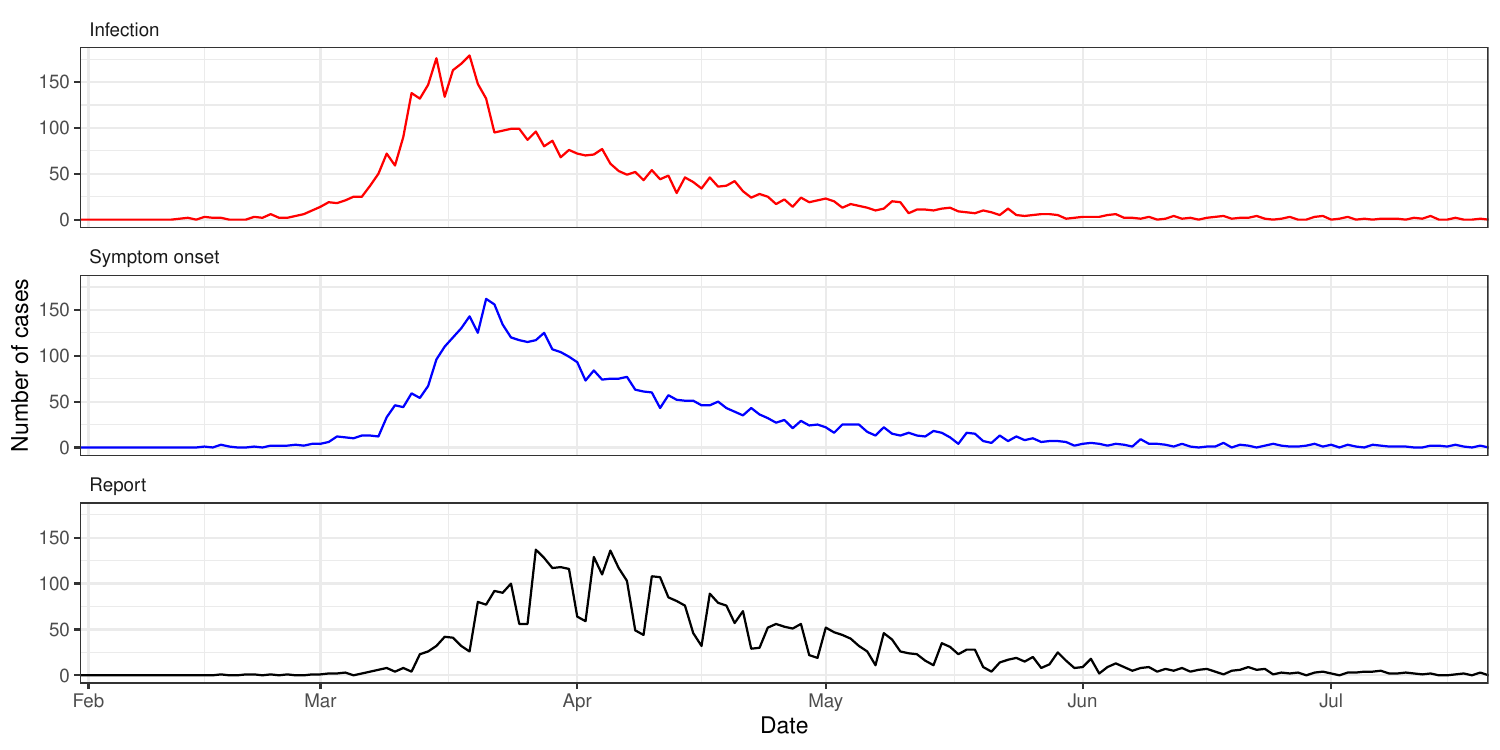}
    \caption{\textbf{Example of a simulated trajectory for the first wave scenario.} Shown is the number of eventually reported cases by date of infection (red), date of symptom onset (blue), and date of report (black). The first wave scenario starts with a small number of infections and exponential growth, and then transitions into exponential decline after 70 days. Differences in reporting delay, \eg reduced reporting on weekends, only affect the time series of cases by date of report.}
    \label{fig:sim_wave1}
\end{figure}

\begin{figure}[!tbh]
    \centering
    \includegraphics[width=\textwidth]{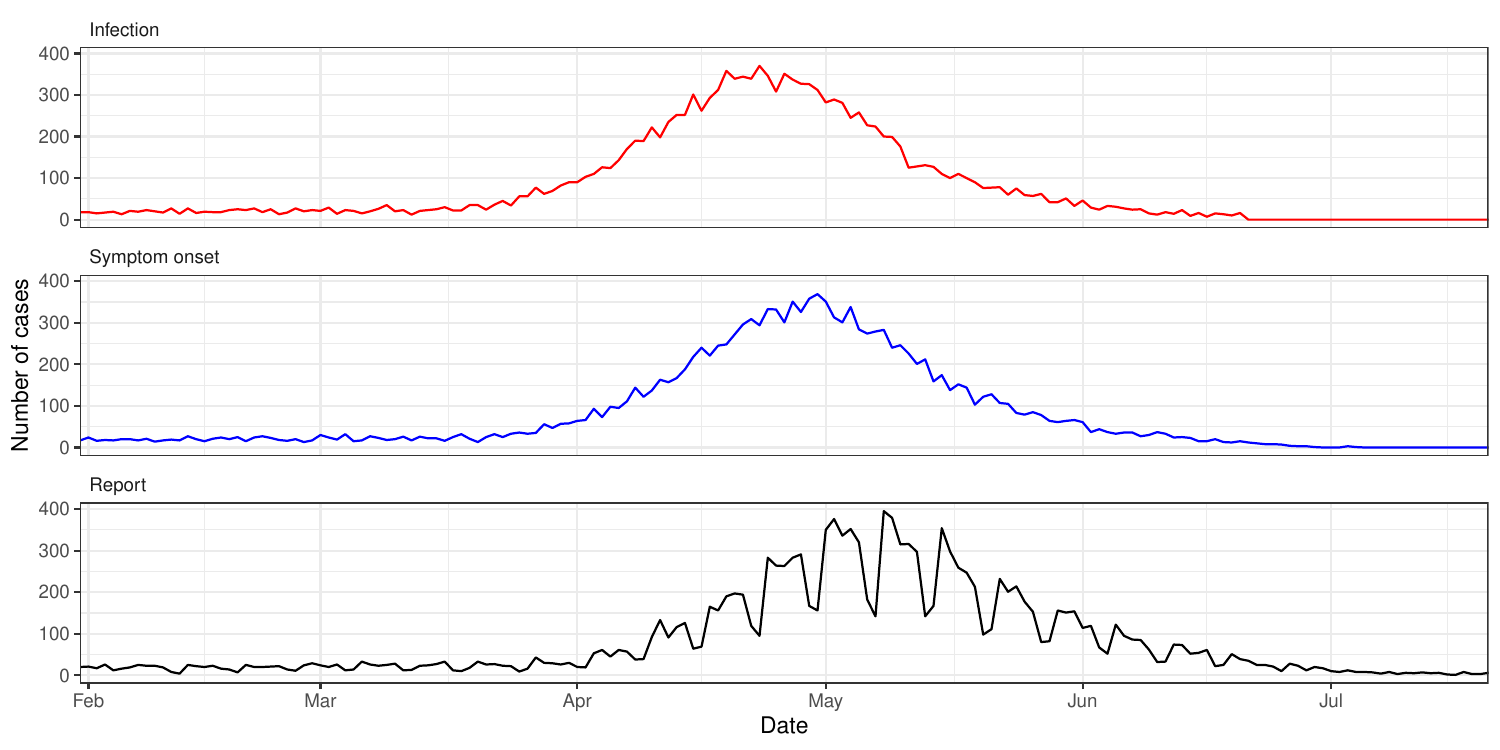}
    \caption{\textbf{Example of a simulated trajectory for the second wave scenario.} Shown is the number of eventually reported cases by date of infection (red), date of symptom onset (blue), and date of report (black). The second wave scenario starts with a higher number of infections and equilibrium-level transmission, which increases to exponential growth after 70 days, and finally transitions into exponential decline. Differences in reporting delay, \eg reduced reporting on weekends, only affect the time series of cases by date of report.}
    \label{fig:sim_wave2}
\end{figure}

\subsection{Performance evaluation}\label{appendix:performance_eval}
We produced 50 different simulation runs for the first wave and second wave scenario, respectively. For each scenario, we then used the different nowcasting approaches on each of the 50 synthetic line lists to obtain nowcasts for different evaluation time periods. These week-long time periods were chosen to cover different phases of the epidemic wave in each scenario. For the first wave scenario, the evaluation time periods were days 64--70 (before peak I), 77--83 (at peak I), 98--104 (after peak I), and 129--135 (suppression). For the second wave scenario, the evaluation time periods were days 64--70 (control), 94--100 (before peak II), 118--124 (at peak II), 142--148 (After peak II). For each of these 7-day periods, we applied the different nowcasting approaches using data observed until the end of the same week (0--6 days lag), one week later (7--13 days lag), and two weeks later (14--20 days lag).

For nowcasting, we used the generation interval and incubation period distributions assumed in the simulation, \ie we assumed these parameters to be correctly known. For the comparison of the different approaches of estimating $R_t$, we used complete synthetic line list data, i.e. no symptom onsets were set to missing. For the comparison of the different approaches of accounting for missing onset dates, we used incomplete synthetic line list data, i.e. missing symptom onsets were simulated as described in \Cref{appendix:sim_linelist}.

For each scenario, the above procedure yielded three differently lagged nowcasts for each approach, for each epidemic phase, and for each of the 50 simulation runs. To evaluate the nowcasting performance of the different approaches, we compared each nowcast with the simulated ground truth from the respective synthetic line list. We assessed performance with regard to nowcasting the number of symptom onsets $N_t$ and nowcasting the effective reproduction number $R_t$, using the weighted interval score (WIS) as described in the main text.

\subsection{Supplementary figures for results on synthetic data}
\begin{figure}[!tbh]
    \vspace{-1.6cm}
    \centering
    \includegraphics[width=\textwidth]{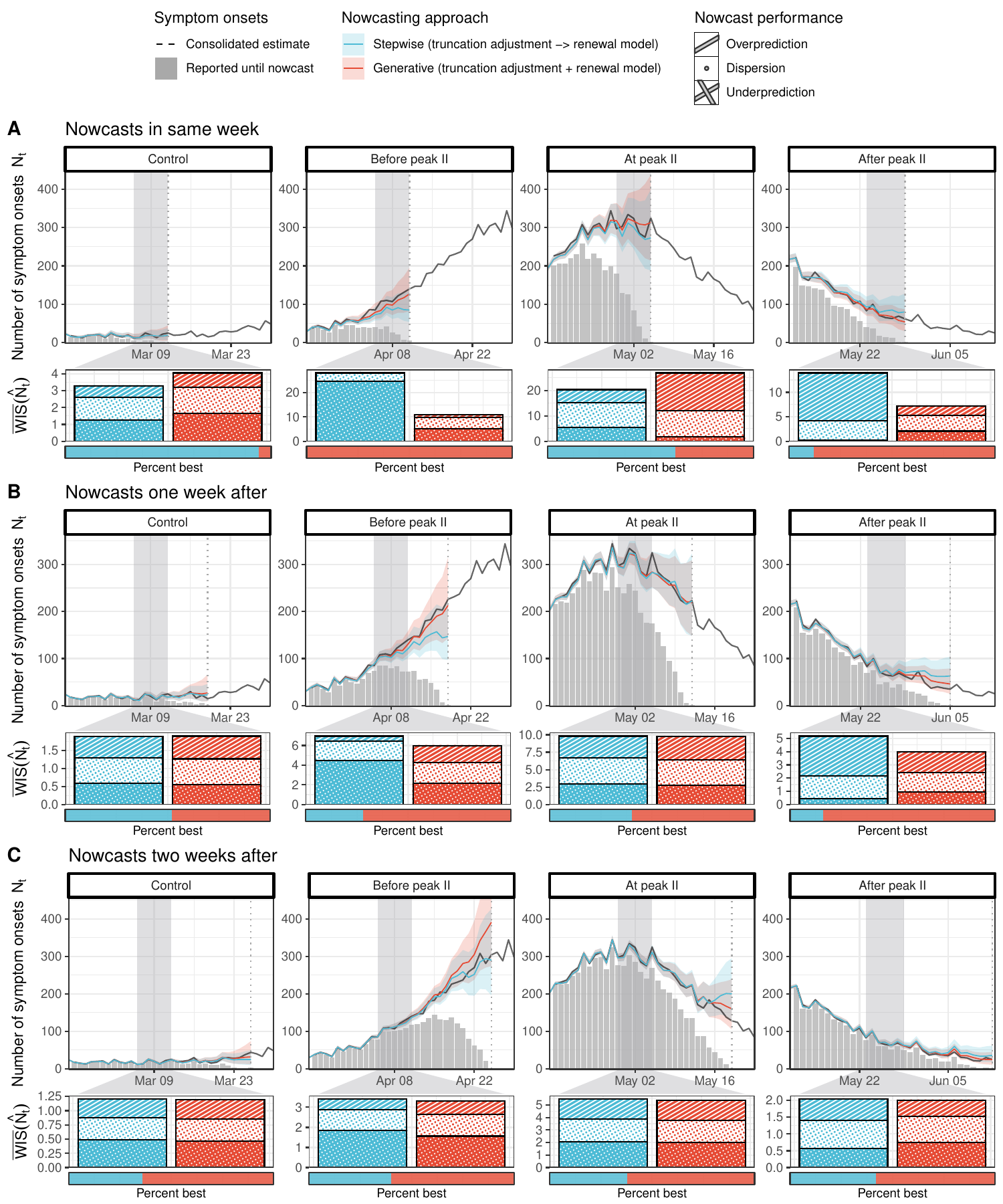}
    \caption{\textbf{Nowcasts of $N_t$ on line list data of a simulated second wave scenario using different approaches of adjusting for right truncation.} Shown are the true number of cases by symptom onset date $N_t$ (black), the number of cases reported until the nowcast date (grey bars), and point nowcasts with 95\% credible intervals (CrI) in four different phases of the epidemic wave, obtained through i) a stepwise approach using cases by date of symptom onset with a truncation adjustment step (blue), and ii) a generative approach using cases by date of symptom onset with an integrated truncation and renewal model (red). The direct approach using cases by date of report does not produce nowcasts of $N_t$ (not shown). Shown below each phase is the weighted interval score (WIS, lower is better) for $N_t$ nowcasts of each approach during a selected week (grey shade) over 50 scenario runs. Colored bars show average scores, decomposed into penalties for underprediction (crosshatch), dispersion (circles), and overprediction (stripes). The horizontal proportion bar below shows the percentages of runs where each approach performed best. Results are shown for nowcasts made at different lags from the selected week (vertical dotted lines), \ie at the end of the selected week (top row), one week later (middle row), and two weeks later (bottom row).}
    \label{fig:wave2_feather_renewal_Nt}
\end{figure}

\begin{figure}[!tbh]
    \vspace{-1.6cm}
    \centering
    \includegraphics[width=\textwidth]{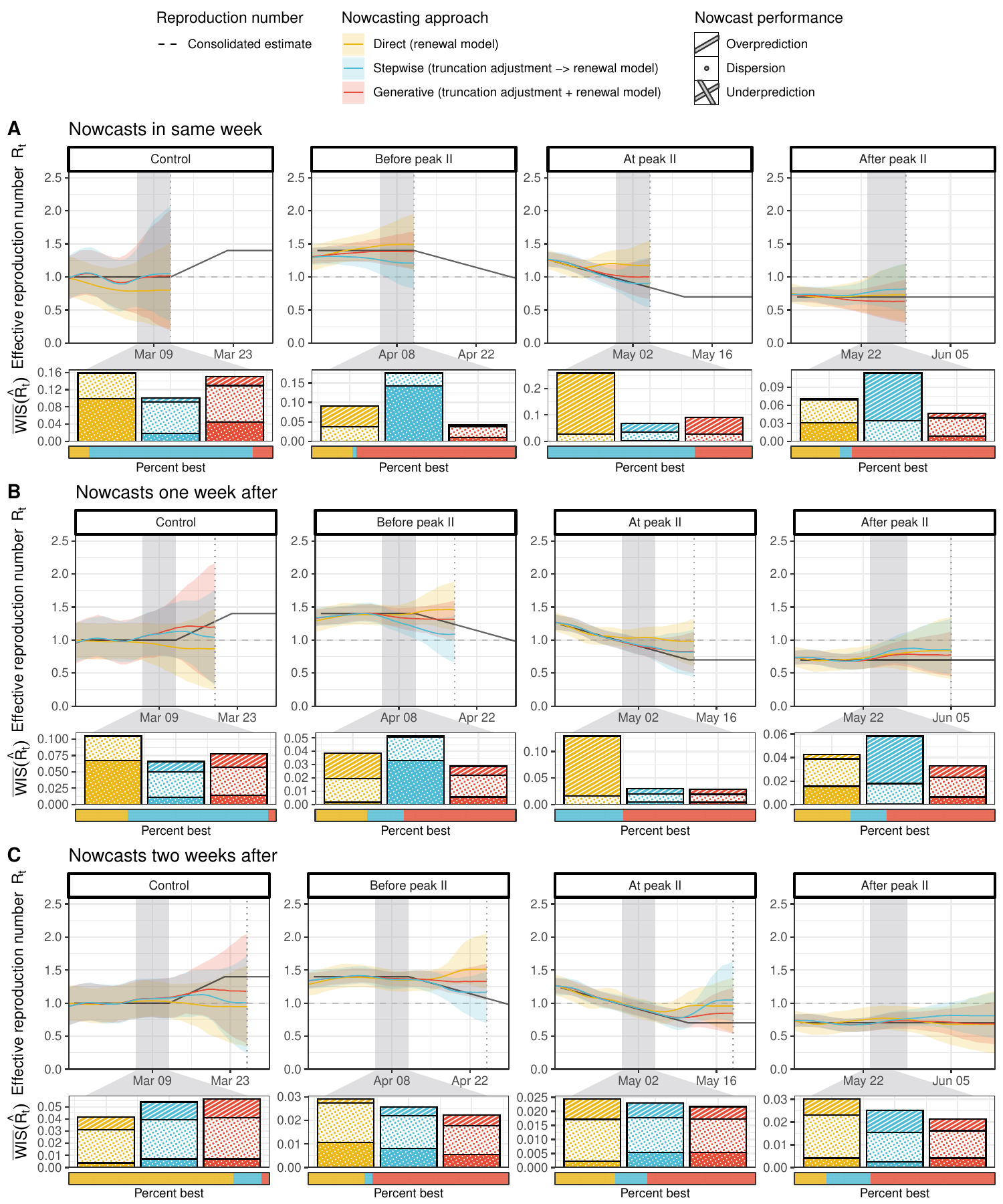}
    \caption{\textbf{Nowcasts of $R_t$ on line list data of a simulated second wave scenario using different approaches of adjusting for right truncation.} Shown are the true $R_t$ (black) and point nowcasts with 95\% credible intervals (CrI) in four different phases of the epidemic wave, obtained through i) a direct approach using cases by date of report with no truncation adjustment (yellow), ii) a stepwise approach using cases by date of symptom onset with a truncation adjustment step (blue), and iii) a generative approach using cases by date of symptom onset with an integrated truncation and renewal model (red). Shown below each phase is the weighted interval score (WIS, lower is better) for $R_t$ nowcasts of each approach during a selected week (grey shade) over 50 scenario runs. Colored bars show average scores, decomposed into penalties for underprediction (crosshatch), dispersion (circles), and overprediction (stripes). The horizontal proportion bar below shows the percentages of runs where each approach performed best. Results are shown for nowcasts made at different lags from the selected week (vertical dotted lines), \ie at the end of the selected week (top row), one week later (middle row), and two weeks later (bottom row).}
    \label{fig:wave2_feather_renewal_Rt}
\end{figure}

\begin{figure}[!tbh]
    \vspace{-1.6cm}
    \centering
    \includegraphics[width=\textwidth]{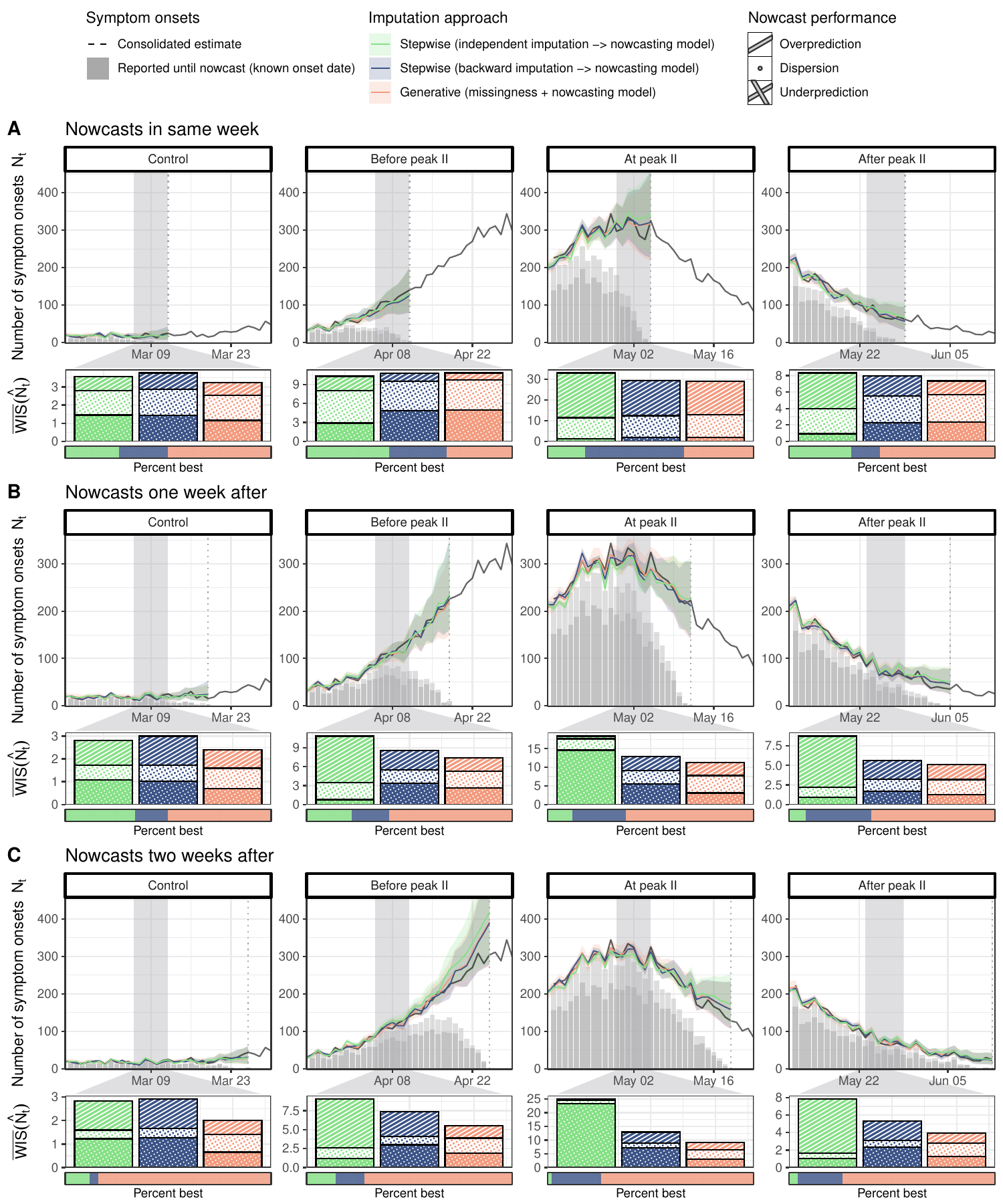}
    \caption{\textbf{Nowcasts of $N_t$ on incomplete line list data of a simulated second wave scenario using different approaches of accounting for missing onset dates.} Shown are the true number of cases by symptom onset date $N_t$ (black), the number of cases reported until the nowcast date (dark grey bars for known, light grey bars for missing onset dates), and point nowcasts with 95\% credible intervals (CrI) in four different phases of the epidemic wave, obtained through i) a stepwise approach using a forward imputation step (green), ii) a stepwise approach using a backward imputation step (blue), and iii) a generative approach using an integrated missingness model (red). All approaches used a generative model for nowcasting. Shown below each phase is the weighted interval score (WIS, lower is better) for $N_t$ nowcasts of each approach during a selected week (grey shade) over 50 scenario runs. Colored bars show average scores, decomposed into penalties for underprediction (crosshatch), dispersion (circles), and overprediction (stripes). The horizontal proportion bar below shows the percentages of runs where each approach performed best. Results are shown for nowcasts made at different lags from the selected week (vertical dotted lines), \ie at the end of the selected week (top row), one week later (middle row), and two weeks later (bottom row).}
    \label{fig:wave2_feather_miss_Nt}
\end{figure}

\begin{figure}[!tbh]
    \vspace{-1.6cm}
    \centering
    \includegraphics[width=\textwidth]{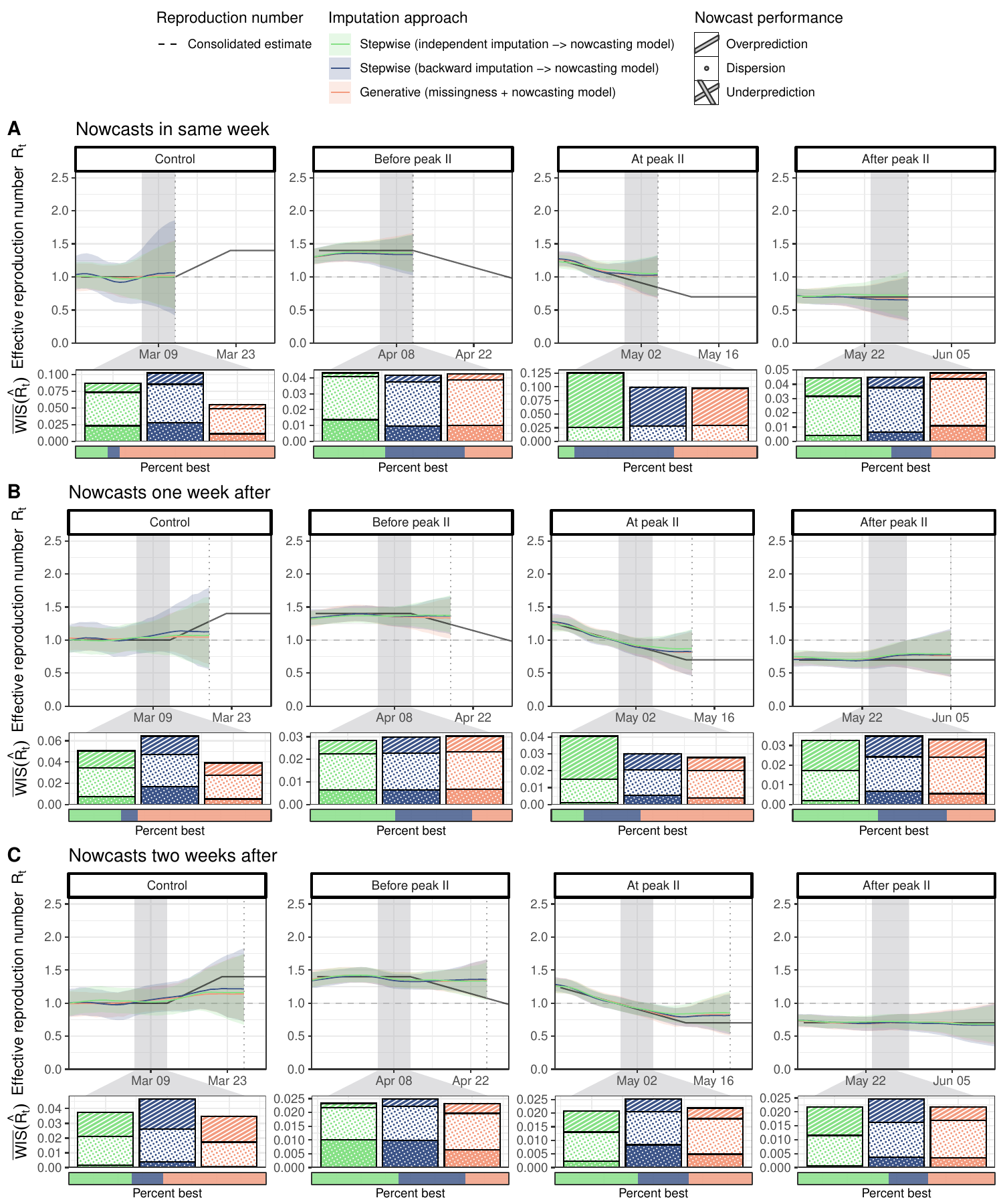}
    \caption{\textbf{Nowcasts of $R_t$ on incomplete line list data of a simulated second wave scenario using different approaches of accounting for missing onset dates.} Shown are the true $R_t$ (black) and point nowcasts with 95\% credible intervals (CrI) in four different phases of the epidemic wave, obtained through i) a stepwise approach using a forward imputation step (green), ii) a stepwise approach using a backward imputation step (blue), and iii) a generative approach using an integrated missingness model (red). All approaches used a generative model for nowcasting. Shown below each phase is the weighted interval score (WIS, lower is better) for $R_t$ nowcasts of each approach during a selected week (grey shade) over 50 scenario runs. Colored bars show average scores, decomposed into penalties for underprediction (crosshatch), dispersion (circles), and overprediction (stripes). The horizontal proportion bar below shows the percentages of runs where each approach performed best. Results are shown for nowcasts made at different lags from the selected week (vertical dotted lines), \ie at the end of the selected week (top row), one week later (middle row), and two weeks later (bottom row).}
    \label{fig:wave2_feather_miss_Rt}
\end{figure}

\begin{figure}[!tbh]
    \vspace{-1.6cm}
    \centering
    \includegraphics[width=\textwidth]{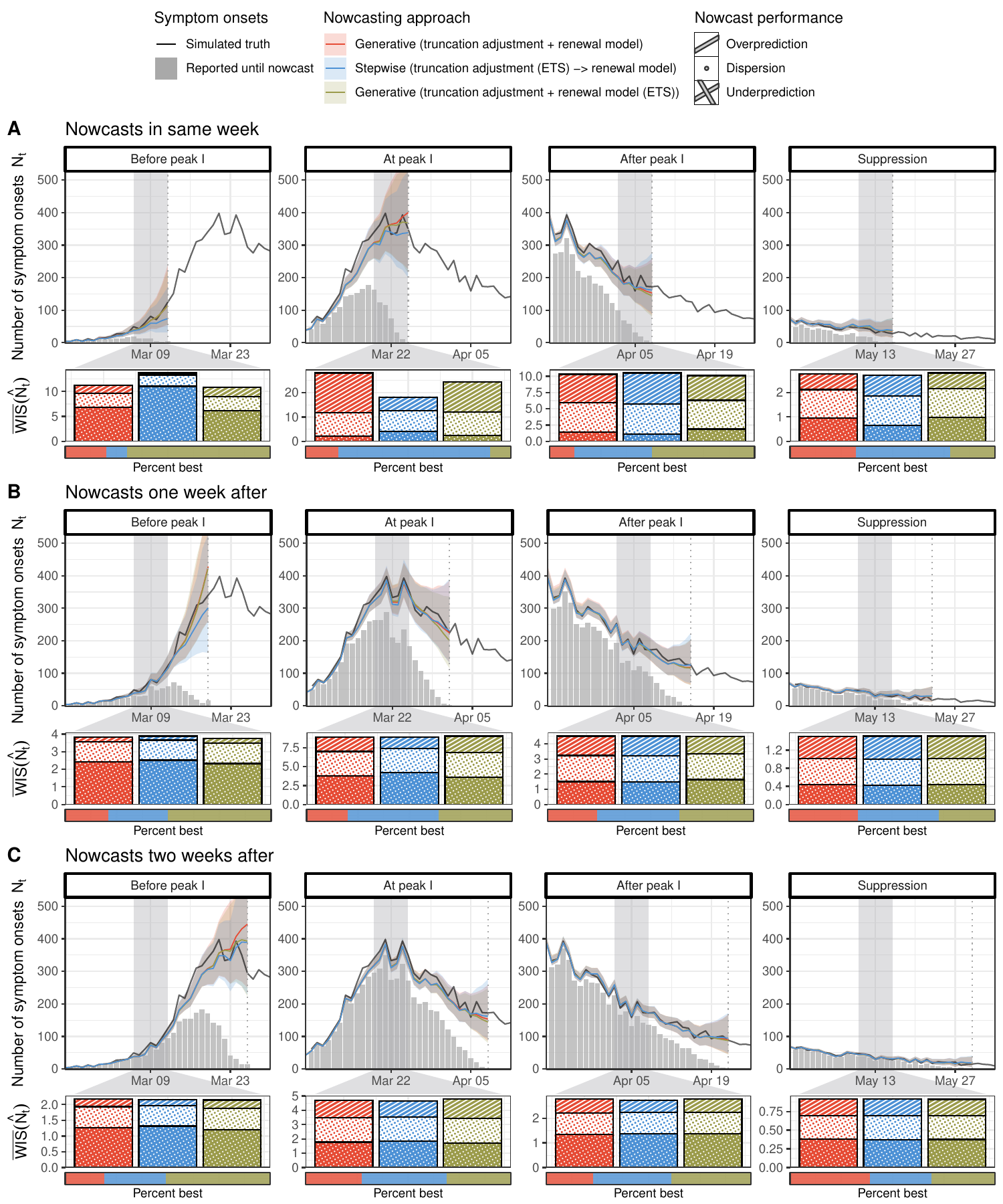}
    \caption{\textbf{Nowcasts of $N_t$ on line list data of a simulated first wave scenario using different approaches of adjusting for right truncation with stationary and non-stationary smoothing priors.} Shown are the true number of cases by symptom onset date $N_t$ (black), the number of cases reported until the nowcast date (grey bars), and point nowcasts with 95\% credible intervals (CrI) in four different phases of the epidemic wave, obtained through i) a generative approach with a stationary random walk prior on $R_t$ (red, same as main text) ii) a stepwise approach with a non-stationary exponential smoothing prior on $\lambda_t$ (blue), and iii) a generative approach with a non-stationary exponential smoothing prior on $R_t$ (beige). Shown below each phase is the weighted interval score (WIS, lower is better) for $N_t$ nowcasts of each approach during a selected week (grey shade) over 50 scenario runs. Colored bars show average scores, decomposed into penalties for underprediction (crosshatch), dispersion (circles), and overprediction (stripes). The horizontal proportion bar below shows the percentages of runs where each approach performed best. Results are shown for nowcasts made at different lags from the selected week (vertical dotted lines), \ie at the end of the selected week (top row), one week later (middle row), and two weeks later (bottom row).}
    \label{fig:wave1_feather_renewal_ets_Nt}
\end{figure}

\begin{figure}[!tbh]
    \vspace{-1.6cm}
    \centering
    \includegraphics[width=\textwidth]{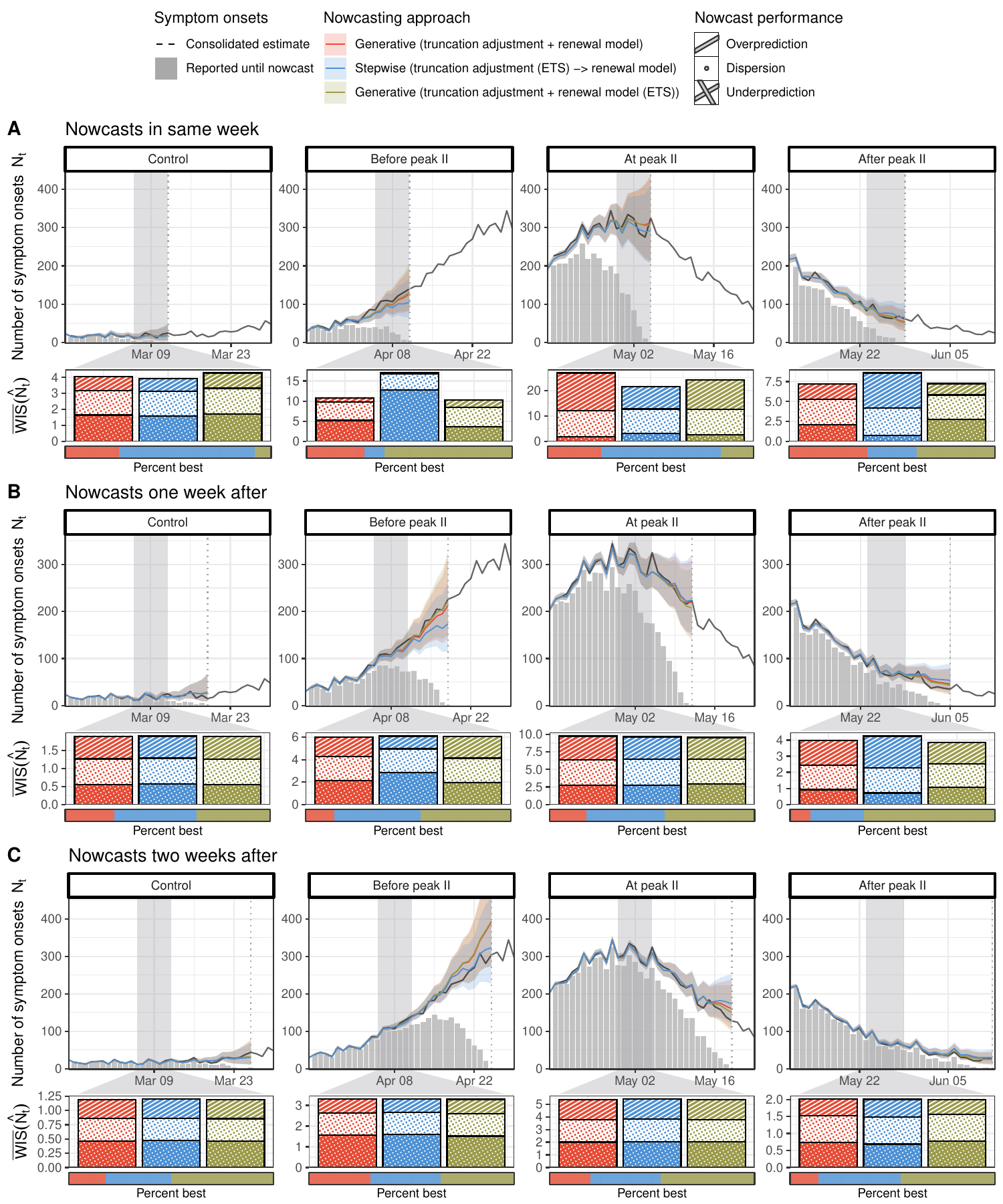}
    \caption{\textbf{Nowcasts of $N_t$ on line list data of a simulated second wave scenario using different approaches of adjusting for right truncation with stationary and non-stationary smoothing priors.} Shown are the true number of cases by symptom onset date $N_t$ (black), the number of cases reported until the nowcast date (grey bars), and point nowcasts with 95\% credible intervals (CrI) in four different phases of the epidemic wave, obtained through i) a generative approach with a stationary random walk prior on $R_t$ (red, same as main text) ii) a stepwise approach with a non-stationary exponential smoothing prior on $\lambda_t$ (blue), and iii) a generative approach with a non-stationary exponential smoothing prior on $R_t$ (beige). Shown below each phase is the weighted interval score (WIS, lower is better) for $N_t$ nowcasts of each approach during a selected week (grey shade) over 50 scenario runs. Colored bars show average scores, decomposed into penalties for underprediction (crosshatch), dispersion (circles), and overprediction (stripes). The horizontal proportion bar below shows the percentages of runs where each approach performed best. Results are shown for nowcasts made at different lags from the selected week (vertical dotted lines), \ie at the end of the selected week (top row), one week later (middle row), and two weeks later (bottom row).}
    \label{fig:wave2_feather_renewal_ets_Nt}
\end{figure}

\begin{figure}[!tbh]
    \vspace{-1.6cm}
    \centering
    \includegraphics[width=\textwidth]{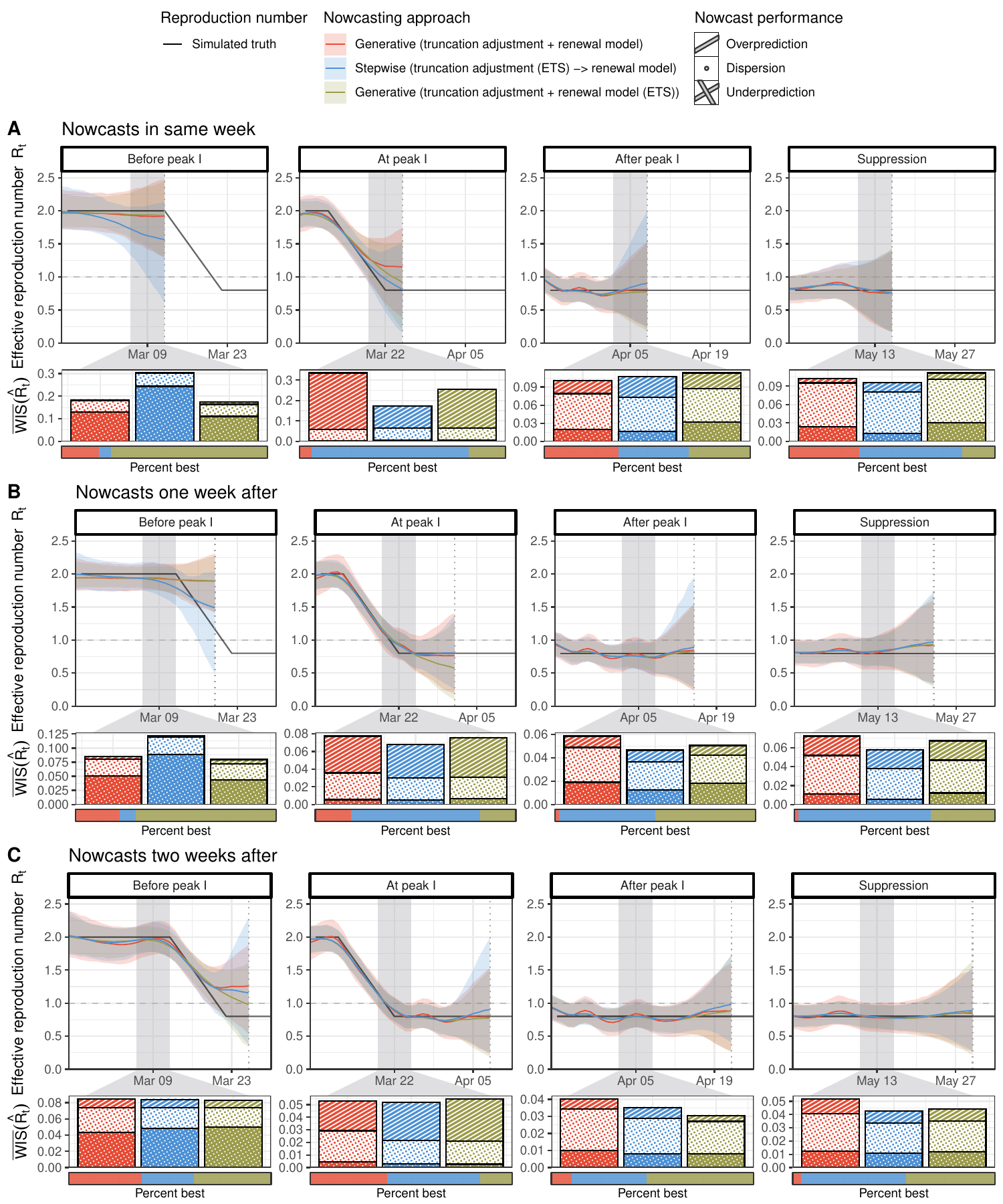}
    \caption{\textbf{Nowcasts of $R_t$ on line list data of a simulated first wave scenario using different approaches of adjusting for right truncation with stationary and non-stationary smoothing priors.} Shown are the true $R_t$ (black) and point nowcasts with 95\% credible intervals (CrI) in four different phases of the epidemic wave, obtained through i) a generative approach with a stationary random walk prior on $R_t$ (red, same as main text) ii) a stepwise approach with a non-stationary exponential smoothing prior on $\lambda_t$ (blue), and iii) a generative approach with a non-stationary exponential smoothing prior on $R_t$ (beige). Shown below each phase is the weighted interval score (WIS, lower is better) for $R_t$ nowcasts of each approach during a selected week (grey shade) over 50 scenario runs. Colored bars show average scores, decomposed into penalties for underprediction (crosshatch), dispersion (circles), and overprediction (stripes). The horizontal proportion bar below shows the percentages of runs where each approach performed best. Results are shown for nowcasts made at different lags from the selected week (vertical dotted lines), \ie at the end of the selected week (top row), one week later (middle row), and two weeks later (bottom row).}
    \label{fig:wave1_feather_renewal_ets_Rt}
\end{figure}

\begin{figure}[!tbh]
    \vspace{-1.6cm}
    \centering
    \includegraphics[width=\textwidth]{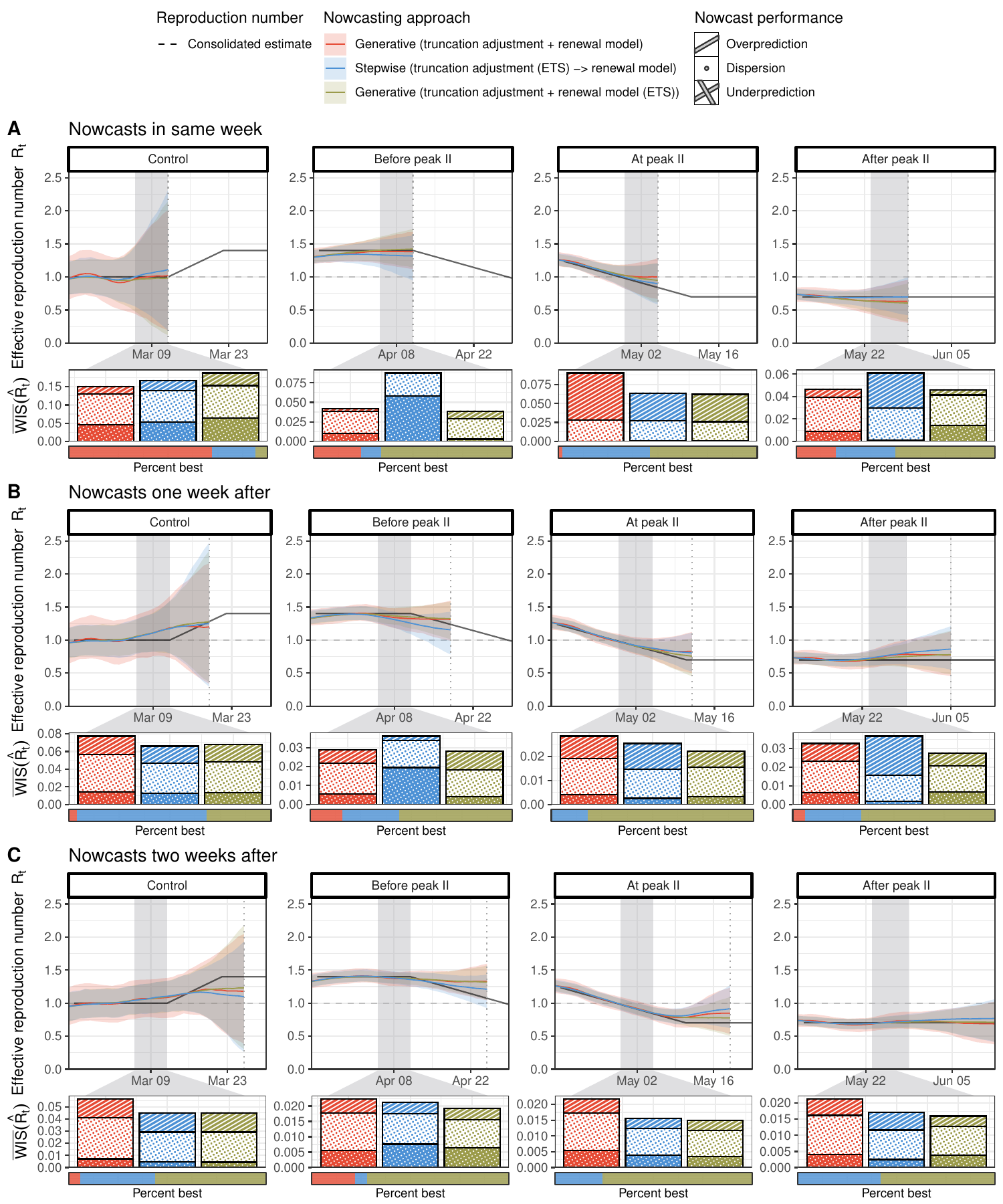}
    \caption{\textbf{Nowcasts of $R_t$ on line list data of a simulated second wave scenario using different approaches of adjusting for right truncation with stationary and non-stationary smoothing priors.} Shown are the true $R_t$ (black) and point nowcasts with 95\% credible intervals (CrI) in four different phases of the epidemic wave, obtained through i) a generative approach with a stationary random walk prior on $R_t$ (red, same as main text) ii) a stepwise approach with a non-stationary exponential smoothing prior on $\lambda_t$ (blue), and iii) a generative approach with a non-stationary exponential smoothing prior on $R_t$ (beige). Shown below each phase is the weighted interval score (WIS, lower is better) for $R_t$ nowcasts of each approach during a selected week (grey shade) over 50 scenario runs. Colored bars show average scores, decomposed into penalties for underprediction (crosshatch), dispersion (circles), and overprediction (stripes). The horizontal proportion bar below shows the percentages of runs where each approach performed best. Results are shown for nowcasts made at different lags from the selected week (vertical dotted lines), \ie at the end of the selected week (top row), one week later (middle row), and two weeks later (bottom row).}
    \label{fig:wave2_feather_renewal_ets_Rt}
\end{figure}

\begin{figure}[!tbh]
    \vspace{-1.6cm}
    \centering
    \includegraphics[width=\textwidth]{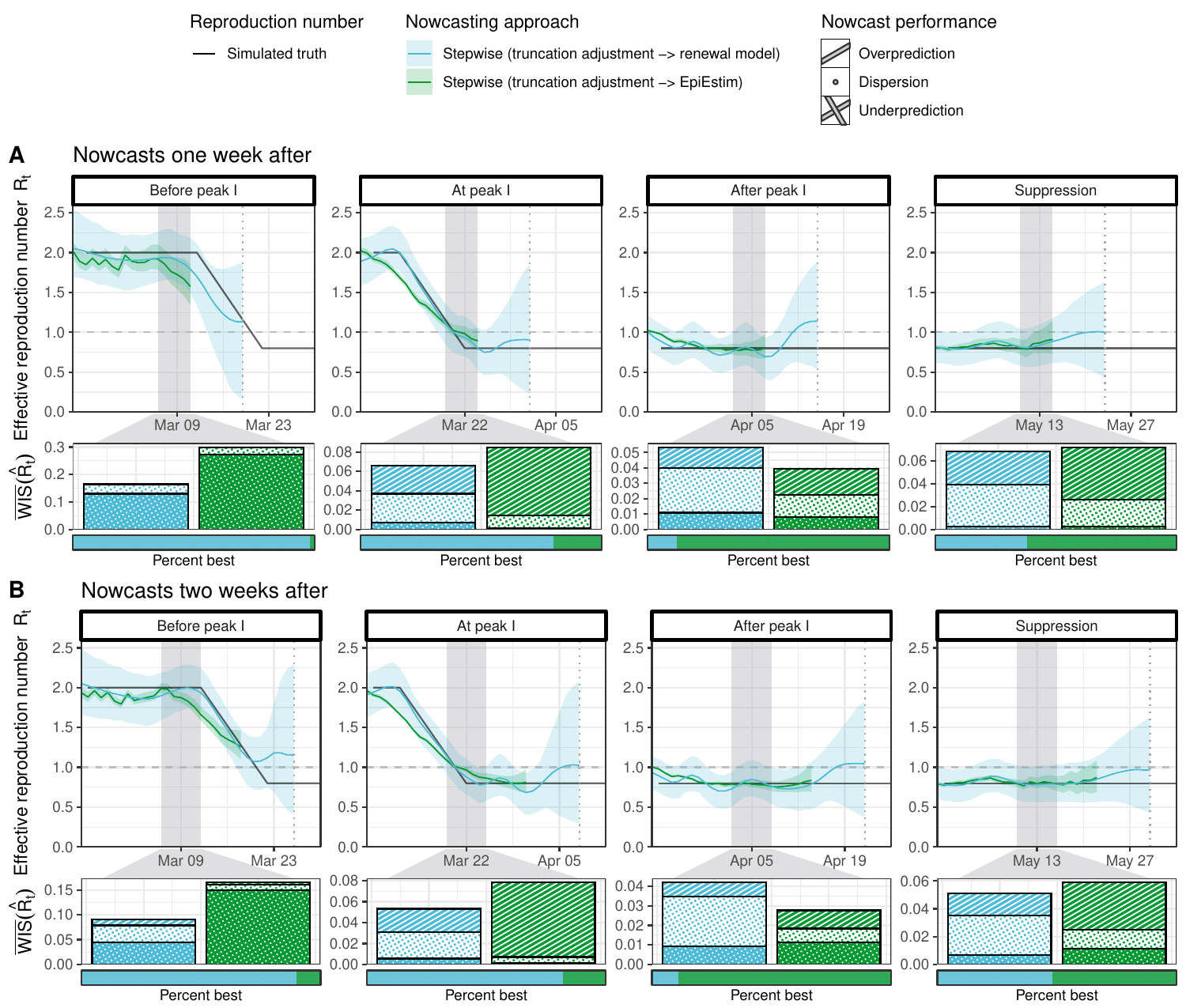}
    \caption{\textbf{Nowcasts of $R_t$ on line list data of a simulated first wave scenario using a stepwise approach with a renewal model and EpiEstim for $R_t$ estimation.} Shown are the true $R_t$ (black) and point nowcasts with 95\% credible intervals (CrI) in four different phases of the epidemic wave, obtained through i) a stepwise approach using a renewal model for $R_t$ estimation (blue), and ii) a stepwise approach using EpiEstim with a smoothing window of 7 days for $R_t$ estimation (green). Shown below each phase is the weighted interval score (WIS, lower is better) for $R_t$ nowcasts of each approach during a selected week (grey shade) over 50 scenario runs. Colored bars show average scores, decomposed into penalties for underprediction (crosshatch), dispersion (circles), and overprediction (stripes). The horizontal proportion bar below shows the percentages of runs where each approach performed best. Results are shown for nowcasts made at different lags from the selected week (vertical dotted lines), \ie one week later (middle row), and two weeks later (bottom row). Nowcasts using EpiEstim were shifted backward in time to account for the incubation period and to center the smoothing window of EpiEstim, and are therefore only available at a lag of 8 days or longer.}
    \label{fig:wave1_feather_epiestim_Rt}
\end{figure}

\begin{figure}[!tbh]
    \vspace{-1.6cm}
    \centering
    \includegraphics[width=\textwidth]{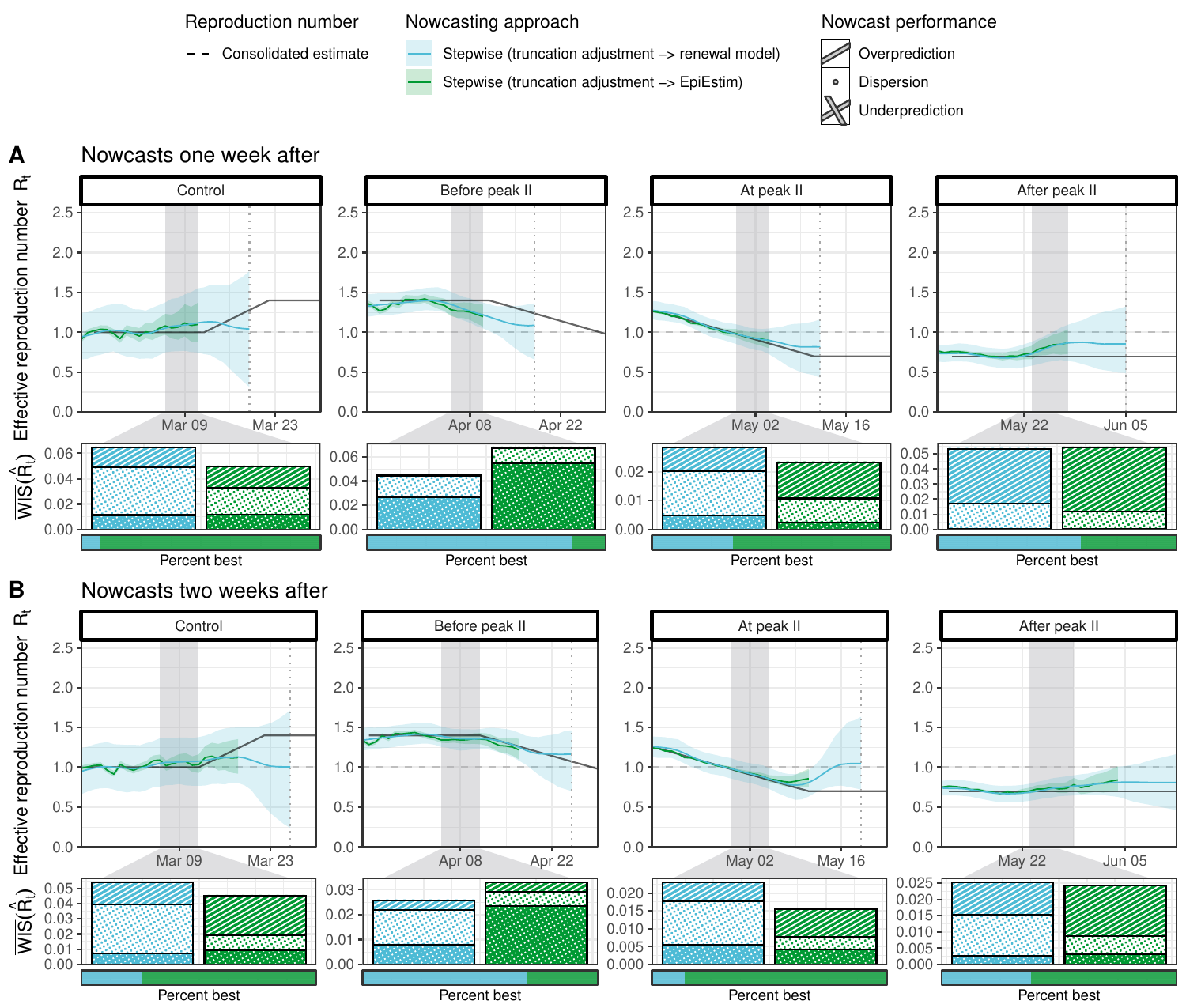}
    \caption{\textbf{Nowcasts of $R_t$ on line list data of a simulated second wave scenario using a stepwise approach with a renewal model and EpiEstim for $R_t$ estimation.} Shown are the true $R_t$ (black) and point nowcasts with 95\% credible intervals (CrI) in four different phases of the epidemic wave, obtained through i) a stepwise approach using a renewal model for $R_t$ estimation (blue), and ii) a stepwise approach using EpiEstim with a smoothing window of 7 days for $R_t$ estimation (green). Shown below each phase is the weighted interval score (WIS, lower is better) for $R_t$ nowcasts of each approach during a selected week (grey shade) over 50 scenario runs. Colored bars show average scores, decomposed into penalties for underprediction (crosshatch), dispersion (circles), and overprediction (stripes). The horizontal proportion bar below shows the percentages of runs where each approach performed best. Results are shown for nowcasts made at different lags from the selected week (vertical dotted lines), \ie at the end of the selected week (top row), one week later (middle row), and two weeks later (bottom row).}
    \label{fig:wave2_feather_epiestim_Rt}
\end{figure}

In the following, we provide supplementary figures for the results on synthetic data. These include the results of the main analyses for the second wave scenario (Figures~\ref{fig:wave2_feather_renewal_Nt} -- \ref{fig:wave2_feather_miss_Rt}), results for models with exponential smoothing priors (Figures~\ref{fig:wave1_feather_renewal_ets_Nt} -- \ref{fig:wave2_feather_renewal_ets_Rt}), and results for the stepwise approach using EpiEstim (\Cref{fig:wave1_feather_epiestim_Rt,fig:wave2_feather_epiestim_Rt}).

% Results for complete line list data (Rt estimation and truncation adjustment), second wave scenario

% Results for incomplete line list data (missing onset date imputation), second wave scenario

%Results for models with exponential smoothing priors

% Results for the stepwise approach using EpiEstim}

\FloatBarrier
\section{Application to COVID-19 in Switzerland}

\subsection{Hospitalization line list data}
\begin{figure}[!tbh]
    \centering
    \includegraphics[width=\textwidth]{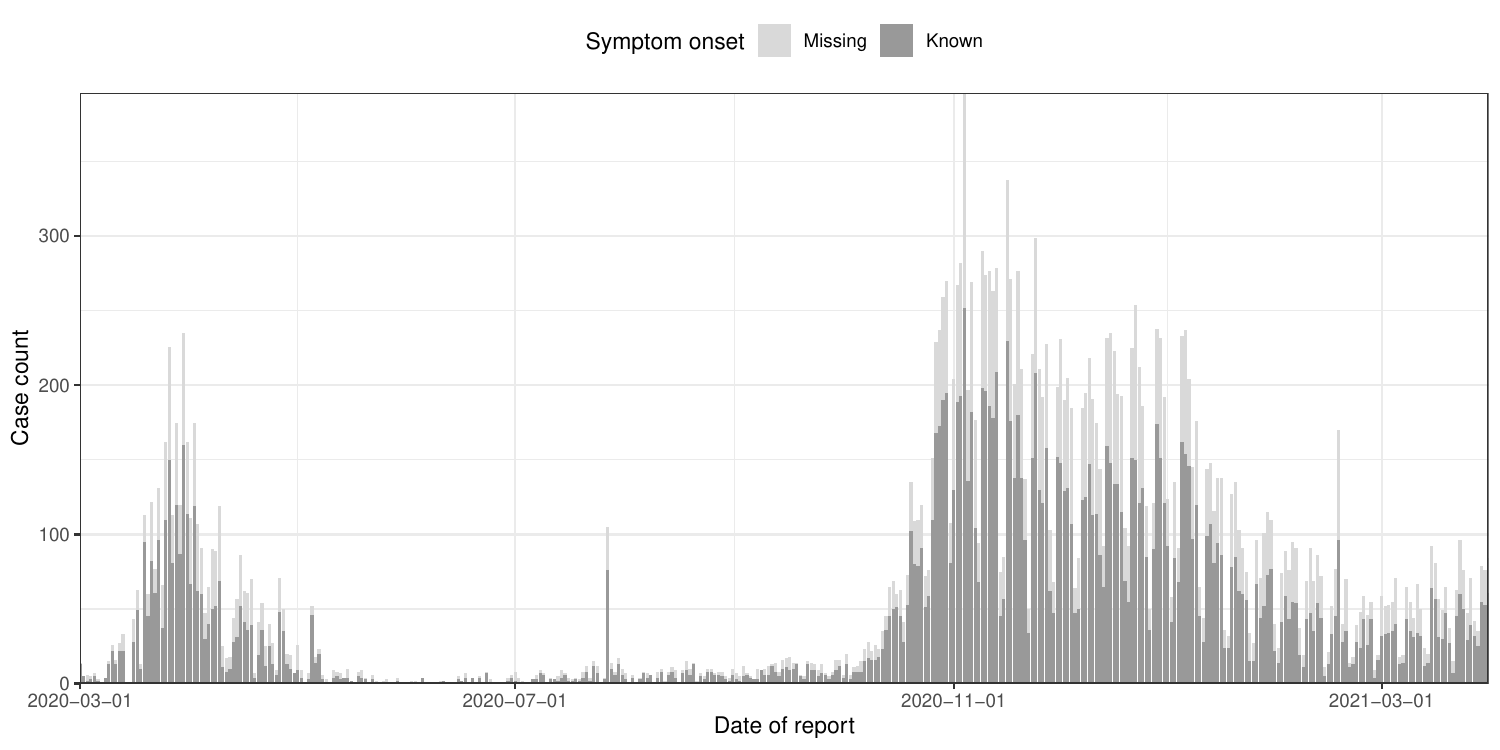}
    \caption{\textbf{Cases of hospitalization with COVID-19 in Switzerland, March~1, 2020~--~March~31, 2021.} Shown is the number of hospitalized patients who tested positive for COVID-19 by date of report of the hospitalization, during the first and second wave of COVID-19 in Switzerland. The case counts are disaggregated into cases with known (dark grey bars) and missing (light grey bars) symptom onset date.}
    \label{fig:switzerland_date_of_report}
\end{figure}

\begin{figure}[!tbh]
    \centering
    \includegraphics[width=\textwidth]{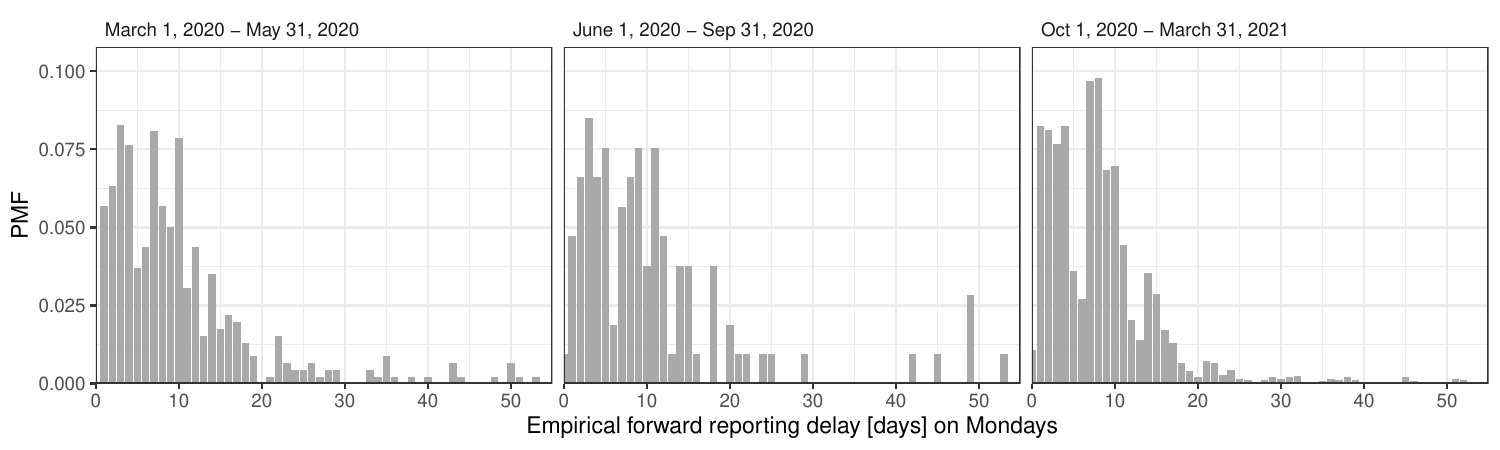}
    \caption{\textbf{Empirical reporting delay distribution of cases hospitalized with COVID-19 in Switzerland, March~1, 2020~--~March~31, 2021.} Shown is the empirical distribution of the forward delay from symptom onset to report of hospitalization during different time periods of the first and second wave of COVID-19 in Switzerland. To highlight the patterns of weekly seasonality, only cases with symptom onset on a Monday were included.}
    \label{fig:switzerland_emp_delay}
\end{figure}

Hospitalization line list data of patients who tested positive for COVID-19 was provided in anonymized form by the Swiss Federal Office of Public Health (FOPH). Cases were stratified by date of symptom onset, date of hospitalization, and date of report of the hospitalization to FOPH. When nowcasting, it is essential to use the date when a case effectively became available for analysis, which was the date of report of the hospitalization in our case. If we instead used the date of hospitalization as report date, any right truncation resulting from the delay between hospitalization and reporting of the hospitalization would be neglected, and the nowcasts would still be downward biased. The modeled delays in this study therefore implicitly included both the delay between symptom onset and hospitalization and the delay between hospitalization and reporting of the hospitalization. The date of hospitalization itself was not explicitly modeled. \Cref{fig:switzerland_date_of_report} shows the number of cases over time by date of report of the hospitalization, stratified by cases with known and with missing symptom onset date. The weekly moving average percentage of cases with missing symptom onset date in the line list varied between 16\% and 63\%. \Cref{fig:switzerland_emp_delay} shows empirical delays between symptom onset and report of hospitalization in different time periods of the COVID-19 epidemic in Switzerland. The assumed maximum delay of 56 days covered 97.03\% of cases. Of these, the mean (standard deviation) of delays was 10.12 (9.23) days from March 1, 2020~--~May 31, 2020, 11.05 (10.91) days from June 1, 2020~--~Sep 31, 2020, and 8.19 (6.85) days from Oct 1, 2020~--~March 31, 2021, respectively.

\subsection{Time-varying epidemiological parameters}\label{appendix:epi_params}
To account for the introduction and spread of the alpha variant of SARS-CoV-2 in Switzerland, we used different incubation period and generation interval distributions for the nowcasting models over time (parameters as provided in main text). All distributions were discretized and truncated at 21 days. To model the transition from introduction to dominance (\ie over 90\% of total sequences per week) of the variant B.1.1.7 (20I) alpha, we used the time period from December 21, 2020~--~March 29, 2021\cite{hodcroft_covariants_2023}. During the transition phase, we used a mixture of the distribution for the wildtype variant and the distribution for the alpha variant, increasing the weight for the alpha variant from 0 to 1 according to a logistic function. Note that for simplicity, we did not include the time-varying incubation period and generation interval distributions explicitly in the renewal model but simply supplied different parameters to the nowcasting models over time. 

% ------------------
% Bibliography
\newpage
\bibliography{references}